\pdfoutput=1
\newcommand*{\ATLASLATEXPATH}{./}
\documentclass[cernpreprint,texlive=2011,txfonts,subfigure=true,maketitle=true,UKenglish]{\ATLASLATEXPATH atlasdoc}
\usepackage{\ATLASLATEXPATH atlaspackage}
\usepackage{\ATLASLATEXPATH atlasphysics} 
\usepackage{\ATLASLATEXPATH atlascover}
\usepackage{\ATLASLATEXPATH atlasbiblatex}

\usepackage{mathrsfs}
\usepackage{graphicx}
\usepackage{rotating}
\usepackage{lineno}
\usepackage{numcompress}
\usepackage{xspace}
\usepackage{multirow}
\usepackage{comment}
\usepackage{placeins}
\usepackage{amsmath}
\usepackage{dcolumn}
\usepackage[normalem]{ulem}
\usepackage[centering,scale=0.75]{geometry}

\hypersetup{colorlinks,breaklinks,pdftitle={ATLAS Draft Cover},pdfauthor={ATLAS Collaboration}}
\hypersetup{linkcolor=blue,citecolor=blue,filecolor=black,urlcolor=blue}

\AtlasTitle {Search for a high-mass Higgs boson decaying
to a $W$ boson pair in $pp$ collisions at $\sqrt{s} = 8$ TeV with the ATLAS
detector}

\AtlasRefCode{HIGG-2013-19}

\AtlasNote{ATL-COM-PHYS-2014-1485}

 \PreprintIdNumber{CERN-PH-EP-2015-185}    

\AtlasJournal{JHEP}

\AtlasAbstract{ %
A search for a high-mass Higgs boson $H$ is performed in the \hwwlnln{}
and \hwwlnqq{} decay channels using $pp$ collision data corresponding to an 
integrated luminosity of \lumi\ fb$^{-1}$ collected at
$\sqrt{s}=$ 8 TeV by the ATLAS detector at the Large Hadron
Collider. No evidence of a high-mass Higgs boson is found. Limits
on $\sigma_H\times \mathrm{BR}(\hww)$ as a function of the Higgs boson mass $m_H$ are
determined in three different scenarios: one in which the heavy Higgs boson has a 
narrow width compared to the experimental resolution, one for a width increasing with the boson mass 
and modeled by the complex-pole scheme following the same behavior as in the Standard Model, 
and one for intermediate widths.  The upper range of the search is $m_H=1500 \GeV$ 
for the narrow-width scenario and $m_H=1000\GeV$ for the other two scenarios.  
The lower edge of the search range is $200$--$300\GeV$ and depends on the analysis
channel and search scenario.  
For each signal interpretation, individual and combined limits from the two $WW$ decay channels are presented.  
At $\mH = 1500\GeV$, the highest-mass point tested, $\sigma_H\times\mathrm{BR}(H\rightarrow WW)$ for a narrow-width Higgs boson
is constrained to be less than 22 fb and 6.6 fb at 95\% CL for the gluon fusion and vector-boson fusion production modes, respectively.
}

\AtlasCoverSupportingNote{ATL-COM-PHYS-2014-193}{https://cds.cern.ch/record/1667586}
\AtlasCoverSupportingNote{ATL-COM-PHYS-2014-920}{https://cds.cern.ch/record/1746185}
\AtlasCoverSupportingNote{ATL-COM-PHYS-2015-130}{https://cds.cern.ch/record/1993835}     % supporting note for combination
\AtlasCoverCommentsDeadline{17 November 2015}

\AtlasCoverAnalysisTeam{
Hass~AbouZeid, 
Olivier~Arnaez, 
Padraic~Calpin,
Sara~Diglio (*), 
Gunar~Ernis,
Pamela~Ferrari,
Jelena~Jovicevic, 
Lashkar~Kashif (*), 
Haifeng~Li (*),
Thomas~Maier, 
Christian~Meineck, 
Corrinne~Mills (*), 
Mark~Neubauer (*), 
Ben~Pearson, 
Jianming~Qian, 
Christian~Riegel,
Rikard~Sandstroem,
David~Shope, 
Michael~Strauss (*), 
Pierre~Taylor, 
Ilya~Tsukerman,
Dmitri~Tsybychev, 
Wolfgang~Wagner,
Christopher~Walker, 
Sau~Lan~Wu, 
Zhiqing~Zhang}

\AtlasCoverEdBoardMember{Luciano Mandelli (*), Andrey Kiryunin, Philip Clark, Kristin Lohwasser}

\AtlasCoverEgroupEditors{atlas-HIGG-2013-19-editors@cern.ch}

\AtlasCoverEgroupEdBoard{atlas-HIGG-2013-19-editorial-board@cern.ch}

\title{  Search for a high-mass Higgs boson decaying to a $W$ boson pair in $pp$ collisions at $\sqrt{s} = 8$ TeV with the ATLAS detector}

\author{The ATLAS Collaboration}
\date{\today}

\newcommand{\lumi}{20.3}      % Luminosity
\def\to{\hskip.05cm\rightarrow\hskip.05cm}
\def\dbline{\noalign{\vskip 0.10truecm\hrule\vskip 0.05truecm\hrule\vskip 0.10truecm}}
\def\sgline{\noalign{\vskip 0.10truecm\hrule\vskip 0.10truecm}}
\def\TeV      {\ensuremath{\mathrm{\,Te\kern -0.1em V}}\xspace}
\def\GeV      {\ensuremath{\mathrm{\,Ge\kern -0.1em V}}\xspace}
\def\MeV      {\ensuremath{\mathrm{\,Me\kern -0.1em V}}\xspace}

\newcommand{\hww}{\ensuremath{H{\rightarrow}WW}}
\newcommand{\hwwlnln}{\ensuremath{H~{\rightarrow}~WW~{\rightarrow}~\ell\nu\ell\nu}}

\newcommand{\wwlnln}{\ensuremath{WW{\rightarrow}\ell\nu\ell\nu}}
\newcommand{\Wg}{\ensuremath{W\gamma}\xspace}

\newcommand{\ZDY}{\ensuremath{Z/\gamma^\ast}}%^\ast{\to}\tau\tau}} %{\to}\ell\ell^\prime+\textrm{neutrinos}}}

\newcommand{\Wjets}{\ensuremath{W{+}\,\textrm{jets}}}
\newcommand{\Zjets}{\ensuremath{Z{+}\,\textrm{jets}}}

\newcommand{\hwwlnqq}{\ensuremath{H~{\rightarrow}~WW~{\rightarrow}~\ell\nu{qq}}}
\newcommand{\HWWlvqq}{\ensuremath{H~{\rightarrow}~WW~{\rightarrow}~\ell\nu{qq}}}

\newcommand{\mlvjj}{\ensuremath{m_{\ell\nu{jj}}}}

\newcommand{\ZeroJet}{\ensuremath{N_\textrm{jet}\,{=}\,0}}
\newcommand{\OneJet}{\ensuremath{N_\textrm{jet}\,{=}\,1}}
\newcommand{\TwoJet}{\ensuremath{N_\textrm{jet}\,{\ge}\,2}}

\newcommand{\ZeroOneJetSimple}{\ensuremath{N_\textrm{jet}\,{\le}\,1}}

\newcommand{\AllJet}{\ZeroJet, 1 and ${\ge}\,2$}
\newcommand{\vMET}{\ensuremath{{\bf E}_{\rm T}^{\rm miss}}}
\newcommand{\METcalo}{\ensuremath{E_{\mathrm{T, calo}}^{\mathrm{miss}}}}
\newcommand{\vMPT}{\ensuremath{{\boldsymbol{p}}_{\mathrm{T}}^{\mathrm{miss}}}}
\newcommand{\MPT}{\ensuremath{p_{\mathrm{T}}^{\mathrm{miss}}}}
\newcommand{\MPTRel}{\ensuremath{p_{\mathrm{T, rel}}^{\mathrm{miss}}}}
\newcommand{\metrel}{\ensuremath{E_{\rm T,rel}^{\rm miss}}}
\newcommand{\mT}{\ensuremath{m_{\mathrm T}}}
\newcommand{\vpT}{\ensuremath{\boldsymbol{p}_{\mathrm{T}}}}

\newcommand{\pTll}{\ensuremath{p_{\rm T}^{\ell\ell}}}

\newcommand{\vpTll}{\ensuremath{{\bf p}_{\rm T}^{\ell\ell}}}

\newcommand{\detall}{\ensuremath{\Delta\eta_{\ell\ell}}\xspace}

\newcommand{\mll}{\ensuremath{m_{\ell\ell}}\xspace}

\newcommand{\pTtot}{\ensuremath{p_{\rm T}^{\rm tot}}\xspace}

\newcommand{\Njet}{\ensuremath{N_\textrm{jet}}}
\newcommand{\Nbjet}{\ensuremath{N_\textrm{$b$-jet}}}

\newcommand{\Dyjj}{\ensuremath{\,\Delta{y}_{jj}\,}}

\newcommand{\Mjj}{\ensuremath{m_{jj}}}
\newcommand{\ptll}{\pTll}   % rename
\newcommand{\GGTOWW}{G\textsc{G2WW}}
\newcommand{\HDECAY}{H\textsc{decay}}
\newcommand{\MCATNLO}{\textsc{MC@NLO}}

\DeclareGraphicsExtensions{.pdf}

\begin{document}

\tableofcontents

\clearpage
\section{Introduction}
\label{sec:intro}

The boson discovered in 2012 by the ATLAS~\cite{ATLASHiggsDiscovery} and CMS~\cite{CMSHiggsDiscovery}
collaborations at the LHC matches the predictions for a Standard Model (SM) Higgs boson 
within the precision of current measurements~\cite{atlas:combined-paper-run1,cms:comb-paper-run1}.
Several extensions of the SM predict heavy neutral scalars in addition to a low-mass scalar compatible with the
discovered boson. Examples include 
generic models in which a low-mass Higgs boson mixes with
a heavy electroweak singlet~\cite{Hill:1987ea,Veltman:1989vw,Binoth:1996au,Schabinger:2005ei,Patt:2006fw,Heinemeyer:2013tqa} to
complete the unitarisation of $WW$ scattering at high energies.

This paper reports the results of a search for a heavy neutral scalar by the ATLAS Collaboration in the decay mode into
two $W$ bosons. Two final states are used: \hwwlnln\ and \hwwlnqq\ ($\ell=e,\mu$). In these final states, ATLAS has
previously reported the results of searches for heavy Higgs bosons using 4.7~\ifb{} of data collected at a centre-of-mass
energy of 7 TeV~\cite{ATLAS-4.7fbHWW,Aad:2012me}. In the \hwwlnln\ final state, a SM Higgs boson in the mass
range \mbox{$133\GeV <\mH< 261\GeV$} was excluded at 95\% confidence level (CL), 
while the \hwwlnqq\ final state was not sensitive to a SM Higgs boson of any mass with the $\sqrt{s} = 7\TeV$ dataset. The CMS Collaboration has
performed a search for a heavy Higgs boson in the \hww\ and \ensuremath{H{\rightarrow}ZZ} channels~\cite{cms:highmass-wwzz-paper-run1}. 
From a combination of the two channels, a hypothetical second Higgs boson with couplings identical to those
predicted by the Standard Model is excluded in the mass range \mbox{$145\GeV <\mH< 1000\GeV$}. 

The analyses reported here improve the results in Refs.~\cite{ATLAS-4.7fbHWW,Aad:2012me} by 
using an integrated luminosity corresponding to $\lumi\,\ifb$ of $pp$ collision data
at $\sqrt{s}=8\TeV$ collected by the ATLAS detector. 
Both analyses are designed to be sensitive to a heavy Higgs boson produced through either or both of the gluon-fusion (ggF)
or vector-boson fusion (VBF) processes.  Both also use a profile-likelihood fit to a distribution in which the hypothetical
signal is peaked but background is monotonically decreasing in the search range in order to test for the presence of signal.
The \hwwlnln\ analysis uses the dilepton transverse mass distribution for the discriminant because the two neutrinos in the 
final state result in insufficient kinematic information to reconstruct the invariant mass of the $WW$ system.  The \hwwlnqq\
analysis uses as the discriminant the invariant mass of the $WW$ system, reconstructed using the $W$ mass as a kinematic constraint
to recover the neutrino momentum up to a twofold ambiguity.
The results of the searches are interpreted in three scenarios:
\begin{enumerate}
 
  \item A Higgs boson with the couplings predicted by the SM for a Higgs boson at high mass and a width 
 	correspondingly increasing with $\mH$, and the lineshape modeled by the complex-pole scheme (CPS)
	for most mass hypotheses, as explained in Sec.~\ref{sec:models}.  Accordingly, this is 
	referred to as the CPS scenario.
  \item A Higgs boson with a narrow width: labelled as narrow-width approximation (`NWA').
  \item An intermediate-width (`IW') scenario, motivated by the electroweak singlet model.
  
\end{enumerate}

Section~\ref{sec:models} of this paper discusses the CPS lineshape model. 
Section~\ref{sec:data} describes the ATLAS detector, the data sample and physics object
reconstruction. Section~\ref{sec:mc} summarises the simulation of signal and
background samples. The event selection and background estimation techniques used in the analyses are
described in Sections~\ref{sec:lvlv_mt} and~\ref{sec:lvqq}. Systematic uncertainties affecting the analyses are
discussed in Section~\ref{sec:syst}. 
Distributions of the discriminants are shown in
Section~\ref{sec:results}. Section~\ref{sec:interp} presents the interpretations of the results from the \hwwlnln\
and \hwwlnqq\ final states, as well as from their combination, in the scenarios listed above. Conclusions of the
study are given in Section~\ref{sec:conc}.

\section{CPS lineshape model for a heavy Higgs boson}
\label{sec:models}

Narrower widths are allowed in general for Higgs bosons in extensions to the Standard Model, but
to explore the implications of the width of the additional Higgs boson, the data are also
interpreted using a signal hypothesis with a lineshape and width identical to a SM Higgs boson. 
The width of a SM Higgs boson increases with increasing mass.  For example, it is $\sim 30\GeV$ at $\mH = 400\GeV$,
and increases to $\sim 650\GeV$ at $\mH = 1000\GeV$.   
Up to $m_H\sim$ 400 GeV, the lineshape of the $WW$ invariant mass ($m_{WW}$) distribution is
well described by a Breit--Wigner distribution with a running width, meaning that
the Higgs boson propagator is calculated for each event based on $m_{WW}$ as described in 
Ref.~\cite{Anastasiou:2012hx}.  For \mbox{$m_H \ge 400\GeV$}, 
the complex-pole scheme~\cite{Kauer:2012hd,Passarino:linesh,Passarino:2012ri} provides a
more accurate description. The CPS propagator is therefore used to describe the lineshape of the Higgs boson produced via both the 
ggF and VBF processes for $m_H \ge 400\GeV$~\cite{Campbell:2011cu,Kauer:2012ma,Bonvini:2013jha}.
The limits using this signal hypothesis are labeled ``CPS scenario'' even though a Breit--Wigner 
distribution is used for $m_H< 400\GeV$.  For that mass range the distributions are similar,
so this is a minor simplification.

For a Higgs boson with a large width, the production cross section as well as the shapes of kinematic variables are affected by the
interference between signal and non-resonant $WW$ background. The interference is small for $m_H< 400\GeV$, 
but is significant at higher masses, since it increases with increasing Higgs boson width. The effect of the interference is included
in the signal samples which use the CPS lineshape, i.e. $m_H \ge 400\GeV$. The interference calculations are described in Section~\ref{sec:mc}.

\section{Data sample and object reconstruction}
\label{sec:data}

The ATLAS detector~\cite{atlas-det} is a general-purpose
particle detector used to investigate a broad range of physics
processes. It includes inner tracking devices surrounded by a
superconducting solenoid, electromagnetic (EM) and hadronic calorimeters
and a muon spectrometer with a toroidal magnetic field. The inner
detector (ID) consists of a silicon pixel detector, a silicon microstrip
detector, and a straw tube tracker that also has transition radiation
detection capability. The ID provides precision tracking of charged
particles with pseudorapidity\footnote{ATLAS uses a right-handed
coordinate system with its origin at the nominal interaction point
(IP) in the centre of the detector and the $z$-axis along the beam
pipe. The $x$-axis points from the IP to the centre of the LHC ring,
and the $y$-axis points upward. Cylindrical coordinates $(r,\phi)$
are used in the transverse plane, $\phi$ being the azimuthal angle
around the beam pipe. The pseudorapidity is defined in terms of the
polar angle $\theta$ as $\eta=-\ln\tan(\theta/2)$.} $|\eta| < 2.5$. The calorimeter
system covers the pseudorapidity range $|\eta| < 4.9$. It is composed
of sampling calorimeters with either liquid argon or
scintillator tiles as the active medium. The muon spectrometer 
provides muon identification and measurement for $|\eta| < 2.7$. During Run
1 of the LHC, the ATLAS detector used a three-level trigger system to 
select events for offline analysis.

Owing to the high LHC luminosity and a bunch separation of 50
ns, the number of proton--proton interactions occurring in the same 
bunch crossing is large (on average 20.7 in 2012).  Proton--proton interactions 
in nearby bunch crossings also affect the detector response.
These additional interactions are collectively referred to 
as event ``pile-up''\footnote{Multiple $pp$ collisions occurring in the
same (nearby) bunch crossing are referred to as in-time (out-of-time)
pile-up.} and require the use of dedicated algorithms and
corrections to mitigate its effect on particle identification, energy calibrations,
and event reconstruction. 

The triggers used in these analyses are listed in
Table~\ref{tab:trigger_lvlv}, together with the minimum transverse
momentum ($\pT$) requirements at the different levels. Both the \hwwlnln\ and the
$\HWWlvqq$ analyses use the single-lepton triggers while the dilepton
triggers are used only by the \hwwlnln\ analysis.  The lepton trigger
efficiencies are measured using $Z$ boson candidates as a function of
lepton $\pt$ and $\eta$. The single-lepton trigger efficiencies are approximately $70\%$ for muons
with $|\,\eta\,|\,{<}\,1.05$, $90\%$ for muons in the range
$1.05\,{<}\,|\,\eta\,|\,{<}\,2.40$, and $\ge 95\%$ for electrons in
the range $|\,\eta\,|\,{<}\,2.40$. Dilepton triggers increase the
signal acceptance for the \hwwlnln\ analysis by enabling lower lepton $\pt$ 
thresholds to be used.

\begin{table}[tb!]
%\raggedright%
\caption{
  The minimum transverse momentum ($\pT$) requirements used at the
  different levels of the trigger. An ``i'' next to the threshold value indicates 
  an isolation requirement that is less restrictive than the isolation requirement
  used in the offline selection. The single-lepton triggers with higher-$\pT$ thresholds
  are more efficient at high lepton \pt than the lower-\pt triggers because of
  this isolation requirement.  For dilepton triggers, the pair of
  thresholds corresponds to the leading and subleading lepton,
  respectively. The $0\GeV$ in the line describing the dimuon trigger 
  indicates that only one muon is required at Level 1. 
  %The ``and'' and ``or'' are logical.
}
\label{tab:trigger_lvlv}
{\small
  \centering
%--------------------------------------------------------------------------------
\begin{tabular*}{1\columnwidth}{
  p{0.250\columnwidth}
  p{0.350\columnwidth}
  l
  @{\extracolsep{\fill}}*{1}{l}
}
\dbline
  Name
& Level-1 trigger
& High-level (software) trigger
\\
\sgline
\multicolumn{3}{l}{Single lepton} \\
\quad$e$          & $18$ OR $30\GeV$ & $24$i OR $60\GeV$ \\ 
\quad$\mu$        & $15\GeV$         & $24$i OR $36\GeV$ \\
\sgline
\multicolumn{3}{l}{Dilepton} \\
\quad$e$, $e$     & $10$ AND $10\GeV$  & $12$ AND $12\GeV$ \\
\quad$\mu$, $\mu$ & $15$ AND $0\GeV$  & $18$ AND  $8\GeV$ \\
\quad$e$, $\mu$   & $10$ AND $6 \GeV$  & $12$ AND  $8\GeV$ \\
\dbline
\end{tabular*}
%--------------------------------------------------------------------------------
}
\end{table}

Events are required to have a primary vertex consistent with the known interaction
region, with at least three associated tracks with $\pt>$~0.4
GeV. If multiple collision vertices are reconstructed, the vertex with
the largest summed $p_{\rm T}^2 $ of the associated tracks is selected
as the primary vertex. Data quality criteria are applied to events to
suppress non-collision backgrounds such as cosmic-ray muons,
beam-related backgrounds or noise in the calorimeters. The resulting
integrated luminosity is 20.3 fb$^{-1}$ at $\sqrt{s}$ = 8 TeV.

Electron candidates are required to have a well-reconstructed track in
the ID pointing to a cluster of cells with energy depositions in the 
EM calorimeter. They are required to be in the range
$|\,\eta\,|\,{<}\,2.47$, excluding the range $1.37\,{<}\,|\,\eta\,|\,{<}\,1.52$
which corresponds to the transition region between the barrel and the 
endcap calorimeters.  Only electrons with $\ET>15\GeV$ are used in the 
analysis.  The fine lateral and longitudinal segmentation of the calorimeter and the
transition radiation detection capability of the ID allow
for robust electron reconstruction and identification in the high
pile-up environment. Criteria including the calorimeter 
shower shape, the quality of the match between the track and the cluster, and 
the amount of transition radiation emitted in the ID, are used to 
define a set of identification
criteria~\cite{atlas:el-id-paper,atlas:el-id-note,atlas:el-energy-calib}.
The ``tight'' criteria, which have the best background rejection,
are used in the $\HWWlvqq$ analysis.  The $\hwwlnln$ analysis uses
the ``medium'' selection, which is more efficient but admits more 
background, for electrons with $\ET > 25\GeV$. 
For electrons with $15\GeV < \ET < 25\GeV$, a likelihood-based
electron selection at the ``very tight'' operating point is used
for its improved background rejection.  

Muon candidates are identified by matching tracks reconstructed in the
ID with tracks reconstructed in the muon
spectrometer~\cite{atlas:muon-id-paper}. The muon spectrometer track
is required to have a track segment in each of the three layers of the
spectrometer, while the ID track must have a minimum
number of associated hits in each subdetector. In the
\hwwlnln\ analysis, muons are required to have
$|\,\eta\,|\,{<}\,2.5$ and $\pT>15\GeV$.  For the $\HWWlvqq$ analysis, muons must
satsify $|\,\eta\,|\,{<}\,2.4$ and $\pT>25\GeV$, since the sole lepton in the 
event must be within the acceptance of the trigger.

Additional selection criteria on the lepton isolation and
impact parameter are used to reduce backgrounds from non-prompt leptons and lepton-like
signatures produced by hadronic activity. These requirements are identical
for the $\hwwlnln$ and $\HWWlvqq$ analyses.  Lepton 
isolation is defined using track-based and calorimeter-based
quantities. The track isolation is based on the scalar sum $\Sigma\pt$
of all tracks with $\pt > 0.4\GeV$ in a cone in $\eta$--$\phi$ space
around the lepton, excluding the lepton track. The cone size is
$\Delta R = \sqrt{(\Delta\phi)^2 + (\Delta\eta)^2} = 0.3$.
The track isolation requires that $\Sigma\pt$
divided by the electron transverse energy $\ET$ (muon $\pt$) be less
than $0.10$ ($0.12$) for $\ET (\pT) > 20\GeV$.  For electrons (muons)
with $15\GeV < \ET (\pT) < 20\GeV$, the threshold is $0.08$. 

The calorimeter isolation selection criterion is also based on a ratio. 
For electrons, it is computed as the sum of the transverse energies, 
$\Sigma \ET$, of surrounding energy deposits (topological clusters) in the EM and hadronic
calorimeters inside a cone of size $\Delta R = 0.3$ around the candidate
electron cluster, divided by the electron $\ET$. The cells within
$\eta\times\phi = 0.125\,\times\,0.175$ around the cluster barycentre
are excluded. The pile-up and underlying event contribution to the
calorimeter isolation is estimated and subtracted
event-by-event~\cite{Cacciari:2007fd}. Electrons with $\ET > 20\GeV$ are
required to have relative calorimeter isolation less than $0.28$.  For
$15\GeV < \ET < 20\GeV$, the threshold decreases to $0.24$.

For muons, the relative calorimeter
isolation discriminant is defined as $\Sigma \ET$ of EM and hadronic calorimeter
cells above a noise threshold inside a cone of size $\Delta R = 0.3$
around the muon direction divided by the muon $\pt$. All calorimeter
cells within a cone of size $\Delta R = 0.05$ around the muon candidate are
excluded from the sum. A correction based on the number of
reconstructed primary vertices in the event is applied to $\Sigma \ET$
to compensate for extra energy due to pile-up. Muons with $\pT > 25\GeV$ are
required to have relative calorimeter isolation less than $0.30$.  Below
that $\pT$ value the threshold decreases in steps with decreasing $\pT$, 
with a minimum value of $0.12$.

The significance of the transverse impact parameter, defined as the
transverse impact parameter $d_0$ divided by its estimated
uncertainty, $\sigma_{d_0}$, of tracks with respect to the primary
vertex is required to satisfy $|d_0|/\sigma_{d_0} <$ 3.0. The
longitudinal impact parameter $z_0$ must be $|z_0|\sin\theta <$ 0.4 mm
for electrons and $|z_0|\sin\theta <$ 1.0 mm for muons.

Jets are reconstructed from topological clusters of calorimeter
cells~\cite{atlas:jet-paper,Lampl:1099735,atlas:jet-pileup-note} using
the anti-$k_{t}$ algorithm with a radius parameter
of 0.4~\cite{AntiKt}. The jet energy dependence on pile-up is
mitigated by applying two data-derived corrections. One is based on the
product of the event $\pT$ density and the jet
area~\cite{Cacciari:2007fd}. The second correction depends on the number of
reconstructed primary vertices and the mean number of expected
interactions. After these corrections, an energy- and $\eta$-dependent
calibration is applied to all jets. Finally, a residual correction
from \emph{in situ} measurements is applied to refine the jet
calibration. In both analyses, jets are required to have
$\pt\,{>}\,25\GeV$ if they have $|\,\eta\,|\,{<}\,2.4$. For jets with
$2.4\,{<}\,|\,\eta\,|\,{<}\,4.5$, the $\pT$ threshold is raised to $30\GeV$.
The increased threshold in the
forward region reduces the contribution from jet candidates produced
by pile-up. To reduce the pile-up contribution further, jets within
the inner detector acceptance are required to have
more than 50\% of the sum of the scalar \pt\ of their associated
tracks due to tracks coming from the primary vertex.  

Very heavy Higgs bosons give large momenta to their decay products. 
In the $\HWWlvqq$ analysis, the dijet system produced by the 
$W$ boson from such a decay is highly boosted and the jets overlap in the 
calorimeter, so they cannot always be resolved with the standard anti-$k_{t}$ algorithm.
Therefore, in this analysis the hadronic $W$ decay can also be reconstructed as a single
jet found by the Cambridge/Aachen algorithm~\cite{Dokshitzer:1997in},
built from topological clusters with a radius parameter of 1.2, referred to as large-R jets. 
These jets can mitigate the loss of signal efficiency, and background
can be reduced by selecting those with features typical of jets originating
from two hard partons. These jets are selected using 
a mass-drop filter algorithm~\cite{Butterworth:2008iy}.

Jets containing $b$-hadrons are identified using a multivariate
$b$-tagging
algorithm~\cite{atlas:btagcalib-2014-note,atlas:btagperf-2014-note} 
which combines impact parameter information of tracks and the
reconstruction of charm and bottom hadron decays. These analyses use 
a working point with an efficiency of 85\% for $b$-jets and a
mis-tag rate for light-flavour jets of 10.3\% in simulated
\ttbar\ events.  High $b$-jet tagging efficiency maximises top-quark
background rejection, which is important for the sensitivity of
analysis categories that require one or more jets.

In the \hwwlnln\ analysis, two different definitions of missing
transverse momentum are used. The calorimeter-based
definition, $\METcalo$, is the magnitude of the negative vector
sum of the transverse momenta of muons, electrons,
photons, and jets.  Clusters of calibrated calorimeter cells 
that are not associated with any of these objects are also included~\cite{MET}. This 
definition takes advantage of the hermeticity of the
calorimeters and their ability to measure energies of neutral
particles. However, the resolution of the calorimeter-based quantity is
degraded by the significant event pile-up. The resolution can be
improved by using track-based measurements of the momenta of
particles not associated with an identified object to replace the calorimeter cell based measurements. 
The tracks are required to have $\pt >$ 0.5 GeV and must originate from the primary
vertex.  In practice, the \pt\ of these tracks replace the $\ET$ of calorimeter cells
not associated with identified objects.  
The accurate primary-vertex association makes the track-based measurement
more robust against pile-up than the calorimeter-based measurement.
The quantity thus formulated is referred to as $\MPT$.

Using the direction of $\MPT$ relative to leptons and jets improves
the rejection of Drell--Yan backgrounds in the \hwwlnln\ final state. A
quantity $\MPTRel$ is defined as follows: 
\begin{equation}
  \begin{array}{ll}
  \multirow{2}{*}{$\MPTRel$ =\ \bigg\{ }
    &\!\!\!\!\MPT\ \sin\Delta\phi_{\mathrm{near}}\quad\textrm{if $\Delta\phi_{\mathrm{near}}<\pi/2$}\\
    &\!\!\!\!\MPT\ \phantom{\sin\Delta\phi_{\mathrm{near}}}\quad\textrm{otherwise,}
  \end{array}
\label{eqn:METRel}
\end{equation}
\noindent
where $\Delta\phi_{\mathrm{near}}$ is the azimuthal distance of the $\MPT$ and
the nearest high-$\pt$ lepton or jet. A calorimeter-based quantity
$\metrel$ is defined similarly. In Drell--Yan events, in which \MET\ arises from
mismeasurement of the $\ET$ or \pt\ of objects, these quantities tend to have small
values, while in events with genuine \MET\, they have larger values on average.
Selection using these quantities therefore rejects Drell--Yan events in preference
to signal events.

\section{Signal and background simulation}
\label{sec:mc}

This section describes the signal and background Monte Carlo (MC) generators used in the 
analyses, the different signal models used in the hypothesis tests, 
and the cross-section calculations used to normalise backgrounds.

For most processes, separate MC programs are used to generate the
hard scattering and to model the parton showering (PS), hadronisation, and
underlying event (UE). $\PYTHIA8$~\cite{pythia8}, $\PYTHIA6$~\cite{pythia},
$\HERWIG$~\cite{herwig} and $\SHERPA$~\cite{Gleisberg:2008ta} are used for the
latter three steps for the signal and for some of the background processes. When
\HERWIG\ is used for the hadronisation and PS, the UE is modelled using
\JIMMY~\cite{jimmy}.

The parton distribution function (PDF) set from CT10~\cite{Lai:2010vv}
is used for the \POWHEG~\cite{Nason:2009ai} and \SHERPA samples, while
CTEQ6L1~\cite{cteq6} is used for the \ALPGEN~\cite{alpgen}, \HERWIG, \GGTOWW~\cite{gg2WW},
$\PYTHIA6$ and $\PYTHIA8$ samples. Acceptances and efficiencies are
obtained from a full simulation~\cite{atlassim} of the ATLAS detector
using either $\GEANT4$~\cite{GEANT4}, or $\GEANT4$ combined with a parameterised 
calorimeter simulation~\cite{atlfast2}.
The simulation incorporates a model of the event pile-up conditions in
the data, including both in-time and out-of-time pile-up.

\subsection{Simulation and normalisation of signal processes}

The $\POWHEG$ generator combined with $\PYTHIA8$ is used to model all signal processes.  
Heavy Higgs boson production via the ggF and VBF processes are considered in both the
$\hwwlnln$ and $\hwwlnqq$ analysis channels.
Contributions from Higgs-strahlung and $t\bar{t}H$
production mechanisms are not considered owing to their very small
cross sections at high Higgs boson masses. For leptonic $W$ decays,
the small contribution from
leptonic $W \rightarrow \tau\nu \rightarrow \ell\nu\nu\nu$ decays is
included. 

The ggF signal cross-section calculation includes corrections 
up to next-to-next-to-leading order (NNLO) in
QCD~\cite{Djouadi:1991tka,Dawson:1990zj,Spira:1995rr,Harlander:2002wh,Anastasiou:2002yz,Ravindran:2003um}. 
Next-to-leading-order (NLO) electroweak (EW) corrections are also
applied~\cite{Aglietti:2004nj,Actis2008}, as well as QCD soft-gluon
resummations up to next-to-next-to-leading logarithmic order
(NNLL)~\cite{Catani:2003zt}. These calculations are described in
Refs.~\cite{Anastasiou:2012hx,deFlorian:2012yg,Baglio:2010ae} and
assume factorisation between the QCD and EW corrections. The VBF
signal cross section is computed with approximate NNLO QCD
corrections~\cite{Bolzoni:2010xr} and full NLO QCD and EW
corrections~\cite{Ciccolini:2007jr,Ciccolini:2007ec,Arnold:2008rz}.
The total width for the CPS scenario follows the SM predictions for high mass
and has been calculated using \HDECAY~\cite{Djouadi:1997yw}. The 
branching fractions for the decay to ${WW}$ as a function of
$m_{H}$ have been calculated using {\sc Prophecy4f}~\cite{Bredenstein:2006rh,Bredenstein:2006ha}.

\subsubsection{Signal samples for CPS scenario}

Simulated Higgs boson samples with the width predicted by the SM as a function
of $\mH$ are generated using $\POWHEG$+$\PYTHIA8$, at $20\GeV$ intervals 
for \mbox{$220\GeV \le \mH \le 580\GeV$}, and at $50\GeV$ intervals 
for \mbox{$600\GeV \le \mH \le 1000\GeV$}.  The CPS-scenario interpretation 
is not performed for $\mH>1000\GeV$ because of the large width of the resonance.
For \mbox{$\mH < 400\GeV$},
ggF and VBF samples are generated with the running-width Breit--Wigner propagator 
described in Section~\ref{sec:models}.  For \mbox{$\mH \ge 400\GeV$}, samples are
generated using a CPS propagator. The calculations using the
Breit--Wigner and the CPS propagators are in good agreement in the mass
range below $400\GeV$. 

Calculations of the interference effect between resonant and non-resonant
$gg\to{WW}$ production are available only at leading-order (LO) accuracy in QCD.
Therefore, this effect is not directly included in the generation of the ggF and VBF 
CPS-scenario signal samples, and is implemented via event weighting 
at particle level.  % \sout{The CPS-scenario 
% Higgs boson signal samples are treated in this way, with the weights applied to the 
% $WW$ invariant mass distribution at particle level. }
The full weighting procedure, including the treatment of
associated uncertainties, is described in detail in
Ref.~\cite{Heinemeyer:2013tqa} and summarised here.

For ggF signal samples, the interference weights are computed at LO
using the \MCFM~\cite{Campbell:2010ff} program, and rescaled
to NNLO following the recommendations given in
Ref.~\cite{Heinemeyer:2013tqa}. EW corrections are also included in the NNLO
result used in the rescaling. The interference changes the total cross section.
For $\mH > 400\GeV$, it increases with increasing $\mH$, with an enhancement of
almost a factor of four for $\mH = 1\TeV$~\cite{Campbell:2011cu}. The interference is negative
below $\mH \approx 400\GeV$, but changes the cross section by 10\% or less.
The weighting procedure has also been
performed with the \GGTOWW\ program; the results show 
good agreement with those using \MCFM. The procedure accounts for
theoretical uncertainties associated with the LO-to-NNLO scaling as
well as those due to missing higher-order terms in the presently
available interference estimation. The weights are applied to the signal samples 
only, because in the absence of signal there is no effect on the background. 
The sum of the weighted signal and the continuum $WW$ background spectra approximately reproduces 
the results of the full calculation.
% a fully consistent treatment of the Higgs signal, the continuum background, and the interference between them.}

For VBF signal samples, the \textsc{REPOLO} tool provided by the
authors of \textsc{VBFNLO}~\cite{Arnold:2012xn} is used to extract the
interference weights. QCD scale and modelling uncertainties associated
with the weights are also estimated using \textsc{REPOLO}. In this
case, the LO-to-N(N)LO differences are expected to be
small~\cite{Figy:2003nv, Ciccolini:2007jr, Ciccolini:2007ec,
  Bolzoni:2010xr, Bolzoni:2011cu}, and no explicit uncertainty is
assigned to take these differences into account.  Because not all of
the information needed for the weight calculation is present in the 
fully reconstructed Monte Carlo samples, the weights are parameterised
as a function of $m_{WW}$ and $\mH$.  A closure test comparing the
signal lineshapes produced by the reweighting compared to the full
calculation for the interference effects shows some differences, which
are largest for $m_{WW}$ far from $\mH$, but do not exceed $10$\%.
These differences are treated as a systematic uncertainty on the signal.

For both ggF and VBF signal, the weights accounting for interference effects are calculated
for each Higgs boson mass at which the samples are simulated, and
applied as a function of $m_{WW}$ in the range 0.5 $<~m_{WW}/m_{H}~<$ 1.5. 
% \sout{The weights are applied to the signal samples only, because in the absence 
% of signal there is no effect on the background. }
The procedure modifies the event kinematics, including the $\mT$ distribution used in the $\hwwlnln$ analysis.
It has been shown that the weights describe the effect
of interference on all kinematic variables used in the analyses~\cite{Heinemeyer:2013tqa}.

\subsubsection{Narrow-width signal samples}

For the narrow-width Higgs boson scenario, signal samples are
generated with $\POWHEG$+$\PYTHIA8$ using a fixed $4.07\MeV$-wide
Breit--Wigner lineshape at $100\GeV$ intervals for
\mbox{$300\GeV \le \mH \le 1500\GeV$}. Owing to the small width, the effect of
interference between signal and continuum background is negligible
over the full mass range explored in the
analyses~\cite{Campbell:2011bn,Campbell:2011cu}, therefore no
interference weights are applied to these samples.

\subsubsection{Signal samples for intermediate-width scenario}

The intermediate-width scenario signal samples are derived by weighting the CPS 
signal samples to modify the width and lineshape and to account for interference. 
The lineshape of the heavy Higgs boson is weighted to one derived
from a running-width Breit--Wigner propagator, and to scale the width down
from the SM width.  The interference weights are derived using 
the \MCFM\ and \textsc{REPOLO} tools respectively for ggF and VBF signals, as in the
CPS scenario, and are computed as a function of the modified width of the heavy scalar.
The interference is a significant effect for \mbox{$\Gamma_{H} \gtrsim 10\GeV$}.
The weights are applied to the $m_{WW}$ distribution and modify the event kinematics
accordingly.  

Intermediate-width signal scenarios are explored for a mass $\mH$ between $200\GeV$ and $1000\GeV$ and
a width in the range \mbox{$0.2\Gamma_{H, \mathrm{SM}} \le \Gamma_{H} \le
0.8\Gamma_{H, \mathrm{SM}}$}, where $\Gamma_{H}$ is the width of the hypothetical particle and
$\Gamma_{H, \mathrm{SM}}$ is the width of a SM Higgs boson for the same mass.
The extremes of a very narrow width and the same width as the SM are covered by the NWA and CPS scenarios.

\subsection{Background processes}

\subsubsection{Background processes for the \hwwlnln{} analysis}

The MC generators
used to simulate the background processes in the \hwwlnln{} analysis, and the cross
sections used to normalise them, are
listed in Table~\ref{tab:cross_bkg}. In this table, 
all $W$ and $Z$ boson decays into leptons ($e,~\mu,~\tau$) are included in the
corresponding products of the cross sections ($\sigma$) and the branching ratios
($\mathrm{BR}$). 

Cross sections for top-quark and diboson processes are computed as follows. 
The $\ttbar$ production cross section is normalised to the NNLO+NNLL computation from 
{\sc TOP++2.0}~\cite{Czakon:2013goa,Czakon:2012pz,Czakon:2012zr}, and
single-top processes are normalised to NNLL calculations of the cross
section~\cite{Kidonakis:2010tc,Kidonakis:2011wy,Kidonakis:2010ux}. The
$WW$ cross section is calculated at NLO accuracy in QCD
using \MCFM. The cross section for non-resonant gluon-fusion
production is calculated at LO accuracy with \GGTOWW, including both
$WW$ and $ZZ$ production and their interference. 

Top-quark event generation
uses {\POWHEG}+$\PYTHIA6$, except for the single-top $t$-channel process $tq\bar{b}$,
for which \textsc{AcerMC}~\cite{Kersevan:2004yg}+$\PYTHIA6$ is used.
The $WW$ background is also modelled using {\POWHEG}+$\PYTHIA6$.
For $WW$, $WZ$ and $ZZ$ backgrounds with two additional jets produced,
the \SHERPA generator is used for event modelling.
The $W(\ZDY)$ process is simulated with {\SHERPA} and {\POWHEG}+$\PYTHIA8$,
with $m_{\gamma^{\ast}}$ extending down to the kinematic threshold and lepton masses
  included in the modeling of the $\gamma^{\ast}$ decay.
The $\Wg$ and Drell--Yan processes are modelled using {\ALPGEN}+\HERWIG with 
merged LO matrix element calculations of up to five jets.
The merged samples are normalised to the NLO
calculation of \MCFM~(for $\Wg$) or the NNLO calculation of
DYNNLO~\cite{Catani:2007vq, Catani:2009sm} (for $\ZDY$). 
A \SHERPA~sample is used to model the $Z\gamma \rightarrow
\ell\ell\gamma$ background. The cross section of this process is normalised to 
NLO using \MCFM.
The $\Wjets$ background shape and normalisation are derived from data, as 
described in Section~\ref{sec:bkg}, so no simulated $\Wjets$ events are used.

\begin{table}[tb!]
 \centering
 \caption{Monte Carlo generators used to model the background processes in the
   \hwwlnln{} analysis. All leptonic decay branching ratios ($e,~\mu,~\tau$) of
   the $W$ and $Z$ bosons are included in the product of cross section
   ($\sigma$) and branching ratio ($\mathrm{BR}$).}
 \label{tab:cross_bkg}
 \begin{tabular}{lcc}
 \dbline
    Background                                               & MC generator       & \hspace*{-5mm}$\sigma\,{\cdot}\,\mathrm{BR}$ (pb) \\
    \sgline
    $\ttbar$                                                 & \POWHEG + $\PYTHIA6$		&  26.6 \\
    $tW$                                                     & \POWHEG + $\PYTHIA6$           	&   2.35 \\
    $tq\bar{b}$                                              & \textsc{AcerMC} + $\PYTHIA6$   	&  28.4 \\
    $t\bar{b}$                                               & \POWHEG + $\PYTHIA6$           	&   1.82 \\
    $q\bar{q}/g\rightarrow WW$                               & \POWHEG + $\PYTHIA6$           	&   5.68 \\
    $gg\rightarrow WW$                                       & \GGTOWW + \HERWIG		&   0.20\\
    QCD $WW + 2$ jets                                        & \SHERPA				&   0.568 \\
    EW $WW + 2$ jets                                         & \SHERPA				&   0.039 \\   	
    $Z/\gamma^{\ast}$+jets ($\mll \ge 10\GeV$)               & \ALPGEN + \HERWIG		& $16.5\times 10^{3}$ \\
    EW $Z/\gamma^{\ast}$ (includes $t$-channel)              & \SHERPA				&   5.36 \\
    $Z^{(\ast)}Z^{(\ast)} \to 4\ell$                         & \POWHEG + $\PYTHIA8$         	&   0.73 \\
    $W(Z/\gamma^{\ast}) (m_{(Z/\gamma^{\ast})} < 7\GeV)$     & \SHERPA				&  12.2 \\ 
    $Z\gamma$($\pT^{\gamma} > 7\GeV$)                        & \SHERPA				& 163 \\
    $W\gamma$                                                & \ALPGEN + \HERWIG		& 369 \\
    Higgs boson ($\mH=125\GeV$)                      	     & \POWHEG + $\PYTHIA8$		&   0.60\\
    \dbline
  \end{tabular}
\end{table}

\subsubsection{Background processes for the \hwwlnqq{} analysis}

Several different Monte Carlo generators are used to simulate the
background to the $\HWWlvqq$ process. The processes used to 
model the background in the analysis are shown in Table~\ref{tab:bkgSamples_lvqq}.
In general, the treatment follows that of the $\hwwlnln$ analysis, 
with the exceptions described here.

The $\Wjets$ background is modelled with the
\SHERPA{} generator version 1.4.1. In order to have enough events for a background prediction at high mass, the
\SHERPA{} samples are generated in multiple bins of $p_{\mathrm{T}}^W$. The bin boundaries
are: $40$--$70\GeV$, $70$--$140\GeV$, $140$--$280\GeV$, $280$--$500\GeV$, and $>500\GeV$.
An inclusive sample is used for $p_{\mathrm{T}}^W<40\GeV$.
Samples of $W$ bosons with only electroweak vertices are also generated to ensure sufficiently
good modelling of this background in the VBF topology.

The top-quark background is modelled using the same generators as in the $\hwwlnln$ analysis.
Events in the $t\bar{t}$ sample are reweighted according to the $\pT$ of the $t\bar{t}$ system 
and the individual top quarks to improve the kinematic agreement between the data and the $\POWHEG$ 
prediction, following the prescription outlined in Ref.~\cite{Aad:2015gra} based
on the measurements of Ref.~\cite{Aad:2014zka}. This treatment is not needed for the $\hwwlnln$ 
analysis because the distributions affected are primarily the number of jets and the jet $\pT$,
and the analysis is not sensitive to either of these because of the normalisation of the top-quark background
individually in each jet bin.
The $Z$+jets background is also generated via
\SHERPA{} and, like the \SHERPA{} $\Wjets$ background, uses samples binned in
$p_{\mathrm{T}}^Z$, with a binning
identical to the $p_{\mathrm{T}}^W$ used for the $\Wjets$ samples.

The $\HERWIG$ generator is used for the $WW$, $WZ$, and $ZZ$ processes. % out
These samples are produced
with inclusive vector boson decays and a single-lepton filter at the event generation
stage. 

\begin{table}[tbp]
   \centering
 \caption{Monte Carlo generators used to model the background processes in the
   \HWWlvqq{} analysis, and the associated cross sections $\sigma$. 
     Leptonic decay branching ratios $\mathrm{BR}$ of
   the $W$ and $Z$ bosons are not included in the number quoted 
   unless explicitly indicated in the process name.}
   \begin{tabular}{lcc}
    \dbline
     Background & MC generator & $\sigma\cdot \mathrm{BR}$~(pb) \\
     \sgline
     $\ttbar$ 		& \POWHEG + $\PYTHIA6$ 	& 252.9 \\ 
     $tW$ 		& \POWHEG + $\PYTHIA6$ 	& 22.4 \\ 
     $tq\bar{b}$	& AcerMC  + $\PYTHIA6$ 	& 28.4 \\ 
     $t\bar{b}$		& \POWHEG + $\PYTHIA6$ 	& 1.82 \\ 
     $W\to \ell \nu$ 	   & \SHERPA & $35.6\times 10^3$ \\
     $W\to \ell \nu$ VBF   & \SHERPA & 12.6 \\ 
     $W\gamma \to\ell\nu\gamma$ 	& \ALPGEN + \HERWIG & 369 \\ 
     $Z\to \ell\ell$ & \SHERPA & $3.62 \times 10^3$ \\
     $Z\gamma \to \ell\ell\gamma$ ($\pT^\gamma > 10\GeV$)  & \SHERPA & 96.9 \\ 
     $WW$ & \HERWIG & 32.5 \\ 
     $WZ$ & \HERWIG & 12.0 \\ 
     $ZZ$ & \HERWIG &  4.69 \\ 
     \dbline
   \end{tabular}
\label{tab:bkgSamples_lvqq}
\end{table}

\section{The \hwwlnln\ analysis}
\label{sec:lvlv_mt}

In the $\hwwlnln$ channel, the final state is two oppositely charged
leptons and two neutrinos, which are reconstructed as missing transverse
momentum.  Additional jets may be present from QCD radiation or from
the scattering quarks in the VBF production mode.  The analysis described here is 
similar to the one designed to study the Higgs boson with $\mH\approx125\GeV$
in the $\wwlnln$ final state~\cite{atlas:ww-paper-run1},
with adaptations made to enhance the sensitivity for a high-mass Higgs boson.

\subsection{Event selection}

The event is required to have two oppositely charged leptons and no additional
lepton with $\pT > 10\GeV$, with the higher- and lower-$\pT$ leptons
respectively satisfying $\pT> 22\GeV$ and $\pT> 10\GeV$. Both leptons must satisfy the 
quality criteria discussed in Section~\ref{sec:data}. Background from
low-mass resonances constitutes a significant contribution, and is rejected by
requiring $\mll > 10\GeV$ in the same-flavour  channel and $\mll > 12\GeV$ in the
different-flavour channel, in which resonances decaying to $\tau\tau$ may contribute. 
In the same-flavour channel, a veto on $Z$ bosons is applied
by requiring $|\mll - m_Z| > 15\GeV$. These criteria form the preselection.

The signal and background compositions depend strongly on the final-state jet
multiplicity ($\Njet$). For $\Njet = 0$, the signal is predominantly from
the ggF process, and $WW$ events dominate the background. For $\Njet = 1$, both the ggF and
VBF signal processes contribute, and the large majority of background events are from
$WW$ and top-quark events, which contribute approximately equally to the background.   
For $\Njet\,{\ge}\,2$, the signal originates mostly from the VBF process and 
top-quark events dominate the background. The analysis is consequently divided into 
$\Njet\,{=}\,0$, 1 and $\,{\ge}\,2$ categories.

The event selection in the various jet multiplicity categories is optimised
using the BumpHunter~\cite{gc:bumphunter} program, maximising the quantity $s/\sqrt{(b+(\Delta b)^{2})}$, 
where $s$ and $b$ are the numbers of signal and background events, respectively, and $\Delta b$ represents
the systematic uncertainty on the background. The value $\Delta b = 10\%$ is used.  
The optimisation has also been performed with $\Delta b = 20\%$ to test for sensitivity 
to the assumed systematic uncertainties, but the resulting selection is not significantly different from
the one adopted.
The optimisation is performed separately for the different- and same-flavour channels. The optimised event
selection criteria that define the signal regions (SRs) in the analysis are summarised in 
Table~\ref{tab:cutsummary1}.

\begin{table}[t!]
  \caption{
    Event selection criteria used to define the signal regions in the \hwwlnln\ analysis. 
    The criteria specific to different-flavour (DF) and same-flavour (SF) channels are noted as such; 
    otherwise, they apply to both. Preselection applies to all $\Njet$ categories. In the 
    $\ge$2 jets category, the rapidity gap is the rapidity range spanned by the two leading
    jets. 
  }
  \label{tab:cutsummary1}
  \centering
  {\scriptsize
  \begin{tabular}{llll}
  \dbline
  Category	    & $\ZeroJet$ & $\OneJet$ & $\TwoJet$ \\
  \sgline
  Preselection     &
  \multicolumn{3}{c}{
  \begin{tabular}{l}
  Two isolated leptons ($\ell\,{=}\,e, \mu$) with opposite charge\\
  $\pT^\textrm{lead}\,{>}\,22$ GeV, $\pT^\textrm{sublead}\,{>}\,10$ GeV \\
  DF: $\mll\,{>}\,10$ GeV\\
  SF: $\mll\,{>}\,12$ GeV, $|\,\mll-\mZ\,|\,{>}\,15$ GeV\\
  \end{tabular}
  }\\
  \sgline
  Lepton $\pT$ & $\pT^\textrm{lead}\,{>}\,60\GeV$ &  $\pT^\textrm{lead}\,{>}\,55\GeV$ &  $\pT^\textrm{lead}\,{>}\,45\GeV$  \\ 
	       & $\pT^\textrm{sublead}\,{>}\,30\GeV$ & $\pT^\textrm{sublead}\,{>}\,35\GeV$ & $\pT^\textrm{sublead}\,{>}\,20\GeV$ \\
  \sgline
  \multirow{3}{*}{\!\!\!\!\!
  \begin{tabular}{l}
  Missing transverse \\
  momentum
  \end{tabular}
  }
  & DF: $\MPT\,{>}\,45$ GeV    & DF: $\MPT\,{>}\,35$ GeV    & DF: $\METcalo\,{>}\,25$ GeV\\
  & SF: $\metrel\,{>}\,45$ GeV & SF: $\metrel\,{>}\,45$ GeV & SF: $\METcalo\,{>}\,45$ GeV\\
  & SF: $\MPTRel\,{>}\,65$ GeV & SF: $\MPTRel\,{>}\,70$ GeV & - \\
  \sgline
  \multirow{3}{*}{General selection}
  		    & - 		& $\Nbjet\,{=}\,0$   & $\Nbjet\,{=}\,0$ \\
  		    & $\ptll\,{>}\,60$ GeV & -		     & $\pTtot\,{<}\,40$ GeV\\
  \sgline
  \multirow{4}{*}{VBF topology}
  		    &-&-& $\Mjj\,{>}\,500$ GeV\\
  		    &-&-& $\Dyjj\,{>}\,4.0$ \\
  		    &-&-& No jet ($\pT\,{>}\,20$ GeV) in rapidity gap\\
  		    &-&-& Both $\ell$ in rapidity gap\\
  \sgline
  \multirow{2}{*}{\!\!\!\!\!
  \begin{tabular}{l}
  $\hwwlnln$ \\
  topology
  \end{tabular}
  }
  		    & $\mll\,{>}\,60$ GeV  & $\mll\,{>}\,65$ GeV  & DF: $\mll\,{>}\,60$ GeV, SF: $\mll\,{>}\,45$ GeV\\
  		    & $\detall\,{<}\,1.35$ & $\detall\,{<}\,1.35$ & $\detall\,{<}\,1.85$ \\
  \dbline
  \end{tabular}
  }
\end{table} 
 
Owing to the topology of \hwwlnln\ events, a selection on the missing transverse
momentum is useful. 
In the different-flavour channel in both the $\ZeroJet$ and $\OneJet$
categories, requirements are imposed on $\MPT$. In the same-flavour channel in these $\Njet$
categories, selections on $\MPTRel$ and $\metrel$ are used since, as explained in
Section~\ref{sec:data}, these quantities efficiently reject Drell--Yan events. In the
$\TwoJet$ category, $\METcalo$ thresholds are used in both the different- and
same-flavour channels. Selection using $\MPTRel$ or $\metrel$ in this category rejects a large fraction of 
signal events and is not optimal; they are therefore not used.

In the $\ZeroJet$ category, additional requirements on the \pT\ of the dilepton system
\pTll and on \mll are applied. In the $\OneJet$ category, a $b$-jet veto is applied
to suppress the top background, and a selection on \mll is imposed. To
orthogonalise the $\ZeroJet$ and $\OneJet$ signal regions with respect to the $WW$ control
regions (Section~\ref{sec:bkg}), the pseudorapidity difference $\detall$ between
the two leptons is required to be smaller than 1.35.

The $\TwoJet$ category is optimised to extract the Higgs boson 
signal produced via vector-boson fusion. The invariant mass \Mjj\ of the two
highest-\pT\ jets, referred to as the tagging jets, is required to be larger than 500
GeV. The magnitude of the rapidity difference between the tagging jets, $\Dyjj$, is required to be
larger than 4.0. In addition, the event must have no additional jets with $\pT>20\GeV$
within the rapidity gap of the tagging jets, 
while both leptons are required to be within this rapidity gap. A $b$-jet veto is applied, and the total
transverse momentum \pTtot\ in the event is required to be smaller than $40\GeV$. The quantity \pTtot\ is defined 
as the magnitude of $\vpT^{\ell1}{+}\vpT^{\ell2}{+}\vMPT{+}\sum\vpT^{\mathrm{jets}}$, where the sum 
is over all jets that pass the nominal analysis jet selection. Selections on \mll are applied as in
the $\ZeroOneJetSimple$ categories, and \detall $<$ 1.85 is required. 
For a Higgs boson with $\mH=300\GeV$ and the ratio of ggF and VBF cross sections
predicted by the SM, $83\%$ of the total signal selected in the $\TwoJet$ category
is produced by the VBF process.  In the $\ZeroJet$ and $\OneJet$ categories, these fractions are 
$2\%$ and $12\%$, respectively. The signal fractions from the VBF process increase with increasing $\mH$.

The discriminant used to derive the final results in this analysis is the
transverse mass \mT, defined as:
\begin{equation}
  \label{eq:mT}
  \mT = \sqrt{(E_{\mathrm T}^{\ell\ell}+\met)^{2} - |\vpTll+\vMET|^{2}},
\end{equation}
where $E_{\mathrm T}^{\ell\ell} = \sqrt{|\vpTll|^{2}+m_{\ell\ell}^{2}}$.

\subsection{Background determination}
\label{sec:bkg}

The major backgrounds in this analysis are top-quark and $WW$ production, with additional contributions from
$W/Z$+jets, multijets, and the diboson processes $WZ$, $W\gamma$, $W\gamma^{\ast}$, and $ZZ$.  
The top-quark and $WW$ backgrounds are normalised to data in control regions (CRs) defined by criteria similar to
those used for the SR, but with some requirements loosened or reversed to obtain signal-depleted samples 
enriched in the relevant backgrounds.  This normalisation is done through a simultaneous fit to the signal
region and all control regions, as described in Section~\ref{subsec:stats}.  This fit uses the complete
background prediction in each region in order to account for the presence of other backgrounds and the potential small 
presence of signal.  
In particular, any background whose normalisation is determined by a control region is scaled by the same 
normalisation factor in all signal and control regions, not just its own control region.
The following subsections describe the methods used to estimate the most important backgrounds, namely, 
$WW$, top-quark events, and $\Wjets$, in more detail. 
The Drell--Yan and non-$WW$ diboson backgrounds are small, and their predictions are computed from simulation.
The small background from the Higgs boson with $\mH \approx 125\GeV$ is also included. The predicted
  cross section, branching ratio, and kinematics for the SM Higgs boson are used.
With few exceptions, the background estimates use the same techniques as Ref.~\cite{atlas:ww-paper-run1}. They
are described there in more detail, and summarized here.

\subsubsection{$WW$ background}

In the $\Njet\,{\le}\,1$ categories, the $WW$ background is normalised using a CR defined with the selection summarised in Table~\ref{tab:WWCR}. To orthogonalise the $WW$ CRs to the
$\ZeroJet$ and $\OneJet$ SRs, the selection on \detall is reversed with respect to the SR definitions: \detall $> 1.35$ is required. Only the
different-flavour final states are used to determine the $WW$ background, and 
the purity is 70.5\% and 40.6\% in the $\ZeroJet$ and $\OneJet$ categories, respectively. 
The normalisation factors obtained from the simultaneous fit to the signal and control regions
are $1.18 \pm 0.04$ for the $\ZeroJet$ CR and $1.13 \pm 0.08$ for the $\OneJet$ CR,
where the uncertainty quoted includes only the statistical contribution.
The high normalisation factor for $WW$ events with zero jets has been studied in
Ref.~\cite{atlas:ww-paper-run1}, and results from poor modelling of the jet veto efficiency.  
The $WW$ prediction in the $\TwoJet$ category is taken from simulation, because it is difficult to isolate a kinematic 
region with a sufficient number of $WW$ events and a small contamination from the top-quark background.

Figure~\ref{fig:CR_WW} shows the \mT\ distributions in the $\ZeroOneJetSimple$ $WW$ CRs.
Normalisation factors obtained from the top CRs as well as from the $WW$ CRs have been applied to these distributions.

\begin{table}[t!]
\caption{
  Event selection criteria for the $\ZeroJet$ and $\OneJet$ $WW$ control regions in the $\hwwlnln$ analysis. The criteria that are different 
  with respect to the SR definition are shown. Only the different-flavour final state is used.
}
\label{tab:WWCR}
\centering
{\small
\begin{tabular}{lll}
\dbline
Category          & $\ZeroJet$ & $\OneJet$ \\
\dbline

\multirow{2}{*}{Lepton transverse momentum} 
 & \multicolumn{2}{c}{$\pT^\textrm{lead}\,{>}\,22\GeV$} \\
 & \multicolumn{2}{c}{$\pT^\textrm{sublead}\,{>}\,15\GeV$} \\
\sgline
Missing transverse momentum & $\MPT\,{>}\,20$ GeV~~~~~~~~ &  $\MPT\,{>}\,35\GeV$ \\
\sgline
\multirow{4}{4cm}{General selection and
  $\hwwlnln$ topology}
& - & $\Nbjet\,{=}\,0$ \\
& $\ptll\,{>}\,35\GeV$ & - \\
& $\mll\,{>}\,75\GeV$  & $\mll\,{>}\,75\GeV$\\
& $\detall\,{>}\,1.35$ & $\detall\,{>}\,1.35$\\
\dbline
\end{tabular}
}
\end{table}

\begin{figure}[t!]
\centering
 \includegraphics[width=0.48\textwidth]{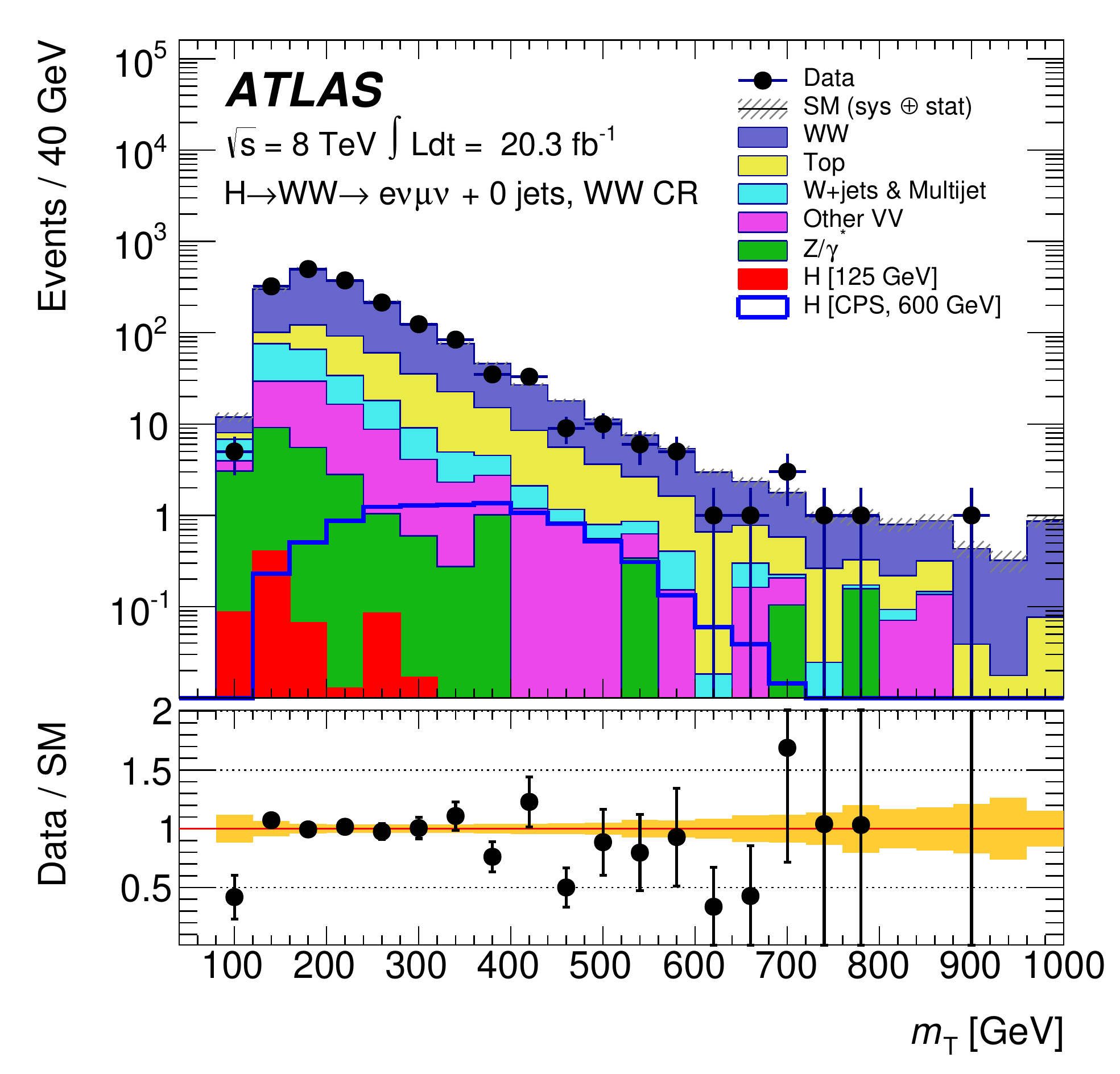}
 \includegraphics[width=0.48\textwidth]{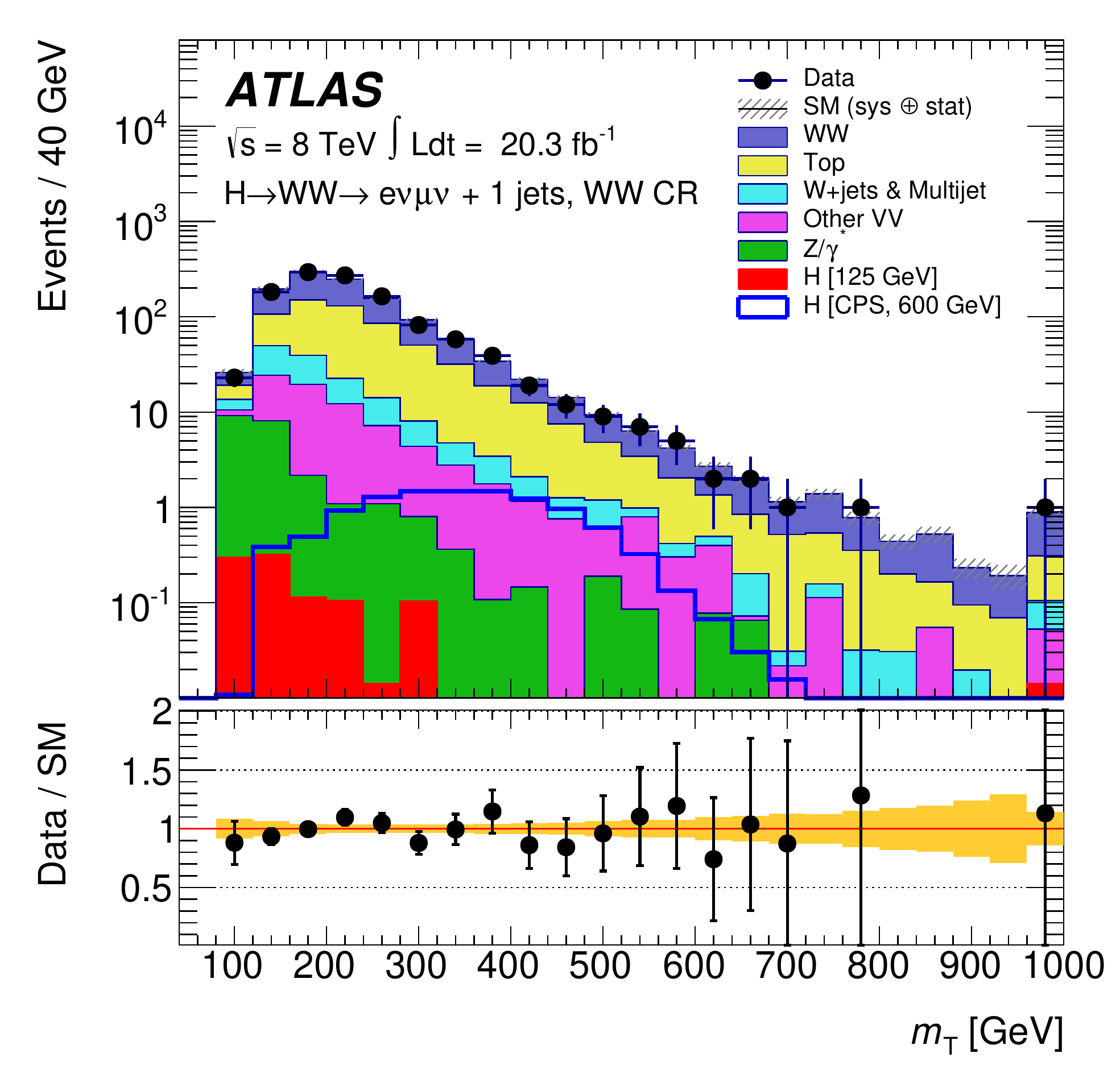}
\caption{Transverse mass distribution in the $\ZeroJet$ (left) and $\OneJet$ (right) $WW$ control regions of the $\hwwlnln$ analysis. 
      	Only the different-flavour final state is
         used. In each figure, the last bin contains the overflow.  The combined statistical and systematic uncertainties on the 
	 prediction are shown by the hatched band in the upper pane and the shaded band in the lower pane.
	 Normalisation factors obtained from a comparison of data and prediction have been applied in these figures.
	}
\label{fig:CR_WW}
\end{figure}

\subsubsection{\ttbar\ and single top background}

Top-quark events can be produced as a $t\bar{t}$ pair, or in association with a $W$ boson or another flavour of quark. In 
the \hwwlnln\ analysis, contributions from $t\bar{t}$ and single-top events are estimated together, with their relative
contributions determined by the predicted cross sections and MC simulation of the acceptances, since it is 
not easy to kinematically separate the two processes and the contribution from single top is relatively small.

% \textbf{\emph{Paragraph rewritten and equation added.}}
Owing to the difficulty of defining reasonably pure control regions in the $\ZeroJet$ category, the top-quark
background in this category is not estimated from the likelihood fit. The jet veto survival
probability (JVSP) procedure, described in more detail in Ref.~\cite{Mellado:2011pc}, is employed instead. 
In this method, the normalisation is derived from the top-quark event yield determined in a control region defined by events
with a different-flavour opposite-sign lepton pair, any number of jets, and $\MPT\,{>}\,45$ GeV.  This sample is
dominated by top-quark events.  The estimated top-quark event yield is the total number of events 
$N_{\textrm{CR}}$ passing this selection minus the expected contribution $B_{\textrm{CR}}$ from other 
processes.  The theoretical cross sections and acceptances from MC simulation are used to calculate $B_{\textrm{CR}}$, except 
the $\Wjets$ background, for which the data-derived estimate described later in this section is used. 
The resulting estimated top-quark event yield is multiplied by the fraction $\epsilon_0$ of top-quark events with no 
reconstructed jets obtained from simulation in the CR. This fraction is corrected using data from a second CR defined like the
first, with the additional requirement of at least one $b$-tagged jet.  The fraction of events in this CR with zero jets
in addition to the $b$-tagged one is measured in both data and simulated top-quark events, denoted $f_0^{\textrm{data}}$ and $f_0^{\textrm{MC}}$, respectively.  
Using these inputs, the estimated number of top-quark background events $N_{\textrm{top}}^{\textrm{est.}}$ in the $\ZeroJet$ 
signal region is estimated as:
\begin{equation}
  \label{eq:jvsp}
  N_{\textrm{top}}^{\textrm{est.}} = (N_{\textrm{CR}} - B_{\textrm{CR}}) \cdot \epsilon_0 \cdot ( f_0^{\textrm{data}} / f_0^{\textrm{MC}} )^2 \cdot \epsilon_{\textrm{rest}} \,, 
\end{equation}
where $\epsilon_{\textrm{rest}}$ is the efficiency of the $\ZeroJet$ selection requirements applied after the jet veto, 
derived from simulated top-quark events.  The theoretical uncertainties on the quantities derived from top-quark MC simulation, 
namely $\epsilon_0$, $f_0^{\textrm{MC}}$, and $\epsilon_{\textrm{rest}}$, are described in Section~\ref{sec:syst}.

In the $\OneJet$ and $\TwoJet$ categories, the normalisation of the top-quark background is determined from control 
regions. As with the $WW$ CR, and unlike the $\ZeroJet$ CRs, these are included in the simultaneous fit with the signal regions. 
  These CRs are defined identically to the respective signal regions, except that the $\MPT$ threshold 
is lowered to $20\GeV$ and the veto on $b$-tagged jets is inverted to require exactly one $b$-tagged jet with
$\pT > 25\GeV$.
The purity is 96.5\% in the $\OneJet$ category and 90.7\% in the $\TwoJet$ category. In the $\OneJet$ category, only the different-flavour final
states are used to obtain the normalisation. In the $\TwoJet$ category same-flavour and different-flavour final states are used 
to increase the number of events and thereby improve the statistical precision. 
The normalisation factors obtained from the simultaneous fit to the signal and control regions 
are $1.05 \pm 0.03$ for the $\OneJet$ CR and $0.92 \pm 0.06$ for the $\TwoJet$ CR,
where the uncertainty quoted includes only the statistical contribution.
Figure~\ref{fig:CR_Top} shows the \mT\ distributions in the $\OneJet$ and $\TwoJet$ top CRs.
The normalisation factors have been applied in these distributions.

\begin{figure}[t!]
\centering
 \includegraphics[width=0.48\textwidth]{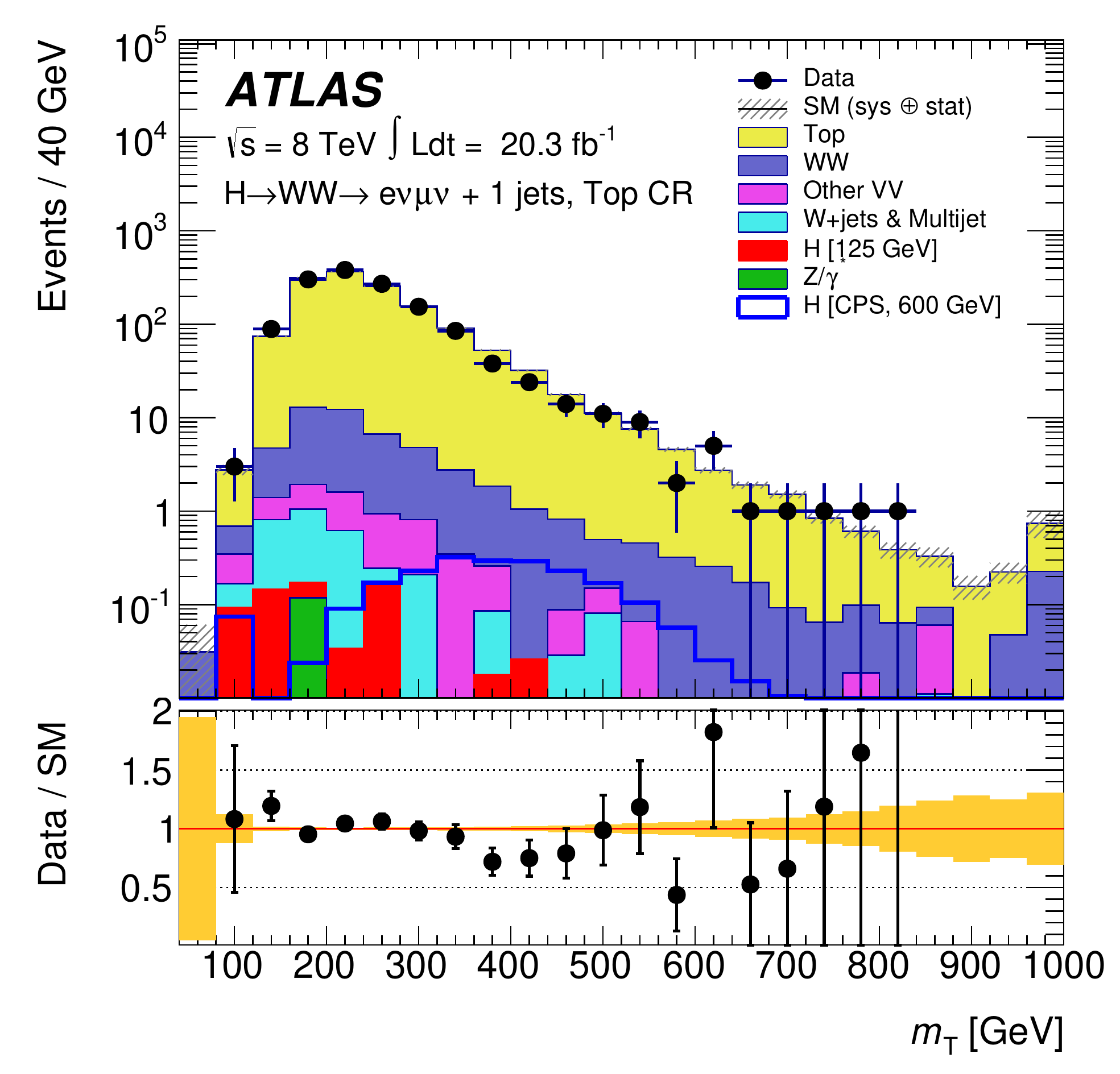}
 \includegraphics[width=0.48\textwidth]{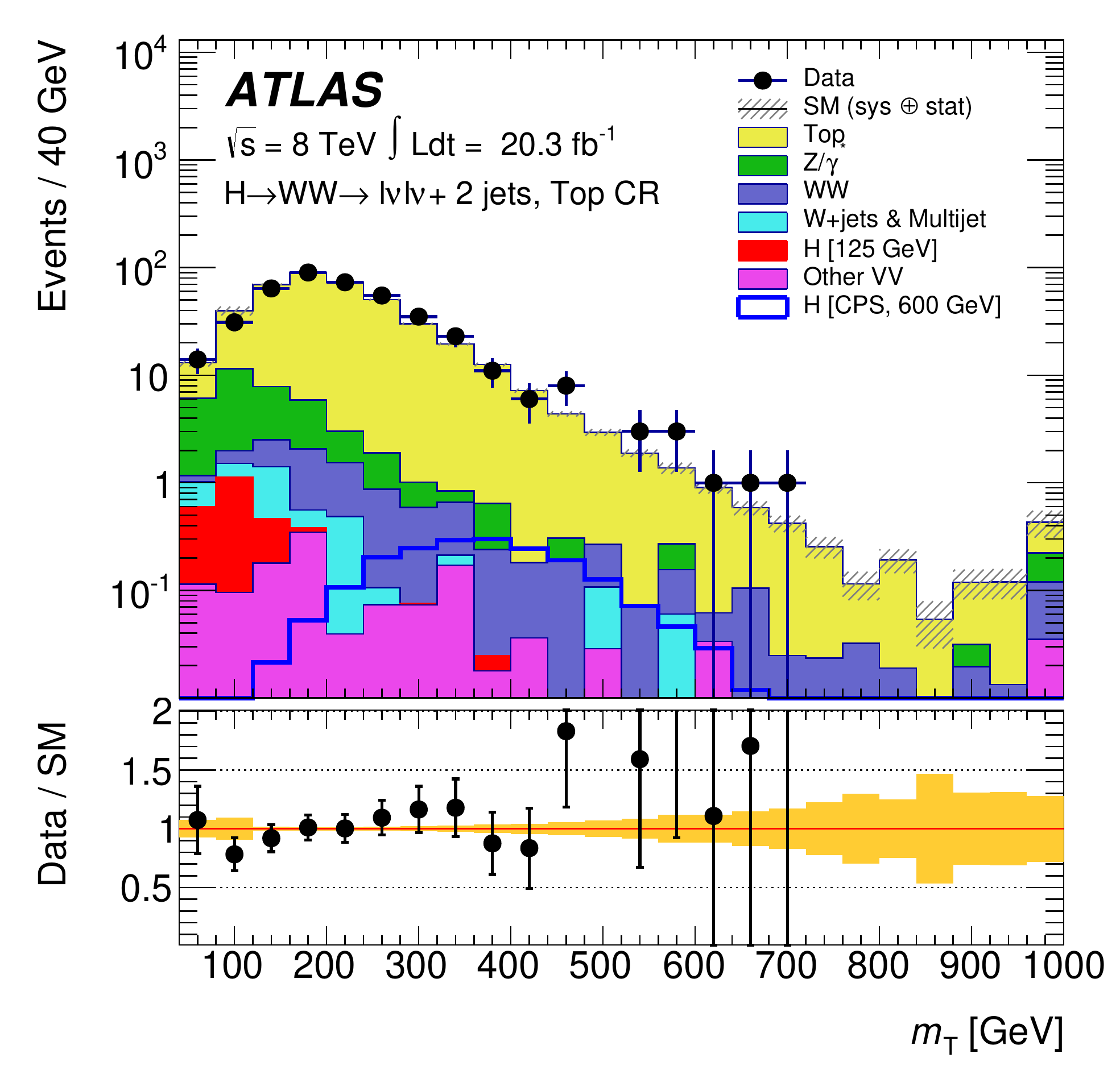}
\caption{Transverse mass distribution in the $\OneJet$ (left) and $\TwoJet$ (right) top control regions of 
        the $\hwwlnln$ analysis. In the $\OneJet$ category only the
        different-flavour final state is used; in the $\TwoJet$ category 
	different-flavour and same-flavour final states are used. In each figure, the
	last bin contains the overflow. The combined statistical and systematic uncertainties on the 
	 prediction are shown by the hatched band in the upper pane and the shaded band in the lower pane. 
	Normalisation factors obtained from a comparison of data and prediction have been applied in these figures.
	}
\label{fig:CR_Top}
\end{figure}

\subsubsection{$W$+jets and multijet background}
\label{subsec:lvlvWjets}

The procedures to estimate the $\Wjets$ and multijet backgrounds using data are described in more detail
in Ref.~\cite{atlas:ww-paper-run1} and summarised here.
The $\Wjets$ background contribution is estimated using a control sample of events in which one of the two lepton 
candidates satisfies the identification and isolation criteria used to define the signal sample (these lepton candidates 
are denoted ``fully identified''), and the other lepton fails to meet these criteria and satisfies a
less restrictive selection (denoted ``anti-identified''). Events in this sample are otherwise required to 
satisfy all of the signal selection criteria. The dominant component of this sample (85\% to 90\%) is $\Wjets$ events 
in which hadronic activity produces an object reconstructed as an anti-identified lepton. It may be either a non-prompt 
lepton from the decay of a hadron containing a heavy quark, or a particle from a jet reconstructed as a lepton candidate.

The $\Wjets$ contamination in the SR is determined by scaling the number of events in the control sample by an extrapolation factor, which is measured in a data
sample of jets produced in association with $Z$ bosons reconstructed in either the $e^+e^-$ or the $\mu^+\mu^-$ final state. 
Kinematic vetoes reduce contamination from $ZZ$ and $WZ$ events, and the expected remaining contribution is subtracted.
The extrapolation factor is the ratio of the
number of fully identified leptons to the number of anti-identified leptons, measured in bins of anti-identified lepton $\pT$ and $\eta$.
To account for differences between the jets associated with $W$ and $Z$ boson production, the extrapolation factors are measured
in simulated $\Wjets$ and $\Zjets$ events, and the ratio of the two extrapolation factors is multiplied by the one measured in the
$\Zjets$ data as a correction.  The central value of the correction factor is close to unity; differences among Monte Carlo generators 
for this ratio of about 20\% are observed and are taken as a systematic uncertainty.  

The background in the SR due to multijets is determined using a control sample that has two anti-identified lepton candidates, but
otherwise satisfies all of the SR selection criteria. An extrapolation factor is estimated using a multijet sample in data and applied
twice to the control sample. The sample used to determine the extrapolation factor is expected to have a similar sample composition in terms of heavy-flavour jets, light-quark jets and gluon jets as the control sample. 
Since the presence of one misidentified lepton in a multijet sample can change the sample composition with
respect to a multijet sample with no lepton selection imposed, corrections to the extrapolation factor 
are made that take into account such correlations.  These are evaluated using a multijet Monte Carlo sample and 
range from $1.0$ to $4.5$ depending on the lepton $\pT$ and flavour.  The uncertainty on these is 30--50\%,
with the dominant contribution being from the heavy-quark cross sections.

\section{The \HWWlvqq{} analysis}
\label{sec:lvqq}

In the $\HWWlvqq$ channel the final state consists of one
$W$ boson decaying into a quark-antiquark pair leading to a pair of jets, 
with the other $W$ boson decaying into a charged
lepton and a neutrino ($W\rightarrow\ell\nu$, with $\ell = e$ or $\mu$). 
This channel is particularly sensitive in searching for a
Higgs boson with a mass greater than twice the $W$ boson mass since 
$m_H$ can be reconstructed on an event-by-event basis and used 
as the discriminant to search for a signal. This event-by-event reconstruction 
is done using kinematic constraints that provide an estimate of the component 
of the neutrino momentum along the beam axis and require signal jets in 
the event to be consistent with coming from a hadronic $W$ decay.

\subsection{Event preselection and categorisation}
\label{sec:lvqq_preevsel}

Events are required to have exactly one reconstructed lepton candidate
($e$ or $\mu$) with $\pt>25\GeV$; no additional lepton with
$\pt>15\GeV$ is allowed. The selected lepton must match the object that
triggered the event. Events in the SR are required to have $\METcalo
> 60\GeV$ in order to suppress multijet processes while retaining
a high signal efficiency.

Jets are used to distinguish between ggF and VBF production as well as to 
reconstruct the hadronic $W$ boson decay.  
Anti-$k_t$ jets are selected with $\pt>30\GeV$ and $\left\vert \eta
\right\vert<4.5$, and large-$R$ jets are selected with $\pt>100\GeV$, 
$\left\vert \eta \right\vert<1.2$, and $m_{J}>40\GeV$, where the $J$ subscript
indicates a large-$R$ jet. Both the anti-$k_{t}$ and large-$R$ jets are required to be 
separated from the charged lepton by $\Delta R > 0.3$.  There is no explicit 
overlap removal between anti-$k_{t}$ and large-$R$ jets.

The momentum of $W$ bosons from the Higgs boson decay increases with 
increasing Higgs boson mass. This feature leads to a progressively 
smaller opening angle between the jets produced by the $W$ boson decay, 
making the jets difficult to distinguish using standard jet reconstruction
algorithms. To mitigate the resulting loss in signal efficiency, the hadronic $W$ decay 
may be reconstructed from either two anti-$k_{t}$ jets or one large-$R$ jet consistent with
originating from a $W$ boson decay.

In hadronic $W$ boson decays reconstructed from two anti-$k_{t}$ jets, 
the best candidate jet pair is referred to as the ``$W$ jets''.
The two jets with an invariant mass closest to the $W$ boson mass
are taken to be the $W$ jets, unless there is more than one jet pair with 
$|m_{jj} - m_W|< 15\GeV$.  In that case, the pair having the highest $\pt$ is chosen. 
Categorisation by production mode is done prior to identification of the hadronic $W$ boson, so
jets identified as the VBF tagging jets according to the procedure described below are excluded.
Hadronic $W$ boson decays are identified using a single large-$R$ jet if there 
is one with $\pT>100\GeV$ 
and a mass closer to the $W$ boson mass than the invariant mass of
the best dijet pair. In this case, the large-$R$ jet replaces the $W$ jets as
the candidate for the hadronically decaying $W$ boson.

Events are classified into two categories designed to distinguish between the 
ggF and VBF production modes, based on the number of jets in the event and the 
properties of those jets.  
In the category designed to be sensitive to the ggF production mode, referred to 
as the ggF selection, all events are required to have at least two anti-$k_{t}$ 
jets or at least one large-$R$ jet, and fail the VBF selection.

In the second category, designed to be sensitive to the VBF production mode, 
events are required to have at least four anti-$k_{t}$ jets or at least two 
anti-$k_{t}$ jets and one large-$R$ jet. Orthogonality between the ggF and VBF 
categories is ensured by identifying the two anti-$k_{t}$ jets $j_1$ and $j_2$ with the largest
invariant mass, and assigning the event to the VBF (ggF)
category if these jets pass (fail) to meet criteria, referred to as the VBF selection,
characteristic of the forward jets produced by the VBF process.  This VBF tagging jet pair is
required to have an invariant mass $m_{j_1,j_2}>600\GeV$, with the leading jet $\pt
> 40\GeV$, and be well separated in rapidity such that $\Delta y(j_1,j_2)=|y_{j_1}-y_{j_2}|>3$.
If the ratio of the ggF and VBF cross sections is as predicted by the SM,
63\% of signal events passing the full VBF preselection are produced
via VBF, and 93\% of signal events passing the ggF preselection
are produced via ggF.

Vetoes, based on the presence of $b$-jets in the event, reject $t\bar{t}$ background.
If both of the $W$ jets are $b$-tagged, the event
is vetoed. If only one of the $W$ jets is $b$-tagged, the event is
kept to maintain signal efficiency since a large fraction of jets from $W\rightarrow c\bar{s}$
decays are $b$-tagged. If any other jet in the event is $b$-tagged,
including the VBF tagging jets, the event is vetoed. If a large-$R$ jet
is used to reconstruct the $W$ boson, events with $b$-tagged jets outside of
$\Delta R = 0.4$ from the axis of the large-$R$ jet are vetoed. 
No flavour tagging is applied to large-$R$ jets.

% \textbf{\emph{Moved this first sentence forward.}}
Further selections are applied to ggF and VBF selected events. In
both categories, each of the $W$ jets is required to have $|\eta|<2.4$
and their invariant mass to be in the range $65\GeV\leq
m_{jj/J} \leq 96\GeV$, that is, close to the $W$ boson mass. 
Additionally, for hadronic $W$ boson candidates reconstructed from two anti-$k_{t}$ 
jets, one of the two $W$ jets is required to have
$\pt>60\GeV$ in both ggF and VBF selected events. Further requirements
are imposed on the azimuthal separation of reconstructed objects which
exploit the decay topology of signal events to improve the expected
sensitivity. A summary of the event preselection is shown in
Table~\ref{tab:preselection}.

The signal region is subdivided into exclusive categories which separate sources 
of signal and background. In addition to the ggF and VBF selection which
separates the two signal production modes, the signal regions are separated
by the flavour of the charged lepton and the sign of its charge.  Electrons
and muons are affected differently by the multijet background, and the categories
with positively-charged leptons have a higher proportion of $\Wjets$ background
because of the charge asymmetry of $W$ production in $pp$ collisions.

\begin{table}[t!]
   \centering
   \caption{Summary of event preselection in the \HWWlvqq~analysis. 
     The ``tagging jets'' $j_1$ and $j_2$ are the pair of anti-$k_t$ jets with the highest 
       invariant mass among all pairs in the event, and $j_1$ is the higher-$\pT$ jet in the pair.
       The decay topology selection differs between events in which the hadronic $W$ boson 
       candidate is reconstructed as a pair of jets $jj$ or as a single large-$R$ jet $J$.
       For the jet-pair topology, the leading jet is denoted $j_{\mathrm{lead}}$ and if only
       a single jet $j$ is referenced, the requirement is applied to both jets.
   }
   \begin{tabular}{lcc}
     \dbline
     Object selection 	& \multicolumn{2}{c}{1 isolated charged lepton ($e$ or $\mu$): $\pT>25\GeV,\ |\eta|<2.4$} \\
     			& \multicolumn{2}{c}{$\METcalo > 60\GeV$}     \\
     			& \multicolumn{2}{c}{jet: $\pT>30\GeV,\ |\eta| < 4.5$} \\
     			& \multicolumn{2}{c}{large-$R$ jet: $\pT>100\GeV,\ |\eta|<1.2$} \\
     \sgline
     VBF selection	& \multicolumn{2}{c}{($\geq$ 4 jets) or ($\geq$ 2 jets + $\geq$ 1 large-$R$ jets)} \\
     			& \multicolumn{2}{c}{$m_{j_1,j_2} > 600\GeV$}   \\
     			& \multicolumn{2}{c}{$p_{\mathrm{T}}^{j_1} > 40\GeV$} \\
     			& \multicolumn{2}{c}{$\Delta y(j_1,j_2) > 3.0$} \\
     \sgline
     ggF selection	& \multicolumn{2}{c}{not VBF tagged and ($\geq$ 2 jets or $\geq$ 1 large-$R$ jet)} \\
     \dbline
     Further selection, hadronic & & 	\\
     	\quad $W$ boson reconstructed as:		& jet pair			& large-$R$ jet	\\
     \sgline
     Decay topology 	& $\pT^{j_{\mathrm{lead}}}>60\GeV$ 	& - \\
     			& $\Delta\phi(jj) < 2.5$		& - \\
     			& $\Delta\phi(j,\ell) > 1.0$		& $\Delta\phi(J,\ell) > 1.0$ \\
     			& $\Delta\phi(j,\METcalo) > 1.0$	& $\Delta\phi(J,\METcalo) > 1.0$ \\
     			& \multicolumn{2}{c}{$\Delta\phi(\ell, \METcalo) < 2.5$} \\
     \sgline
     $b$-tagging &	&	\\
     \quad veto events with:  & both $W$ candidate jets $b$-tagged	& $b$-tagged jet with $\Delta R(j,J) > 0.4$ \\
     			& or any other jet $b$-tagged 	& - \\
     \sgline
     $W$-mass window  &  $65\GeV\leq m_{jj}\leq 96\GeV$	& $65\GeV\leq m_J\leq 96\GeV$   \\
     \dbline
   \end{tabular}
\label{tab:preselection}
\end{table}

\subsection{$WW$ invariant mass reconstruction}
\label{sec:lvqq_mass_recon}

The invariant mass of the $WW$ system is reconstructed 
from the four-momenta of the two $W$ boson candidates.  The reconstructed invariant mass
is denoted $\mlvjj$ regardless of whether the hadronic $W$ boson is reconstructed from 
a jet pair or a large-$R$ jet. 
The reconstruction of the leptonic $W$ boson decay relies on the charged
lepton and neutrino four-momenta.  The complete four-vector is measured
for the charged lepton, and the $\MPT$ provides the transverse components
of the neutrino momentum.  
The neutrino longitudinal momentum $p_z^\nu$ is computed using
the quadratic equation resulting from the mass constraint
$m(\ell\nu)=m(W)$. In the case of two real solutions of this equation,
the solution with the smaller $|p_{z}^{\nu}|$ is taken. In the case of
complex solutions, only the real part of the solution is taken. Based
on signal simulation, this procedure has been shown to give the
correct $p_z^\nu$ solution in $60$--$70$\% of events after the
preselection, depending on the event category and the Higgs boson
mass. The experimental mass resolution of the reconstructed Higgs boson 
varies from $\sim 30\GeV$ for
$\mH = 300\GeV$ to $\sim 60\GeV$ for $\mH = 1\TeV$. For $\mH = 420\GeV$,
the mass resolution is about the same as the width of a SM
Higgs boson at that mass, $\sim 36\GeV$. 

\subsection{Signal region selection}
\label{sec:lvqq_SR_selection}

The sensitivity to a heavy Higgs boson in the $\HWWlvqq$ channel is
improved by applying event selection in addition to the preselection
described in Section~\ref{sec:lvqq_preevsel}, as a function
of the Higgs boson mass hypothesis. 

The mass-dependent optimised selection is based on a set of kinematic 
quantities that discriminate between the signal from a hypothetical CP-even scalar and the background.
For the ggF selection, these are the leading jet $\pT$, the subleading jet $\pT$,
the large-$R$ jet $\pT$, the lepton $\pT$, $\Delta \phi_{jj}$, $\Delta
\phi_{\ell \nu}$ and \MET. For the VBF selection, these are $\Delta
\phi_{jj}$, $\Delta \phi_{\ell \nu}$ and $\pT$ balance, where $\pT$
balance is defined as

\begin{equation}
 p_{\mathrm{T},{\mathrm{balance}}}=\frac{\left(\boldsymbol{p}^{\ell} + \boldsymbol{p}^{\nu} + \boldsymbol{p}^{j_3} + \boldsymbol{p}^{j_4}\right)_{\mathrm{T}}}
   		{\pT^{\ell}+\pT^{\nu}+\pT^{j_3}+\pT^{j_4}}
\end{equation}
  
with $j_3$ and $j_4$ representing the $W$-jets, to distinguish them from
the tagging jets $j_1$ and $j_2$. In the case of large-$R$ jet
events, the terms representing $j_3$ and $j_4$ are replaced with a
single term $\pT^J$ which represents the large-$R$ jet momentum
instead. Also, the $\Delta\phi_{jj}$ selection is only applied to
events in which two resolved jets form the hadronically decaying $W$
boson. Fewer criteria are used for the VBF selection than
the ggF selection because of the smaller event yields in the VBF channel.

The selection is optimised as a function of the Higgs boson mass
through a two-step procedure. In the first step, the selection
that optimises the expected signal significance are
found in $100\GeV$ increments of the mass hypothesis $\mH$. The expected signal significance
is defined as $s/\sqrt{s+b}$, where $s$ is the number of expected
signal events and $b$ is the number of expected background
events. Other estimators for the significance have been tested
($s/\sqrt{b}$ and $s/\sqrt{b+\Delta b}$ with $\Delta b = 10\%$ and $30\%$) and shown to
provide the same optimal selection. Since the majority of signal
events are localised to a region in $\mlvjj$ that is small compared
with the overall fit region, the significance calculation does not
include all signal and background events, but rather only events in
which $\mlvjj$ is within a specified range around 
$\mH$, 
defined as the region that contains $90\%$ of the signal events.
The resulting selection criteria become stricter with increasing $\mH$
as the decay products are produced at higher momenta, allowing greater
background rejection while maintaining good signal efficiency.

In the second step, the selection is slightly relaxed, because the optimal selection,
particularly for low $\mH$, typically causes the peak of the signal $\mlvjj$ distribution to
coincide with that of the expected background, thus reducing the
sensitivity of the analysis because of large systematic uncertainties
in describing the turn-over region in $\mlvjj$. 
Typically, the optimal value for the next lower $\mH$ increment is used. 

\subsection{Background estimation}
\label{sec:lvqq_backgrounds}

In all signal regions, the background is expected to be dominated by
$W$+jets production, with other important contributions from $t\bar{t}$, single top, 
and multijet production that can be selected owing to the presence of leptons from
heavy-flavour decays or jets misidentified as leptons.  Diboson events, including
$WW$, $WZ$, $ZZ$, $W\gamma$, and $Z\gamma$, as well as $Z$+jets events, contribute
at a smaller level and are also accounted for as backgrounds.

The $\Wjets$ and top-quark backgrounds are modelled using 
simulation but their normalisations (one for $W$+jets and one for top-quark
backgrounds) are determined through a simultaneous fit to the signal and control 
regions, similarly to what is done for the $\ell\nu\ell\nu$ final state.  
The profile likelihood fit is described in more detail in Section~\ref{subsec:stats}.
Multijet backgrounds are estimated
using a CR selected in data to be enriched in leptons produced by hadronic
activity. 
The small additional background from $\Zjets$ and dibosons, including $Z\gamma$ and $W\gamma$, 
are estimated using their theoretical cross sections with simulation for the 
event selection acceptance and efficiency.

\subsubsection{$W$+jets and top-quark background}
\label{sec:lvqq_wjetstop}

$W$+jets and top-quark production are the most important
backgrounds in the $\HWWlvqq$ analysis. Their normalisations are set, and the $\mlvjj$ shape 
corrected, using the data observed in the corresponding CRs, defined below.

The $W$ control region (WCR) is defined similarly to the SR, but with the 
signal contributions suppressed by rejecting
events with a dijet mass, or a large-$R$ jet mass, consistent with the
hadronic decay of a $W$ boson. The WCR for the ggF selection is defined
by the upper and lower sidebands to the reconstructed $W$ boson mass,
\begin{eqnarray}
 52\GeV\,<\,m_{jj}\,<\,\phantom{1}65~\GeV\mathrm{,~ggF~lower~sideband}\\
 96\GeV\,<\,m_{jj}\,<\,126~\GeV\mathrm{,~ggF~upper~sideband}
\end{eqnarray}
using the dijet or large-$R$ jet mass closest to the $W$ mass, as described in
Section \ref{sec:lvqq_preevsel}. The corresponding sidebands in the
WCR for the VBF selection are
\begin{eqnarray}
 43\GeV\,<\,m_{jj}\,<\,\phantom{1}65~\GeV\mathrm{,~VBF~lower~sideband}\phantom{.}\\
 96\GeV\,<\,m_{jj}\,<\,200~\GeV\mathrm{,~VBF~upper~sideband}.
\end{eqnarray}
The width of the sidebands is increased in the VBF selection 
compared to the ggF selection to improve the statistical precision
of the background estimate. For both the ggF and VBF selection, the
number of background events is similar in the upper and lower sidebands, so 
no statistical bias is generated by using them jointly to define
a common background normalisation.  Separate CRs are defined
for the ggF and VBF selection, but all lepton flavours and charges as well as 
the large-$R$ and dijet $W$ reconstruction topologies are merged into a single CR.
The CR selection also follows the 
mass-dependent SR selection described in Section~\ref{sec:lvqq_SR_selection}.
The purity of the WCR depends on the particular selection, but varies between about 70\% and over 80\%, 
with higher purities for higher $\mH$ selection and for ggF compared to VBF.
The $\Wjets$ normalisation factors are consistent with unity and stable with respect to the 
$\mH$-dependent selection and ggF vs.~VBF selection.  The normalisation factors resulting
from the simultaneous fit to the signal and control regions range from $0.9 \pm 0.1$ to $1.3 \pm 0.3$.
The quoted uncertainties include statistical and experimental systematic
components.  The highest values are found with the VBF and high-$\mH$ selection.

Figure~\ref{fig:reweight:w} shows the $\mlvjj$ distributions
observed in the ggF and VBF WCRs.
Non-$W$-boson contributions are subtracted from the data, and the resulting
distribution is compared to the prediction from simulated $\Wjets$ events, after
normalising the prediction to the data. The observed distributions differ substantially from those
predicted in both the ggF and VBF WCRs. An alternative MC sample, generated
using $\ALPGEN+\PYTHIA6$, results in better agreement with the data in the WCR,
but does not have enough events for a statistically precise prediction in the 
signal region for large values of $\mH$.

To correct the observed mismodelling, simulated $\Wjets$ events 
are reweighted using the $\mlvjj$ distribution observed in the WCR.
In order to obtain a smooth function to use for the reweighting,
the ratio of the data to the prediction is fit with polynomial functions.
The degree of the polynomial is chosen to have enough flexibility to yield a good fit quality,
the fit range being restricted to $\mlvjj$ values where a statistically meaningful fit can be made.
For the ggF WCR, shown on the left side of Figure~\ref{fig:reweight:w}, a second-order polynomial 
function is used and the fit is extended to $\mlvjj=1.7\TeV$. Above that
value of $\mlvjj$, a constant function at the value of the polynomial 
at $\mlvjj = 1.7\TeV$ is used.  
For the VBF WCR, shown on the right side of Figure~\ref{fig:reweight:w}, a third-order polynomial
function is used, up to $\mlvjj=0.9\TeV$.  Fits extending to higher values of $\mlvjj$
have been attempted but require either a more complex fitting function or have a visibly
poor-quality fit to the data.  For simplicity, a constant function is used for $\mlvjj > 0.9\TeV$,
as illustrated in the figure.  The value of the function is the value of the third-order polynomial 
at $\mlvjj = 0.9\TeV$.

\begin{figure}[t!]
 \centering
 \includegraphics[width=0.49\textwidth]{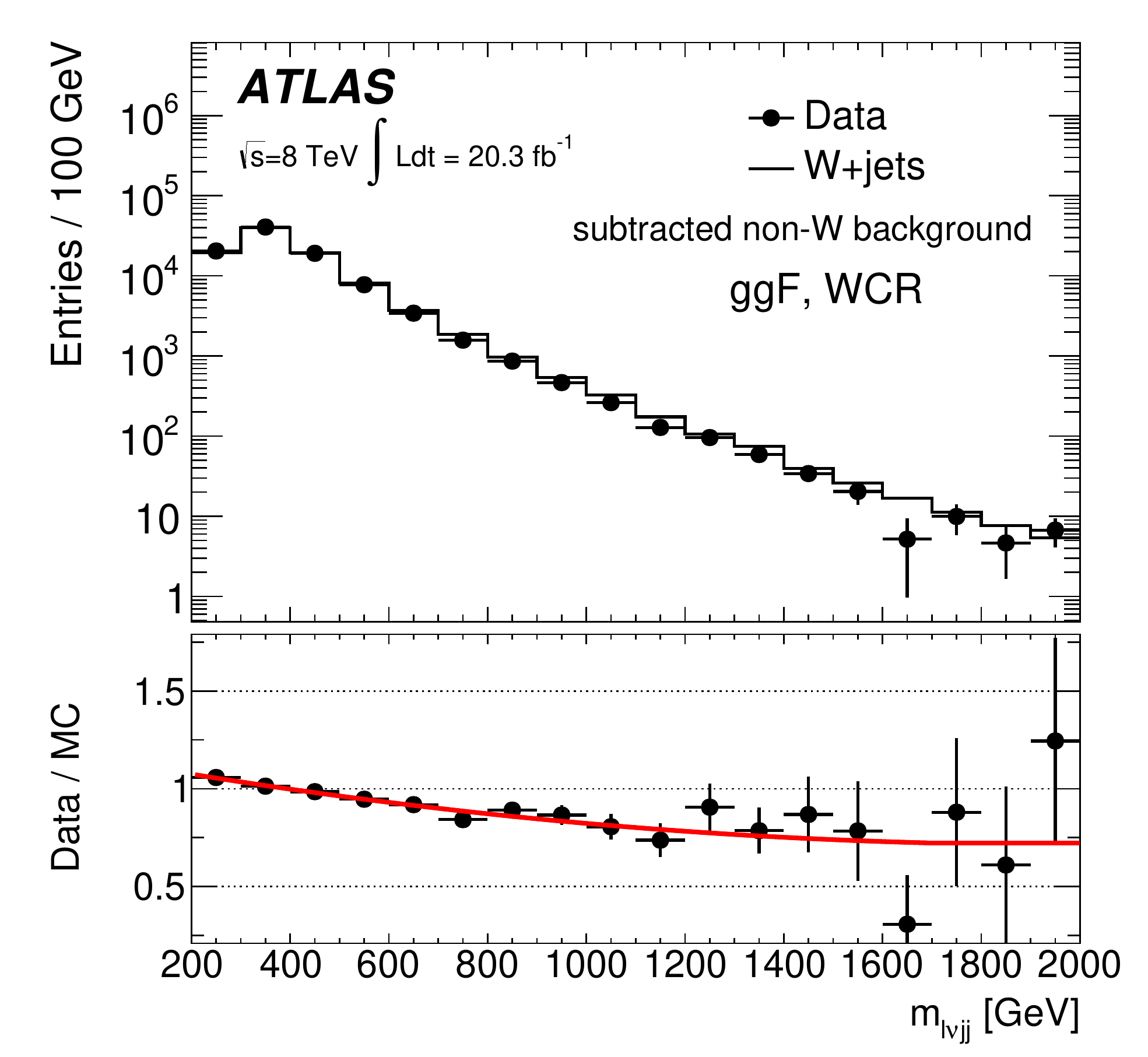}
 \includegraphics[width=0.49\textwidth]{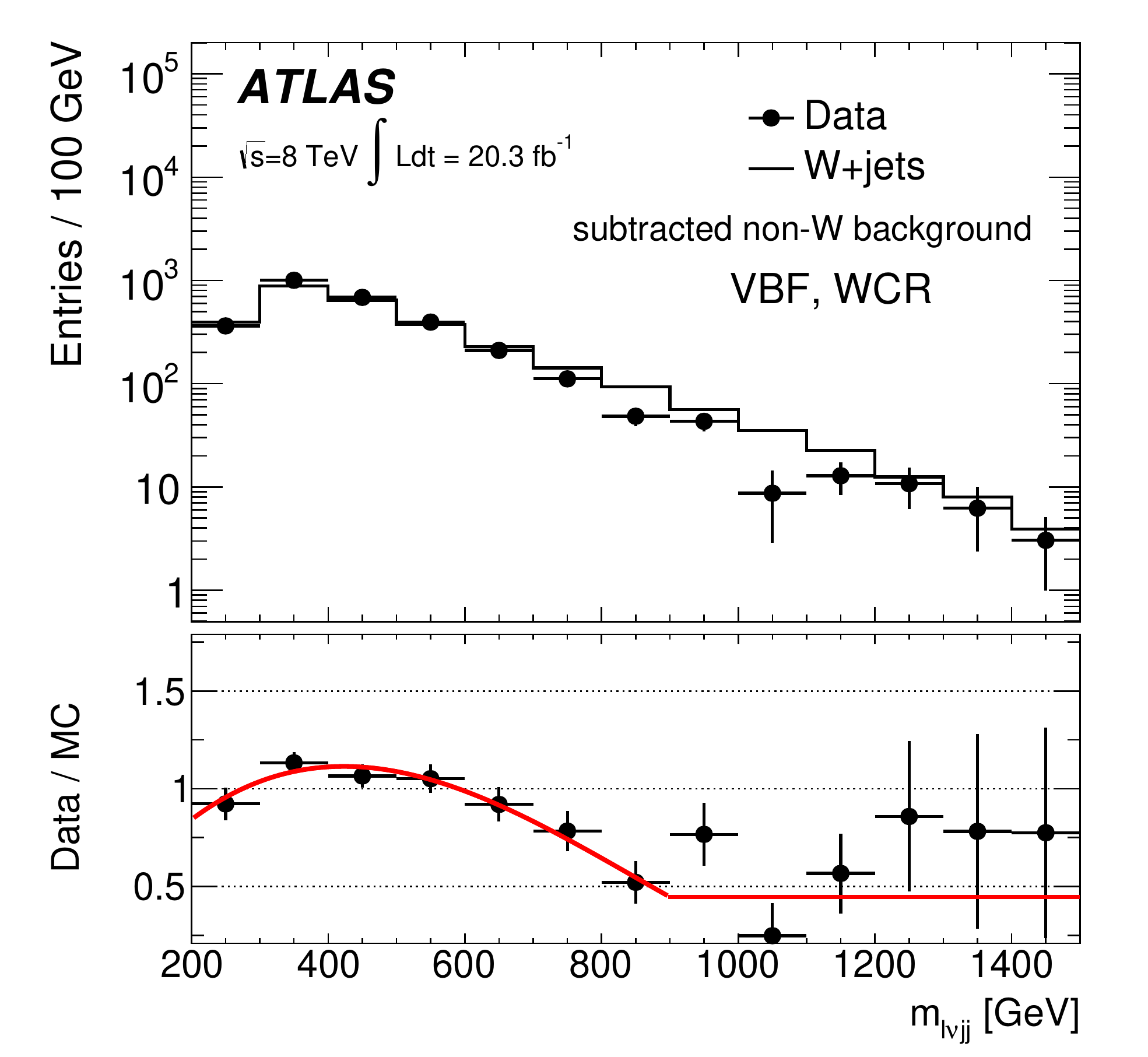}
 \caption{Data and Monte Carlo comparison of the shape of the invariant mass of the $WW$ system
   $\mlvjj$ in the ggF (left) and VBF (right) WCRs after the
   $m_{H}=300\GeV$ selection for the $\HWWlvqq$ analysis. All the lepton flavour and charge
   categories are summed together. To isolate the effects of $W$+jets background
   modelling, other contributions (top,
   diboson, $\Zjets$, multijet) are subtracted from the data. The Monte Carlo distributions are normalised to the remaining data.
   The ratio of the data to the Monte Carlo distribution is shown in the bottom panel, along with a 
   red line showing the resulting weights that are applied to correct the Monte Carlo predictions
   in the rest of the analysis.
 }
 \label{fig:reweight:w}
\end{figure}

The top-quark control region (TopCR) is designed to be as pure as
achievable for the second largest background, $t\bar{t}\to WbWb\to
\ell\nu jj+bb$. The event topology of this background is similar to
that of the Higgs boson signal, but contains two characteristic
$b$-jets. The TopCR is defined to be identical to the SR, but with the 
$b$-jet veto reversed. As with the WCR, the TopCR region selection follows the 
mass-dependent signal region selection described in Section~\ref{sec:lvqq_SR_selection},
and lepton flavours and hadronic $W$ topologies are merged but separate
TopCRs are defined for the ggF and VBF topologies.
The purity of the TopCR is about 80\% and does not depend strongly on the region-specific selection.
Similarly, the value of the resulting normalisation factor is stable with respect to the
kinematic selection and is consistent with unity within the uncertainties. The values of
the normalisation factor found by the simultaneous fit to the signal and control regions range from
$0.9 \pm 0.1$ to $1.3 \pm 0.2$. Both extremes occur in the VBF control region for $\mH$-dependent
selection for $\mH \ge 700\GeV$, which is most subject to statistical fluctuations.

The $\mlvjj$ distributions in the ggF and VBF TopCRs are shown in
Figure~\ref{fig:reweight:top}. As in Figure~\ref{fig:reweight:w}, 
processes other than top-quark single and pair production are subtracted from
the data, and the resulting distribution is compared to the top-quark prediction after normalising the 
prediction to the data. Similarly to the WCR, differences in shape are
observed, and simulated top-quark events are reweighted accordingly as a function 
of $\mlvjj$.  A first-order polynomial function is fit to the data. 
Since the purity of events with a top quark in the TopCR is very high, the $W$+jets contribution 
in Figure~\ref{fig:reweight:top} has no $\mlvjj$ reweighting applied.

\begin{figure}[t!]
 \centering
 \includegraphics[width=0.49\textwidth]{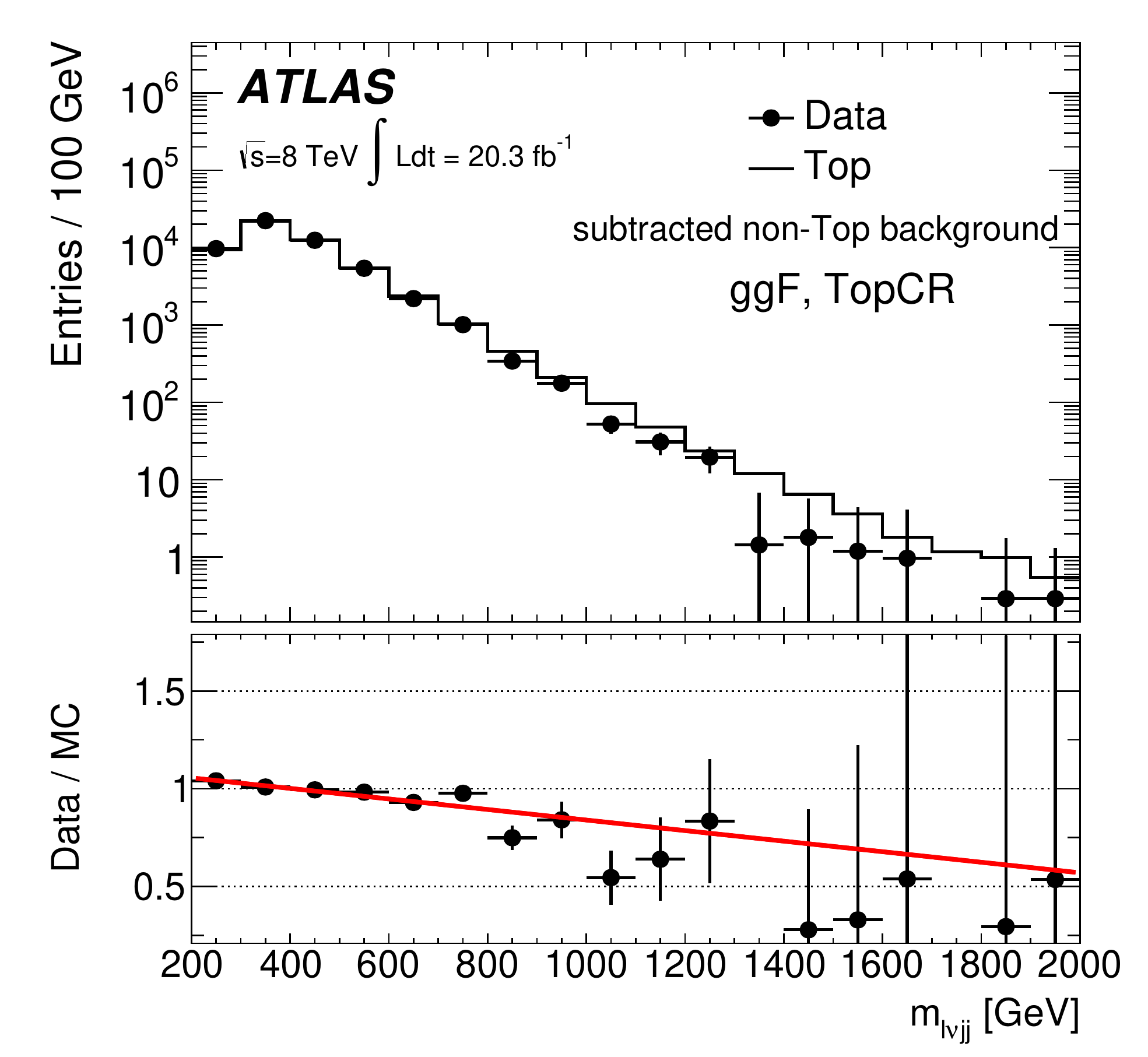}
 \includegraphics[width=0.49\textwidth]{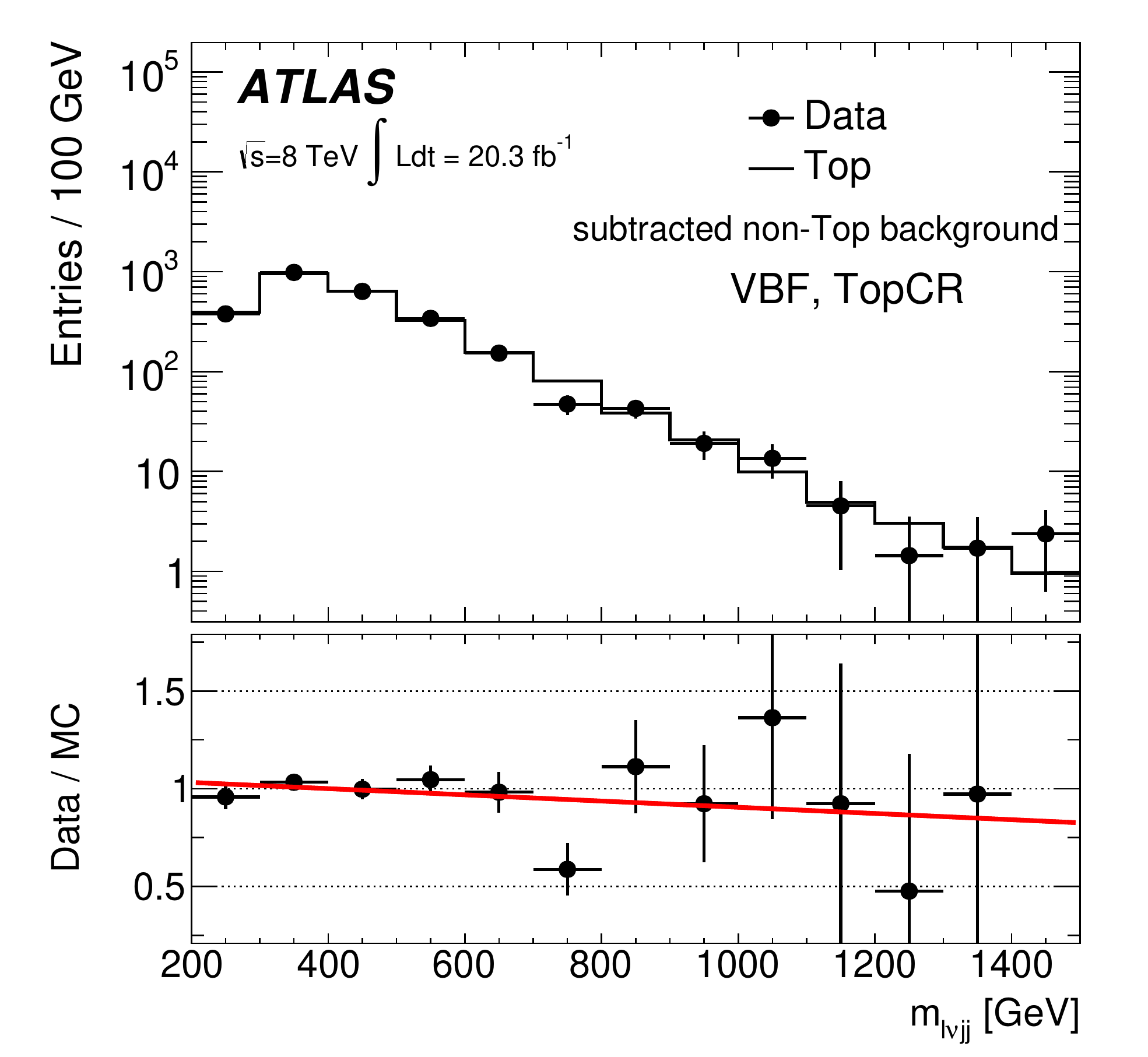}
 \caption{Data and Monte Carlo comparison of the shape of the invariant mass of the $WW$ system
   $\mlvjj$ in the ggF (left) and VBF (right) TopCRs after the
   $m_{H}=300\GeV$ selection for the $\HWWlvqq$ analysis. All the lepton flavour and charge
   categories are summed together. To isolate the effects of top-quark background
   modelling, other contributions ($\Wjets$,
   diboson, $\Zjets$, multijet) are subtracted from the data. The Monte Carlo distributions are normalised to the remaining data.
   The ratio of the data to the Monte Carlo distribution is shown in the bottom panel, along with a 
   red line showing the resulting weights that are applied to correct the Monte Carlo predictions
   in the rest of the analysis.
 }
 \label{fig:reweight:top}
\end{figure}

For both the $\Wjets$ and top-quark backgrounds, the event weights used to
correct the MC simulated events are derived from corresponding CRs after the $m_{H}=300\GeV$
selection, because this is the most inclusive selection. 
The higher-mass criteria select subsets of the events accepted 
by the $300\GeV$ selection.  The agreement between the data and MC distributions
in the $W$ and top-quark CRs when other mass-dependent selection criteria are applied
is consistent with the results from the $300\GeV$ selection.
The reweighting is event-by-event, using the fitted functions
shown in Figures~\ref{fig:reweight:w} and~\ref{fig:reweight:top} and
the $\mlvjj$ value of each simulated $W$+jets or top-quark
background event. The reweighting is applied after the preselection,
and therefore propagates to all signal and control regions. 
For events with $\mlvjj$ above the fitted range, the weight at the
upper boundary of the fit is used.

Half of the difference between the nominal and 
the reweighted $\mlvjj$ distribution is taken as a systematic
uncertainty on the $\mlvjj$ shape of these backgrounds. These
uncertainties are included as Gaussian constraints in the profile
likelihood fit, which allows the fit to adjust the shape of these
backgrounds using the shape of the data in the signal regions.

\subsubsection{Multijet background}
\label{sec:lvqq_multijet}

The shapes of multijet background distributions are modelled using
histograms derived from data samples selected similarly as for signal events, except that the
lepton identification requirements are loosened, while the isolation
requirement is not changed. In  the electron channels, a loosened
identification selection is applied to the data, with a veto for
electrons selected using the standard criteria. In the 
muon channels, the impact parameter significance requirement is reversed.

The normalisation of the multijet background in a given
event category is derived from a standalone template fit (separate from the final simultaneous
fit) to the \met\ distribution without the
\met\ requirement applied. The template for the multijet background in each 
region is taken from data selected with modified lepton selection as described
above, but otherwise following the event selection of that signal or control region.
The relative contributions of backgrounds other than the
multijet background are fixed to their SM expectations.

The multijet background is relatively small, constituting between 1\% and 4\% of
the total expected background depending on the selection used, with 
smaller contributions for the mass-dependent SR selection applied at higher values of $\mH$.
This background is also concentrated at low values of $\mlvjj$, so its
effect is reduced by the $\mlvjj$ shape fit.  

\section{Systematic uncertainties}
\label{sec:syst}

This section describes the systematic uncertainties affecting the analysis results.  Experimental and theoretical 
uncertainties common to the $\hwwlnln$ and $\hwwlnqq$ analyses are described first, followed by a discussion
of uncertainties particular to each channel.

\subsection{Common experimental uncertainties}

The dominant sources of experimental uncertainties on the signal and background yields are the jet
energy scale and resolution, and the $b$-tagging efficiency. Other sources of uncertainty include those on the
scale and resolution of the lepton energy or momentum, lepton identification and trigger efficiencies, the scale and resolution 
of the missing transverse momentum, and the luminosity calibration.  All experimental uncertainties are treated
by varying the object subject to a particular uncertainty and then re-running the full analysis.

The jet energy scale is determined from a combination of test-beam data, simulation, and \emph{in situ}
measurements~\cite{atlas:jet-paper}. Its uncertainty is split into several independent categories:
modelling and statistical uncertainties on the extrapolation of the jet calibration from the central region ($\eta$
intercalibration), high-$\pT$ jet behaviour, Monte Carlo non-closure uncertainties, uncertainties on the jet quark and gluon
compositions and their calibrations, the $b$-jet energy scale uncertainties, uncertainties due to modelling of in-time and
out-of-time pile-up, and uncertainties on \emph{in situ} jet energy corrections. Some of these categories
are further subdivided by the physical source of the uncertainty. For the anti-$k_{t}$ jets used in these analyses, 
the jet energy scale uncertainty ranges from $1\%$ to $7\%$ depending on $\pT$ and $\eta$.
The resolution varies from $5\%$ to $20\%$, and the relative uncertainty on the resolution 
ranges from $2\%$ to $40\%$. The lowest-$\pT$ jets, immediately above the jet selection thresholds, have both 
the poorest resolution and the largest uncertainty.

The evaluation of the $b$-jet tagging efficiency uses a sample dominated by dileptonic decays of top-quark
pairs~\cite{atlas:btagperf-2014-note}. 
To improve the precision,
this method is combined with a second calibration method based on samples containing muons reconstructed in the vicinity
of the jet. The uncertainties related to $b$-jet identification are decomposed into six uncorrelated components using an
eigenvector method~\cite{atlas:btagcalib-2014-note}, the number of components being equal to the number of $\pT$ bins used
in the calibration. The uncertainties range from $<1\%$ to $7.8\%$. The uncertainties
on the misidentification rate for light-quark jets depend on $\pT$ and $\eta$, with a range of $9\%$--$19\%$.
The uncertainties on $c$-jets reconstructed as $b$-jets range between $6\%$--$14\%$ depending on the jet $\pT$.

The reconstruction, identification, isolation, and trigger efficiencies for electrons and muons, as well
as their momentum scales and resolutions, are estimated
using $Z{\to}ee, \mu\mu$, $\Jpsi{\to}ee, \mu\mu$, and $W{\to}e\nu, \mu\nu$ decays~\cite{atlas:muon-id-paper,
atlas:el-id-note, atlas:el-id-paper}. The uncertainties on the lepton identification and trigger
efficiencies are smaller than $1\%$ except for the uncertainty on the electron identification efficiency,
which varies between $0.2\%$ and $2.7\%$ depending on $\pT$ and $\eta$.

The changes in jet energy and lepton momenta due to systematic variations are propagated to $\MET$, such that changes in the
high-$\pT$ object momenta and in $\MET$ are fully correlated. Additional contributions to the $\MET$
uncertainty arise from the modelling of low-energy particle measurements (``soft terms'')~\cite{MET}. Calorimeter measurements of
these particles, used in $\METcalo$, use calibrated clusters of cells not associated with reconstructed
physics objects, with a noise threshold applied.  The longitudinal and perpendicular components of the soft terms, defined with respect to the \MET\
computed using hard objects, are smeared and rescaled to evaluate the associated uncertainties. 
The uncertainties are parameterised as a function of the magnitude of the vector sum $\vec{\pT}$ of the 
high-$\pT$ objects, and are evaluated in bins of the average number of interactions per bunch crossing.
Differences of the mean and width of the soft term components between data and simulation result in variations 
on the mean of the longitudinal component of about $0.2\GeV$.  The uncertainty on the
resolution of the longitudinal and perpendicular components is $2$\% on average.
The systematic uncertainties related to the track-based soft term for $\MPT$ 
are calculated by comparing the properties of $\MPT$ in $Z\,\to\,ee, \mu\mu$
events in data and simulation as a function of the magnitude of
the summed $\vec{\pT}$ of the leptons and jets in the event. The
variations on the mean of the longitudinal component
are in the range $0.3$--$1.4\GeV$ and the uncertainties on the resolution on
the longitudinal and perpendicular components are in the range
$1.5$--$3.3\GeV$, where the lower and upper bounds correspond to the
range of the sum of the hard $\pT$ objects below $5\GeV$ and above
$50\GeV$, respectively.

The uncertainty on the integrated luminosity is $2.8\%$. It is derived following the same methodology as in 
Ref.~\cite{Aad:2013ucp} from a calibration of the luminosity scale derived from beam-separation scans.

\subsection{Common theoretical uncertainties}

Theoretical uncertainties on the signal production cross section affect the $\hwwlnln$ and $\hwwlnqq$ 
analyses in the same way.  These include uncertainties
due to the choice of QCD renormalisation and factorisation scales, the PDF model used to evaluate the cross section and
acceptance, and the underlying event and parton shower
models.  These are described and evaluated as a function of $\mH$ in Refs.~\cite{LHCHiggsCrossSectionWorkingGroup:2011ti,LHCHiggsCrossSectionWorkingGroup:2012vm}. The QCD scale uncertainty on
the inclusive signal cross sections is evaluated to be 8\% for ggF and 1\% for VBF
production. The PDF uncertainty on the inclusive cross sections is 8\% for ggF and 4\% for VBF production.
Uncertainties on the interference weighting of the CPS signal samples, described in Section~\ref{sec:mc}, are also
included.

\subsection{Uncertainties specific to the \hwwlnln\ analysis}

The uncertainties specific to the $\hwwlnln$ analysis arise primarily from the theoretical modelling of
the signal acceptance in jet bins, the theoretical models used in the background predictions, and the 
additional consideration of uncertainties on the $\mT$ shape used in the likelihood fit.
Statistical uncertainties on yields in control regions, and the effect of subtracting other processes
from the control region yield, are also included in the total uncertainty on background yields predicted
using control regions.

Since the analysis is binned by jet multiplicity, large uncertainties from variations of QCD renormalisation and factorisation
scales affect the predicted contribution of the ggF signal in the exclusive jet bins, and can cause event migration
among bins. These uncertainties are estimated using the HNNLO program~\cite{Catani:2007vq,Grazzini:2008tf} and the
method reported in Ref.~\cite{Stewart:2011cf} for Higgs boson masses up to 1500 GeV. The sum in quadrature of the inclusive jet
bin uncertainties amounts to 38\% in the 0-jet category and 42\% in the 1-jet category for $m_H$ = 600 GeV.
For $m_H$ = 1 TeV, these uncertainties are 55\% and 46\%, respectively.

For the backgrounds normalised using control regions, theoretical uncertainties arise from the use of 
simulations of the background used in the extrapolation from the control region to the signal region. For the $WW$ background
in the \ZeroOneJetSimple\ categories and the top-quark background in the $\OneJet$ and $\TwoJet$ jet categories, theoretical uncertainties on the
extrapolation are evaluated according to the prescription of Ref.~\cite{LHCHiggsCrossSectionWorkingGroup:2012vm}. The uncertainties
include the impact of missing higher-order QCD corrections, PDF variations and MC modelling. For backgrounds 
normalised to the theoretical prediction without use of a control region, a similar prescription is followed for
the acceptance uncertainty, and uncertainties on the predicted inclusive cross section also apply.

For the $WW$ background, the uncertainties on the control region extrapolation in the 0- and 1-jet categories 
amount to 4.2\% and 9.7\%, respectively. 
In the $\ge$ 2 jet category, the $WW$ yield is taken from the theoretical expectation. 
The PDF uncertainty on the cross sections of the $q\bar{q} \rightarrow WW$ + 2 jets and
$gg \rightarrow WW$ + 2 jets processes is evaluated to be 4\% and that on the acceptance of these processes to be 2\%. The QCD scale
uncertainty on QCD $WW + 2$ jets is 14\%, while that on the acceptance is 20\%. The modelling uncertainty is derived by comparing
samples generated with {\sc Sherpa} and MadGraph~\cite{Alwall:2007st} generators. This uncertainty amounts to 34\% for QCD $WW + 2$ jets and 7\%
for EW $WW + 2$ jets after the selection on \mll.
In all jet categories, a correction is applied to the $WW$ background to take higher-order EW corrections to the cross section
into account. A conservative uncertainty of 100\% is applied to this correction.

% \textbf{\emph{Added notation from new equation in Sec.~\ref{sec:lvlv_mt}.}}
For the top-quark background estimation in the 0-jet category, a 3.9\% theoretical uncertainty is assigned 
from the use of simulated top-quark events to model the ratio of the signal region
jet veto efficiency to the square of the efficiency of the additional-jet veto in the $b$-tagged control region 
($\epsilon_0 / (f_0^{\textrm{MC}})^2$ in Eq.~\ref{eq:jvsp}).
An additional 4.5\% theoretical uncertainty is used on the efficiency $\epsilon_{\textrm{rest}}$ on the remaining selection that defines the
$\ZeroJet$ signal region, which is also derived from simulated top-quark events.
The most important component of these uncertainties is the variation among predictions by different MC generator
  and parton shower algorithms.  Smaller uncertainties attributable to the QCD scale choice, PDF model, $t\bar{t}$--$Wt$ 
    interference, and the single-top cross section are also included.
For the top-quark background in the 1- and $\geq$ 2 jet categories, the uncertainties are respectively 5.7\% and 9.8\%.

The main uncertainty on the $\Wjets$ and multijet background predictions arise from the extrapolation factors
relating anti-identified to identified leptons.  For $\Wjets$, a modelling uncertainty on the simulation-based correction applied
to the $\Zjets$ extrapolation factor of about 20\% and the statistical uncertainty from the $\Zjets$ data used to 
measure the extrapolation factor contribute in roughly equal proportions.  The main uncertainty on the multijet contribution
arises from an uncertainty on the modelling of the correlation between the extrapolation factors for two anti-identified
leptons in the same event.

In the same-flavour channel in the $\ge$ 2 jet category, the Drell--Yan background is non-negligible. The background is estimated from
simulation in this final state as in the other final states, but a 15\% theoretical uncertainty is assigned to the cross section of the
process using the total relative theoretical uncertainty on the $Z$ + 2 jets cross-section prediction in the high-$m_{jj}$ region, following
Ref.~\cite{Aad:2014dta}.

In addition to the uncertainties on the normalisation of backgrounds, experimental and theoretical uncertainties
on the $\mT$ shape model used in the fit are considered.  
Simulated data are used for the $\mT$ shape for signal and all backgrounds except for $\Wjets$.
Only shape variations which are statistically significant compared to the statistical uncertainty from the simulation
model of the shape are included.  For the background model, uncertainties due to $b$-tagging efficiency, 
lepton identification, trigger, and isolation efficiency scale factors fall into this category, as does the uncertainty on
the extrapolation factors for the $\Wjets$ and multijet background estimates.  
For the ggF signal, the experimental systematic uncertainties from $\met$ are treated 
as $\mT$ shape uncertainties. The theoretical uncertainty from the 
interference weighting are also included as a systematic uncertainty on the $\mT$ shape 
for the ggF CPS signal samples.

\subsection{Uncertainties specific to the \hwwlnqq\ analysis}

Uncertainties in the $\hwwlnqq$ analysis are analogous to those in the $\hwwlnln$ analysis, but
different uncertainties are prominent. The absence of exclusive jet binning means that there are no additional
theoretical uncertainties on the signal acceptance.  Experimental uncertainties, particularly those
relating to the modelling of jet energies, are more important. 

In addition to the common experimental uncertainties, the $\hwwlnqq$ results have an uncertainty
arising from the energy resolution of large-$R$ jets.  This uncertainty is determined
by matching reconstructed jets in simulated events with their associated particle-level jets,
and computing the ratio of the reconstructed energy (mass) to their true values as determined 
after parton showering and hadronisation.
Previous ATLAS studies of large-$R$ jet
energy/mass resolution indicate that the resolution in simulation can vary by up to
20\%~\cite{STDM-2011-19}. Based on these studies, a systematic uncertainty is estimated by smearing 
the jet energies by a factor corresponding to an increase of 20\% in
their resolution in bins of $\pT$ and $|\eta|$.

The dominant uncertainty on the background modelling in the $\hwwlnqq$ analysis is that on 
the shape of the $m_{l\nu jj}$ spectrum for the $\Wjets$ and top-quark backgrounds.
This uncertainty is 50\% of the difference between the reweighted and nominal $m_{l\nu jj}$ distributions,
as described in Section~\ref{sec:lvqq_wjetstop}, and is taken as a systematic uncertainty on the shape of
the background distribution. 

The systematic uncertainty on multijet production is determined 
by comparing the nominal background estimate with an estimate derived using different lepton selection criteria including a
reversed isolation requirement. Uncertainties are derived for different production mechanisms
and decay channels as a function of the Higgs boson mass, and vary from about 10\% to 100\% of the multijet background.

\section{Signal and background predictions compared to data}
\label{sec:results}

\subsection{The \hwwlnln\ analysis}
\label{sec:results_lvlv_mt}

In Table~\ref{tab:lvlvSRyields}, the expected signal and background in the \hwwlnln\ analysis 
are summarized and compared to the number of data events passing the signal region selection.
The CPS scenario is used for the signal, and the SM Higgs boson cross section is used 
to normalise it.  For this comparison only, to give an indication of the sensitivity of the shape 
fit, an $\mT$ window is added to the selection.  The $\mT$ 
requirements are chosen to be about 80\% efficient for the signal, and are:
$180\GeV < \mT < 270\GeV$ for $\mH = 300\GeV$, $250\GeV < \mT < 500\GeV$ for $\mH = 600\GeV$,
and $300\GeV < \mT < 750\GeV$ for $\mH = 900\GeV$.  
The predicted event yields given do not include adjustments resulting from the fit to data.

The systematic uncertainty on the background is derived from the expected uncertainty on the
fit results, by fitting the nominal expected signal and background using the profile likelihood 
described in Section~\ref{sec:interp}.  The total expected uncertainty on the fit result has 
contributions from the theoretical uncertainties on the signal and background predictions, the 
experimental uncertainties, and the statistical uncertainties.  After removing the
signal theoretical uncertainties and the statistical uncertainty in quadrature, the approximate
systematic uncertainty on the background, accounting for all systematic correlations among
the backgrounds, can be extracted.  

The fractional background composition for each $\Njet$ category is given in the bottom half of the table.
The relative contributions of the various sources to the total background does not vary substantially as a 
function of $\mT$, so the $\mT$ selection described above is not applied.

\begin{table*}[tb!]
\centering
\caption{Summary of the expected signal and background in the \hwwlnln\ signal regions.
  The top table compares the observed number of candidate events in data $N_{\mathrm{data}}$ with the expected 
  signal $N_{\mathrm{sig}}$ for several $\mH$ values and the total background $N_{\mathrm{bkg}}$, along with
  its statistical ($\sqrt{N_{\mathrm{bkg}}}$) and systematic ($\delta N_{\mathrm{bkg}}^{\mathrm{sys}}$) uncertainties.
  The predictions are quoted for an $\mT$ interval which is about 80\% efficient for the signal (details in the text).
  The different-flavour and same-flavour final states are combined, and in the top table 
  the $\ZeroJet$ and $\OneJet$ categories are summed. 
  The bottom table shows the composition of the background for the analysis
  categories of different jet multiplicity. The $VV$ background category includes all diboson
processes except for $WW$. The $Wj+jj$ column contains the sum of the $\Wjets$ and multijet backgrounds. }
\vspace{0.3cm}
\label{tab:lvlvSRyields}
\begin{tabular}{rlD{.}{.}{-1}D{.}{.}{-1}r@{}c@{}lrr}
\dbline
	 & \multicolumn{3}{c}{CPS signal expectation} & \multicolumn{3}{c}{Bkg.~expectation} & Observed \\ 
Category & $\mH$  & \multicolumn{1}{c}{$N_{\mathrm{sig}}^{\mathrm{ggF}}$} & \multicolumn{1}{c}{$N_{\mathrm{sig}}^{\mathrm{VBF}}$} 
& $N_{\mathrm{bkg}}$ & $\,\pm\,\sqrt{N_{\mathrm{bkg}}}$ & $\,\pm\,\delta N_{\mathrm{bkg}}^{\mathrm{sys}}$ & $N_{\mathrm{data}}$ \\ 
\sgline
$\ZeroOneJetSimple$	& $300\GeV$  &  144   &  10  & 961 & $\pm$ 31 & $\pm$ 47  & 951   \\ 
    	 		& $600\GeV$  &   29   &   3  & 584 & $\pm$ 24 & $\pm$ 12  & 538    \\ 
			& $900\GeV$  &    5   &   2  & 325 & $\pm$ 18 & $\pm$ \phantom{0}8  & 290     \\ 
\sgline
$\TwoJet$		& $300\GeV$  &    3.1 &  13.5 & 18   & $\pm$ 4.2 & $\pm$ 5\phantom{.2} &  20  \\ 
    			& $600\GeV$  &    0.8 &   4.2 &  9.5 & $\pm$ 3.1 & $\pm$ 1.9         &  15  \\
    			& $900\GeV$  &    0.2 &   2.3 &  5.6 & $\pm$ 2.4 & $\pm$ 1.5         &  10  \\
\dbline
\end{tabular}

\vspace*{0.2cm}

\begin{tabular}{lD{.}{.}{-1}D{.}{.}{-1}D{.}{.}{-1}D{.}{.}{-1}D{.}{.}{-1}D{.}{.}{-1}}
\dbline
 Category 	& \multicolumn{1}{p{1.75cm}}{\centering$WW$}  	
 		& \multicolumn{1}{p{1.75cm}}{\centering$VV$}
		& \multicolumn{1}{p{1.75cm}}{\centering top quark}
		& \multicolumn{1}{p{1.75cm}}{\centering$Wj+jj$}  
		& \multicolumn{1}{p{1.75cm}}{\centering$Z/ \gamma^\ast$} 
		& \multicolumn{1}{p{1.75cm}}{\centering$H[125\GeV]$}       \\
\sgline
$\ZeroJet$  & 54.8\%  &  3.5\%  & 37.3\%  &  2.6\%  &  1.6\%  &  0.1\%  \\
$\OneJet$   & 40.7\%  &  3.5\%  & 49.3\%  &  3.4\%  &  2.9\%  &  0.2\%  \\
$\TwoJet$   & 24.6\%  &  2.3\%  & 36.3\%  &  2.5\%  & 30.7\%  &  3.6\%  \\
\hline\hline
\end{tabular}

\end{table*}

Figure~\ref{fig:SR_low} shows \mT\ distributions in the signal
regions, separately for the different- and same-flavour channels and
for each jet category. No significant data excess is observed in any 
final state. The deficit of data events in the high-\mT\ region 
of the 0-jet different-flavour final state has been investigated
and no underlying systematic experimental or modelling effect has been identified. 
In particular, no correlated kinematic effects are observed in the data, an alternative
$WW$ MC event generator ($\MCATNLO$~\cite{mcatnlo}) does not
qualitatively change the level of disagreement, and the $\mT$ distribution in the 
$WW$ control region (Figure~\ref{fig:CR_WW}) does not show a comparable deficit.

\begin{figure}[hbpt!]
\centering                                                                   
\includegraphics[width=0.43\textwidth]{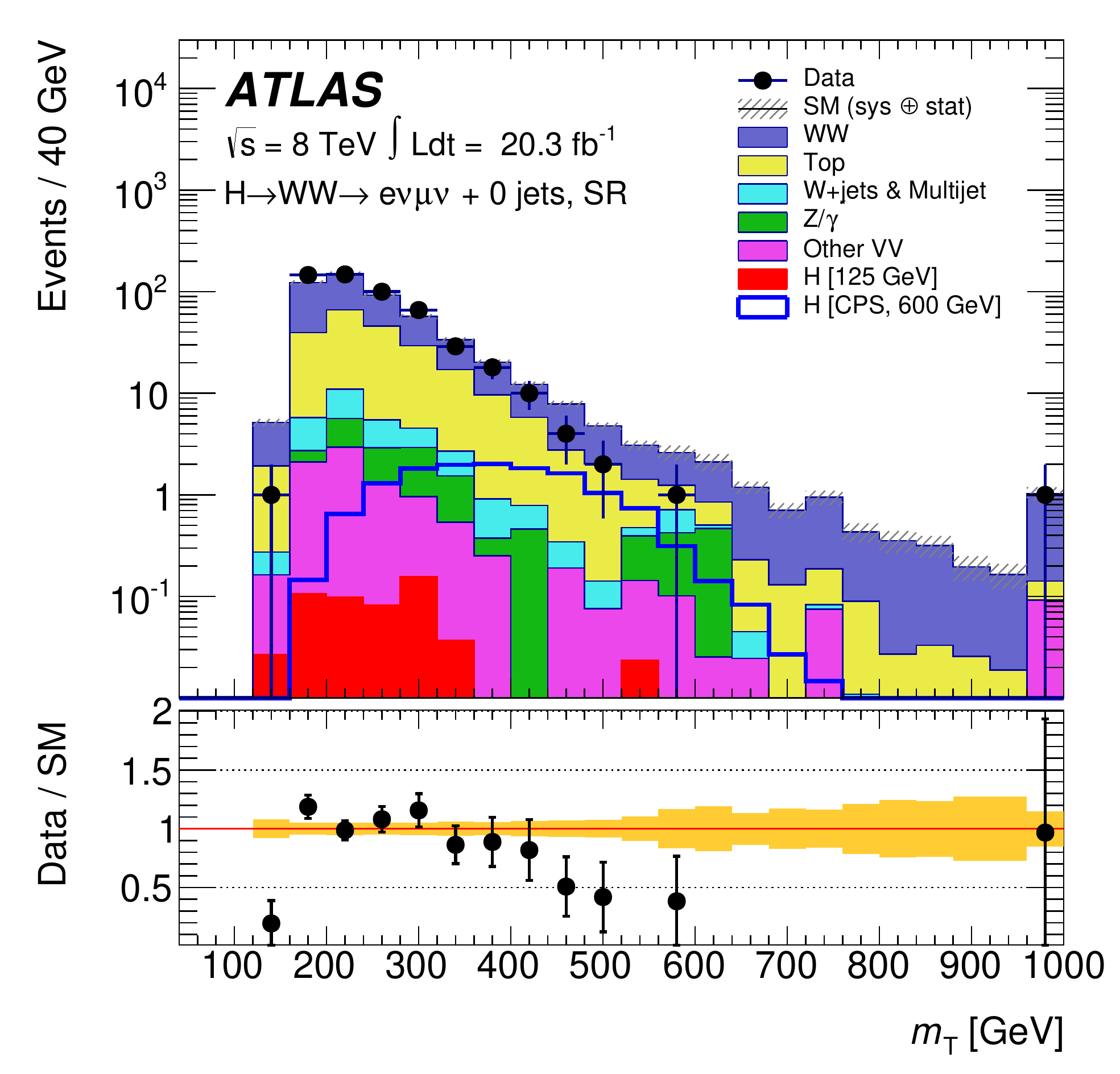}
\includegraphics[width=0.43\textwidth]{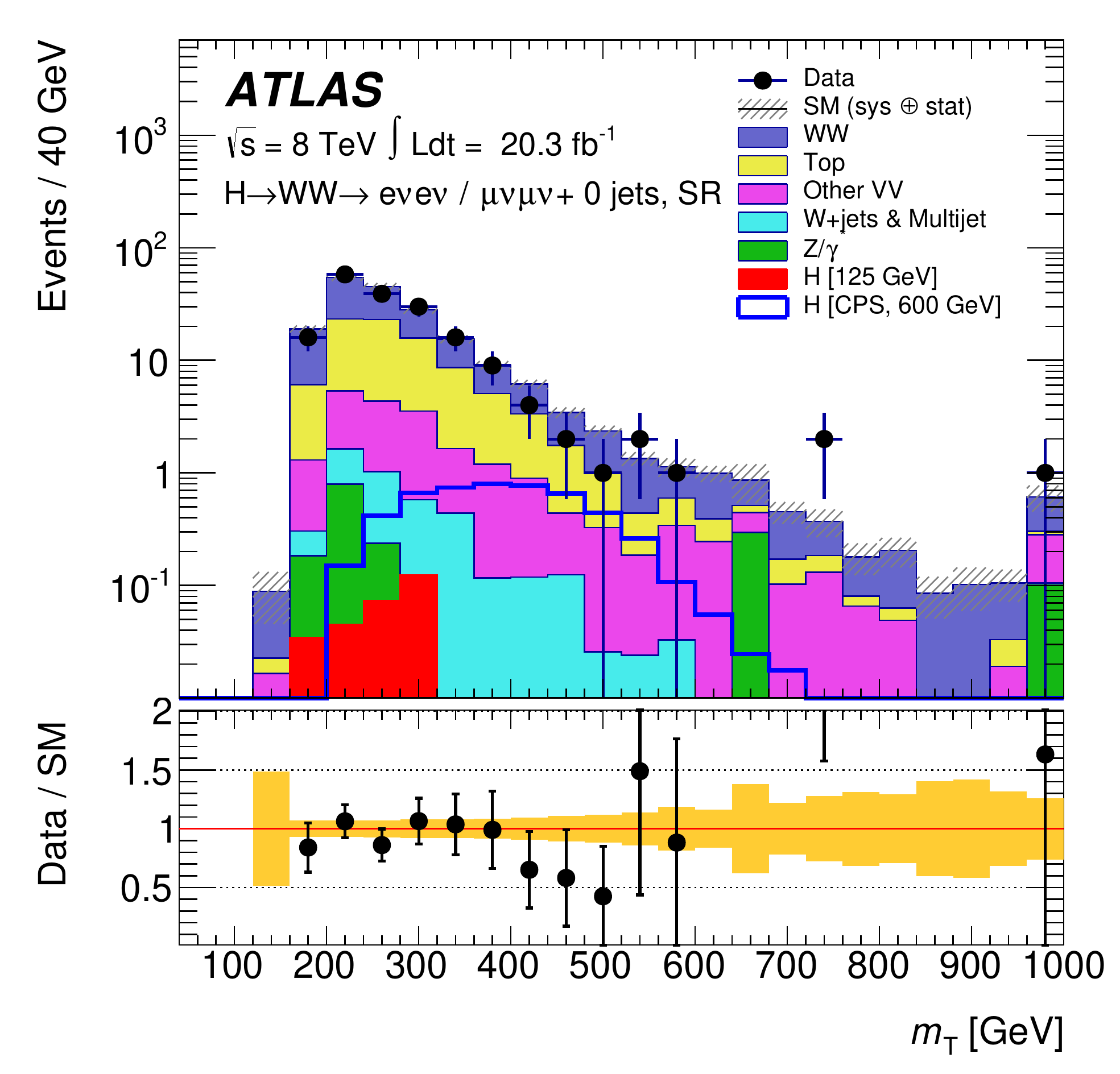}
\includegraphics[width=0.43\textwidth]{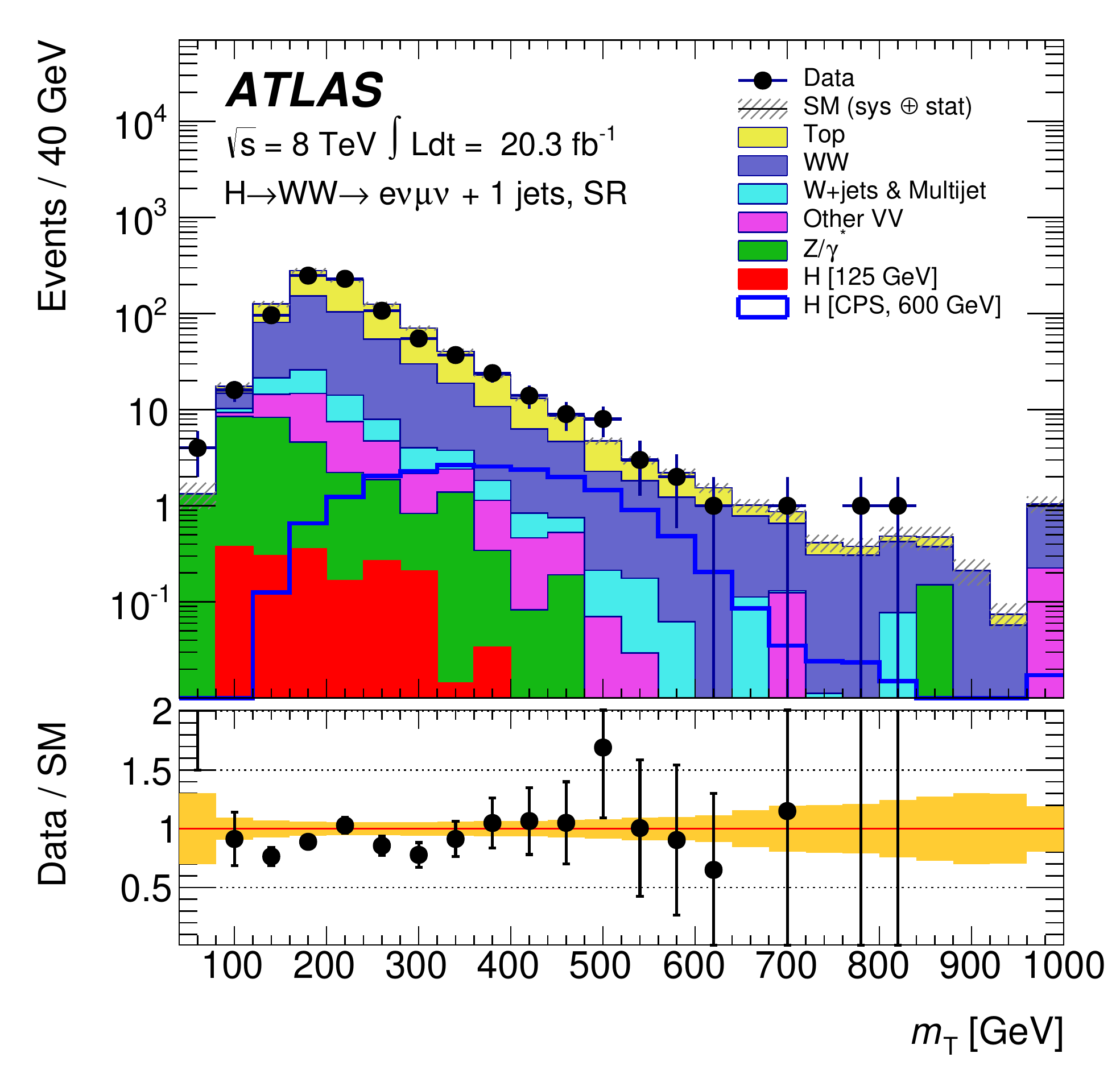}
\includegraphics[width=0.43\textwidth]{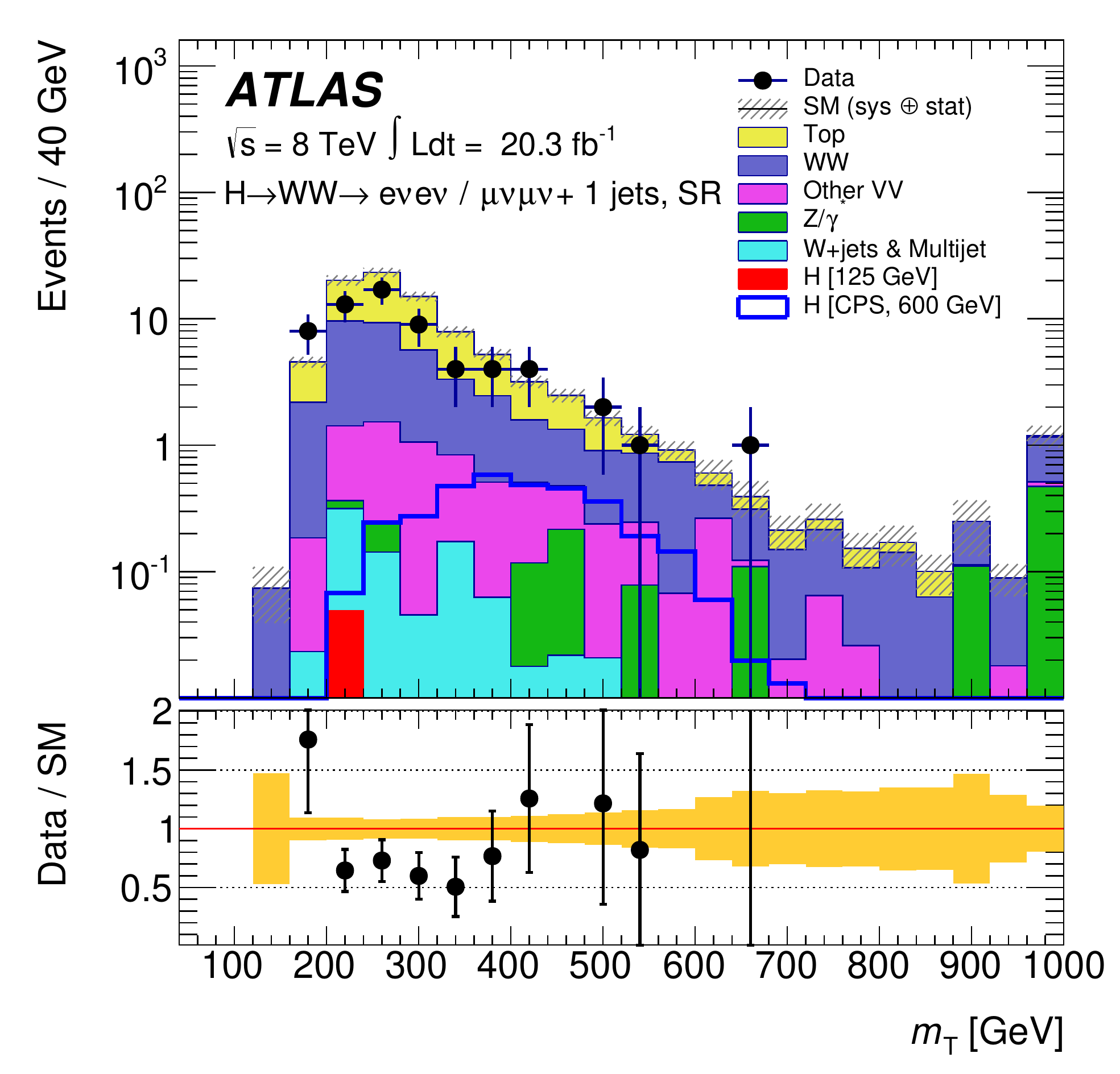}
\includegraphics[width=0.43\textwidth]{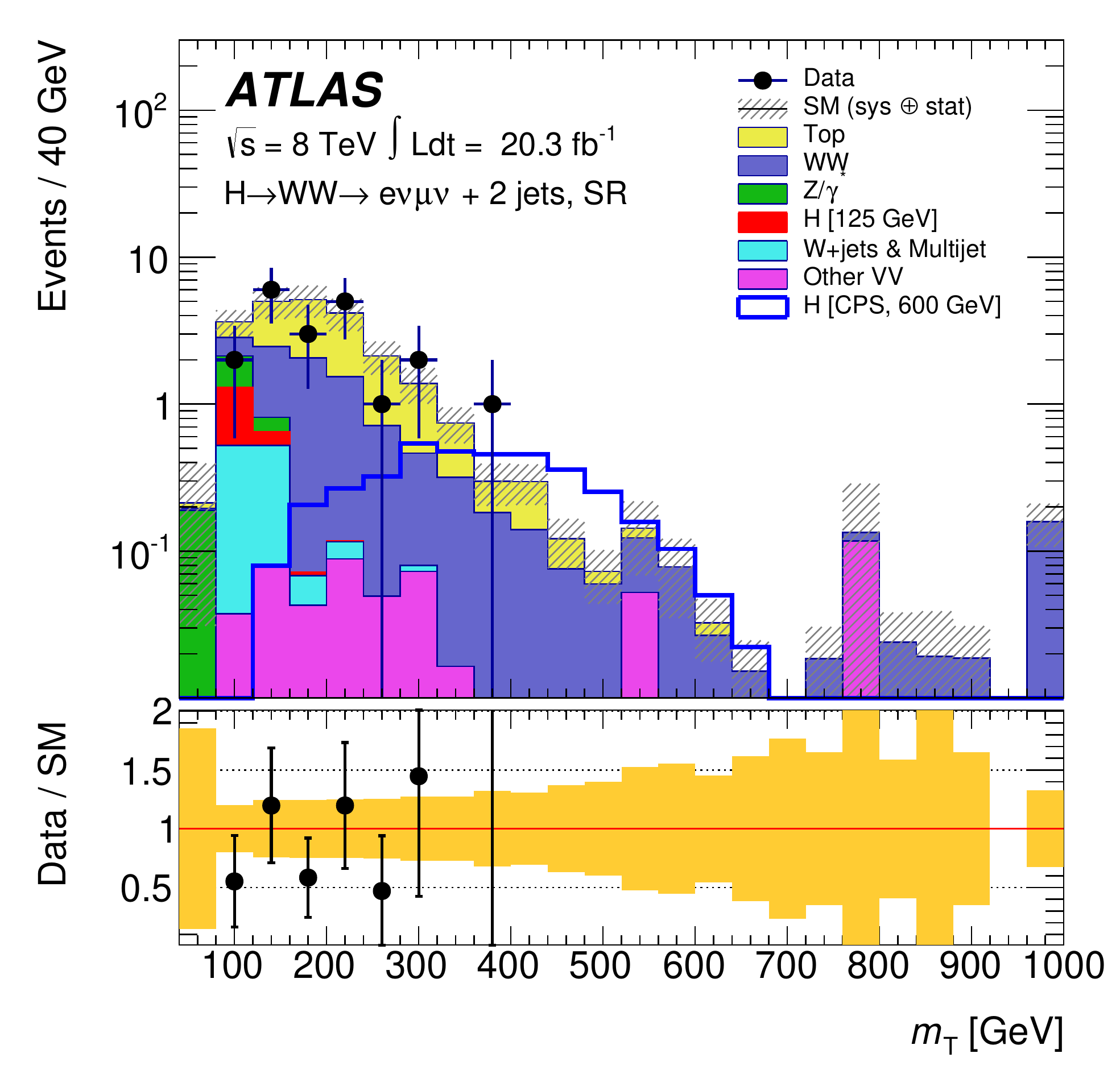}
\includegraphics[width=0.43\textwidth]{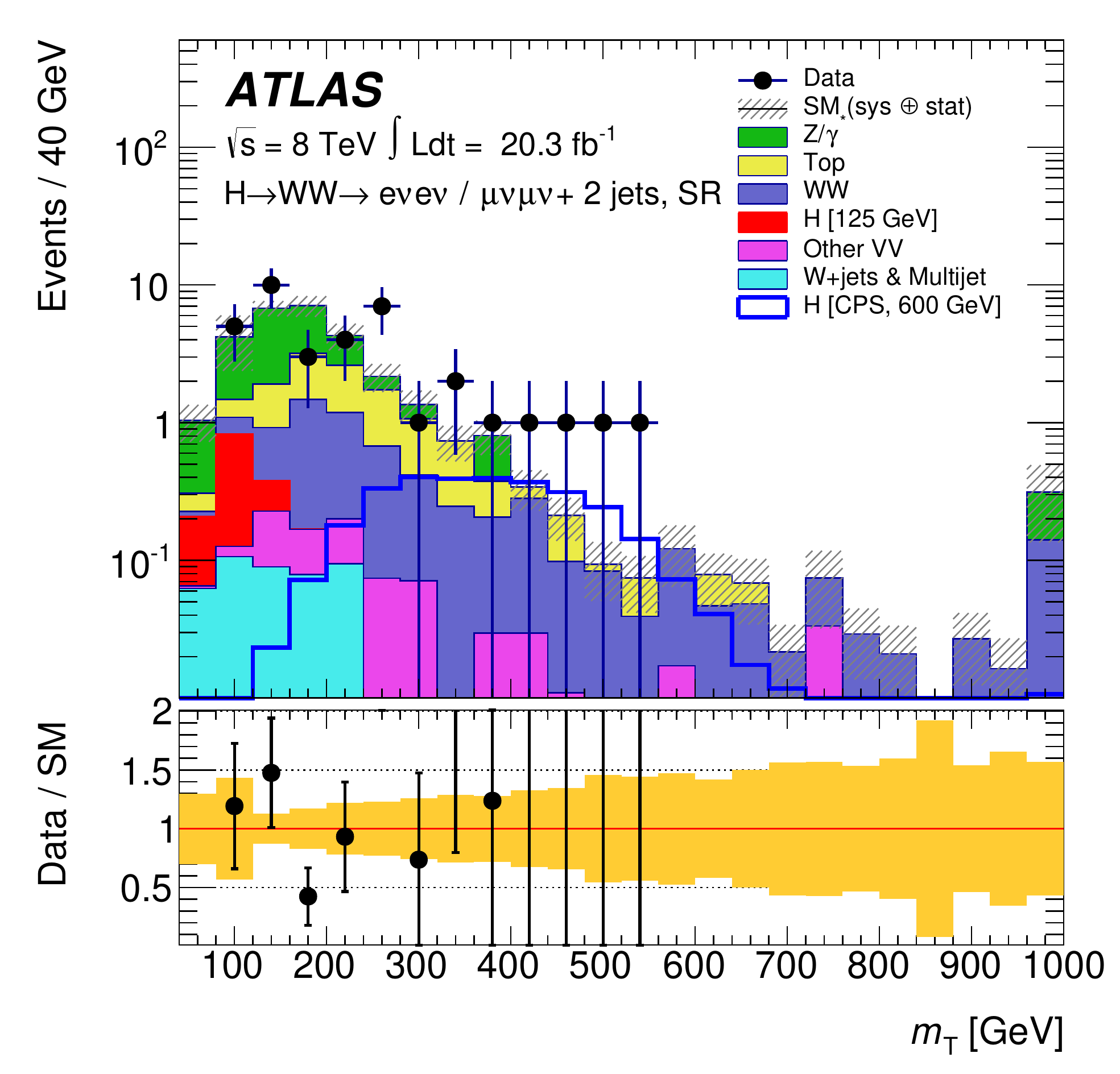}
\caption{Transverse mass distributions in the 0-jet (top), 1-jet (middle) and $\geq$ 2 jet (bottom) categories in the $\hwwlnln$ signal
  regions. Different-flavour (left) and same-flavour (right) final states are shown. The histograms for the
  background processes, including the observed $H[125\GeV]$, are shown stacked, and the distribution for a hypothetical 
  CPS signal process with $\mH=600\GeV$ and the cross section predicted by the SM for that mass is overlaid. 
  The combined statistical and systematic uncertainties on the 
  prediction are shown by the hatched band in the upper pane and the shaded band in the lower pane.
  In each figure, the last bin contains the overflow.
  }
\label{fig:SR_low}
\end{figure}

\subsection{The \HWWlvqq{} analysis}
\label{sec:results_lvqq}

In Table~\ref{tab:lvqqSRyields}, the expected signal and background in the \HWWlvqq\ analysis 
are summarized and compared to the number of data events passing the signal region selection.
For this comparison only, to give an indication of the sensitivity of the shape 
fit, a requirement that $\mlvjj$ be close to the mass hypothesis $\mH$ is added to the 
$\mH$-dependent signal region selection.  The requirement is 
$|\mlvjj - \mH| < 200\GeV)$, except for $\mH = 300\GeV$,
for which the lower bound is $200\GeV$, corresponding to the lower $\mlvjj$ considered in the analysis.
The systematic uncertainty on the background is derived by fitting the nominal expected signal and 
background, following the procedure used for the $\hwwlnln$ analysis.  
The fractional background composition for the ggF and VBF analysis categories and the
$\mH$-dependent selection is shown in the bottom half of the table.  The relative contributions of the various 
sources to the total background do not vary substantially as a function of $\mlvjj$, so the $\mlvjj$ window 
described above is not applied.

\begin{table*}[tb!]
\centering
\caption{
  Summary of the expected signal and background in the \HWWlvqq\ signal regions.
  The top table compares the observed number of candidate events in data $N_{\mathrm{data}}$ with the expected total 
  signal, $N_{\mathrm{sig}}$, for several $\mH$ values and the total background $N_{\mathrm{bkg}}$, along with
  its statistical ($\sqrt{N_{\mathrm{bkg}}}$) and systematic ($\delta N_{\mathrm{bkg}}^{\mathrm{sys}}$) uncertainties.
  The predictions are quoted for an $\mlvjj$ interval around the $\mH$ hypothesis (details in the text).
  The bottom table shows the composition of the background for
  the ggF and VBF analysis categories at several $\mH$ hypotheses.
  The ``other'' column contains the $\Zjets$ and diboson backgrounds. }
\vspace{0.3cm}
\label{tab:lvqqSRyields}
\begin{tabular}{llrrr@{}c@{}lrr}
\dbline
	 &  \multicolumn{3}{c}{CPS signal expectation} & \multicolumn{3}{c}{Bkg.~expectation} & Observed \\ 
Category &  $\mH$ & $N_{\mathrm{sig}}^{\mathrm{ggF}}$ & $N_{\mathrm{sig}}^{\mathrm{VBF}}$ 
& $N_{\mathrm{bkg}}$ & $\,\pm\,\sqrt{N_{\mathrm{bkg}}}$ & $\,\pm\,\delta N_{\mathrm{bkg}}$ & $N_{\mathrm{data}}$  \\ 
\sgline
ggF	& $300\GeV$   & 1320   & 100  & 112890 & $\pm$ 340 		& $\pm$ 460 & 111199  \\ 
	& $600\GeV$   &  440   &  40  &  23680 & $\pm$ 150 		& $\pm$ 131 &  23397  \\ 
	& $900\GeV$   &   40   &  10  &   1940 & $\pm$ \phantom{0}40	& $\pm$ \phantom{0}56 &   1754  \\ 
\sgline
VBF	& $300\GeV$   &   24   &  41  &   2282 & $\pm$ \phantom{0}48	& $\pm$ \phantom{0}33 &   2090  \\ 
	& $600\GeV$   &   24   &  34  &    850 & $\pm$ \phantom{0}29	& $\pm$ \phantom{0}19 &    829  \\ 
	& $900\GeV$   &    3   &  11  &     52 & $\pm$ \phantom{00}7	& $\pm$ \phantom{00}7 &     68  \\ 
\dbline
\end{tabular}

\vspace*{0.2cm}

\begin{tabular}{llrrrr}
\dbline
Category & $\mH$	& \multicolumn{1}{p{1.5cm}}{\centering$\Wjets$}   
  			& \multicolumn{1}{p{1.5cm}}{\centering top}   
			& \multicolumn{1}{p{1.5cm}}{\centering multijet}  
			& \multicolumn{1}{p{1.5cm}}{\centering other}       \\
\sgline
ggF	& $300\GeV$   & 70\%	& 20\%	& 3\%	&  7\%	\\
	& $600\GeV$   & 71\%	& 19\%	& 2\%	&  7\%	\\
	& $900\GeV$   & 73\%	& 14\%	& 3\%	& 10\%	\\
\sgline
VBF	& $300\GeV$   & 58\%	& 33\%	& 5\%	&  4\%	\\
	& $600\GeV$   & 61\%	& 27\%	& 6\%	&  5\%	\\
	& $900\GeV$   & 52\%	& 32\%	& 4\%	& 12\%	\\
\dbline
\end{tabular}

\end{table*}

Figure~\ref{fig:PRESEL_LNUJJ_h_m} show the $\mlvjj$ distributions
and the ratio of data to background expectation for the WCR (top),
TopCR (middle), and SR (bottom) after the ggF
preselection on the left hand side of the figure. Shown on the right are the corresponding distributions for
the VBF preselection.  These distributions do not include the background normalisations applied 
by the fit to the control regions, but the $\mlvjj$ reweighting described in Section~\ref{sec:lvqq_wjetstop} 
is applied.
\begin{figure}[hbpt!]
\centering                                                                   
\includegraphics[width=0.43\textwidth]{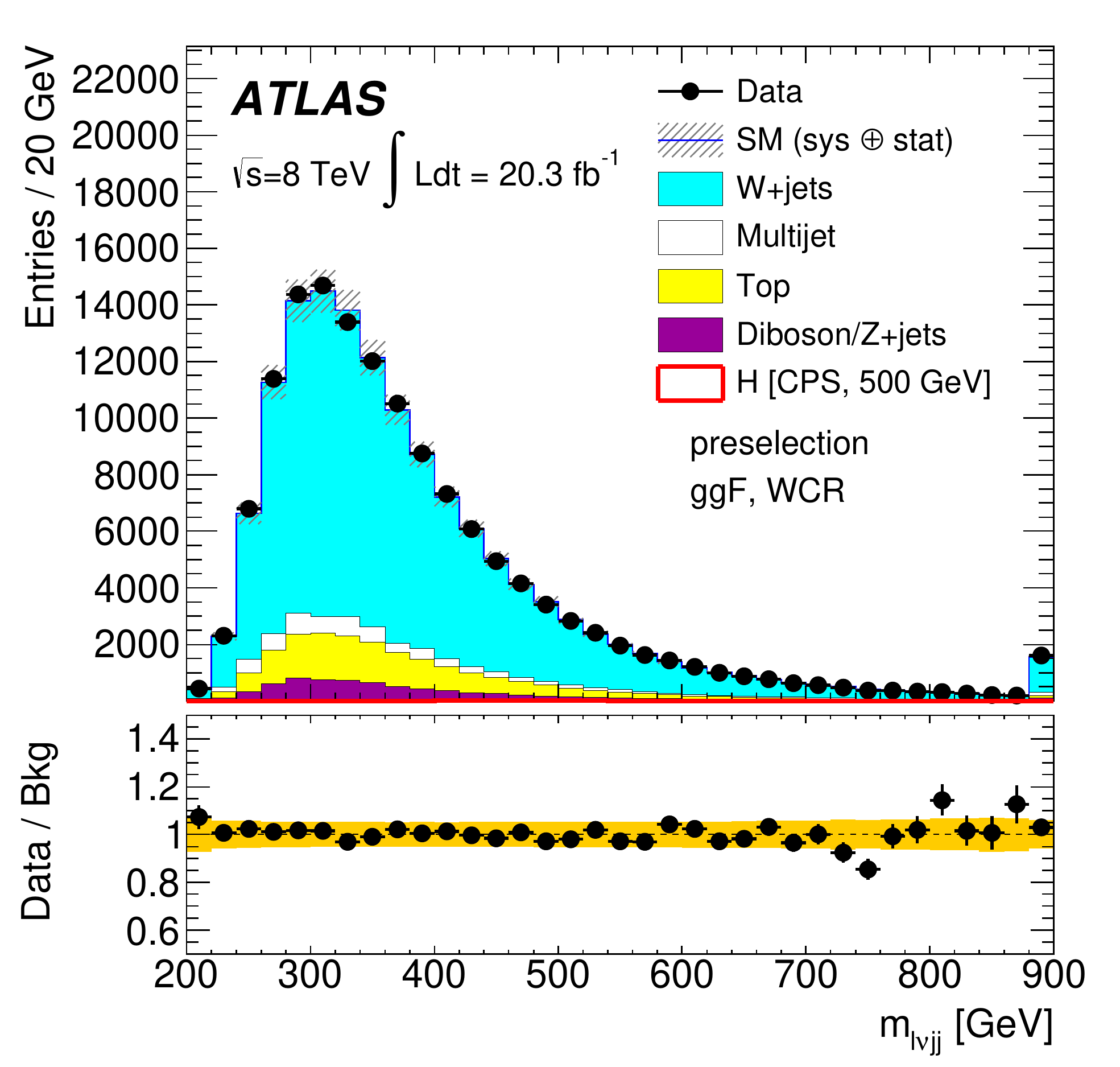}
\includegraphics[width=0.43\textwidth]{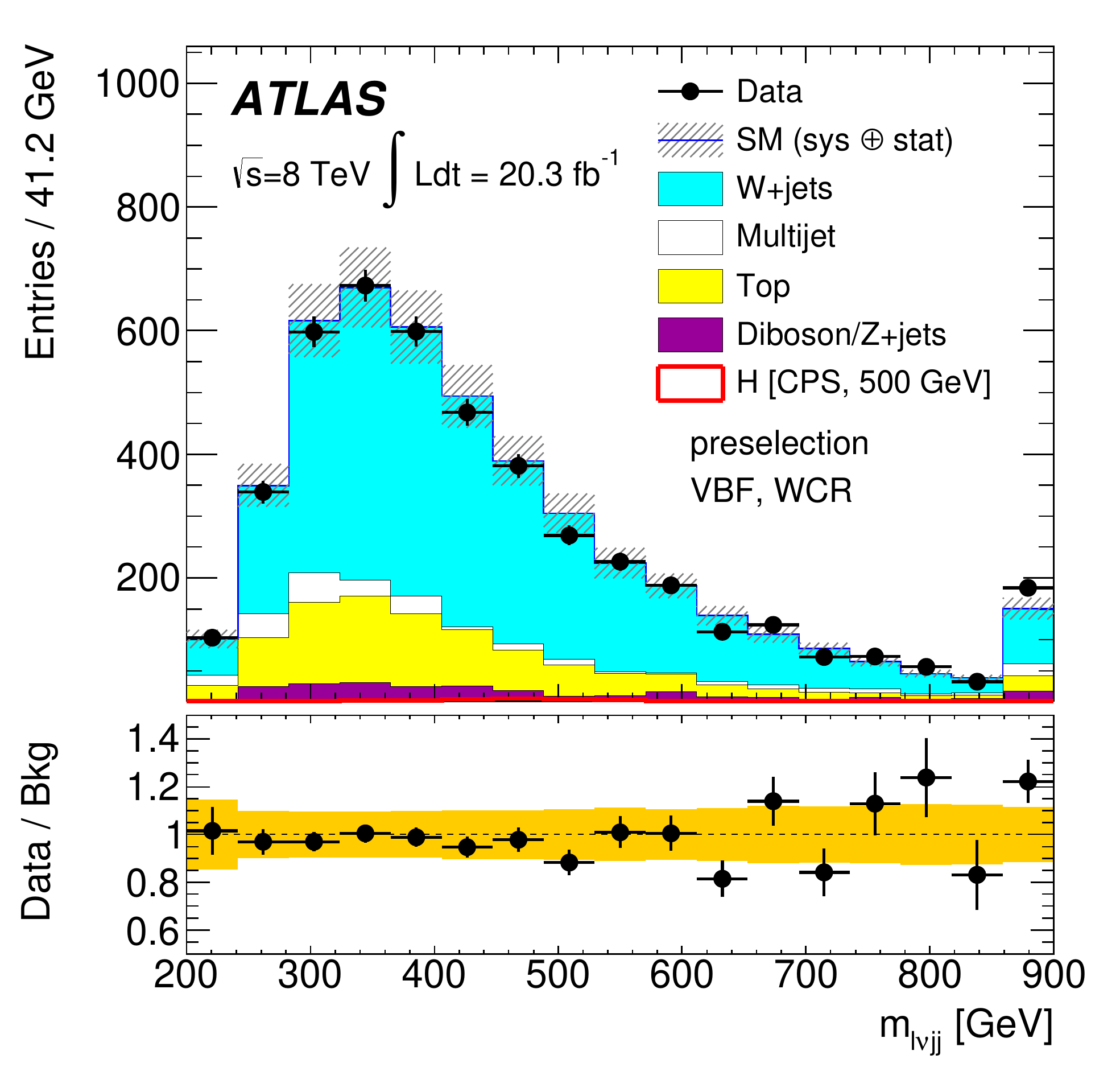}
\includegraphics[width=0.43\textwidth]{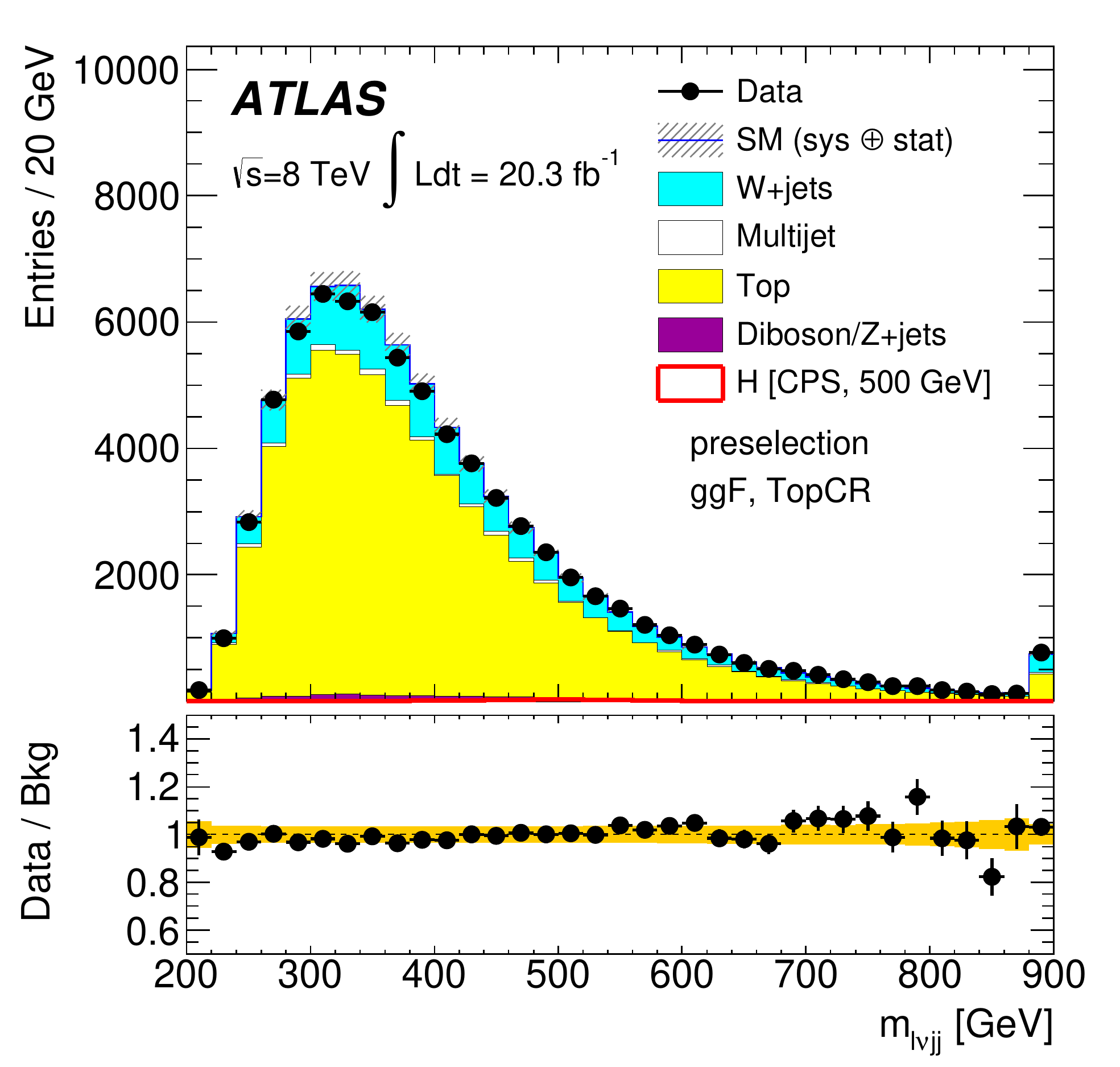}
\includegraphics[width=0.43\textwidth]{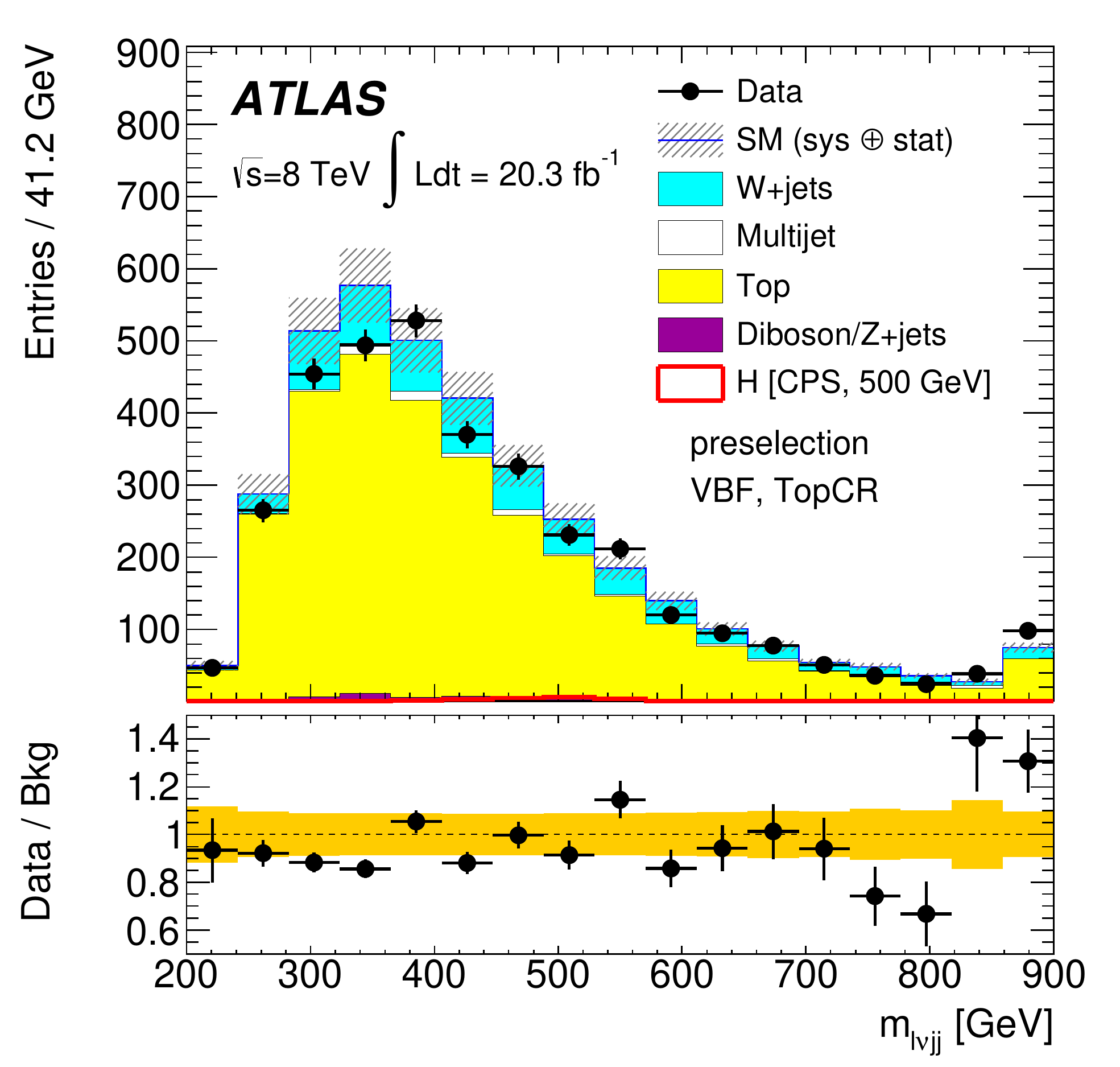}
\includegraphics[width=0.43\textwidth]{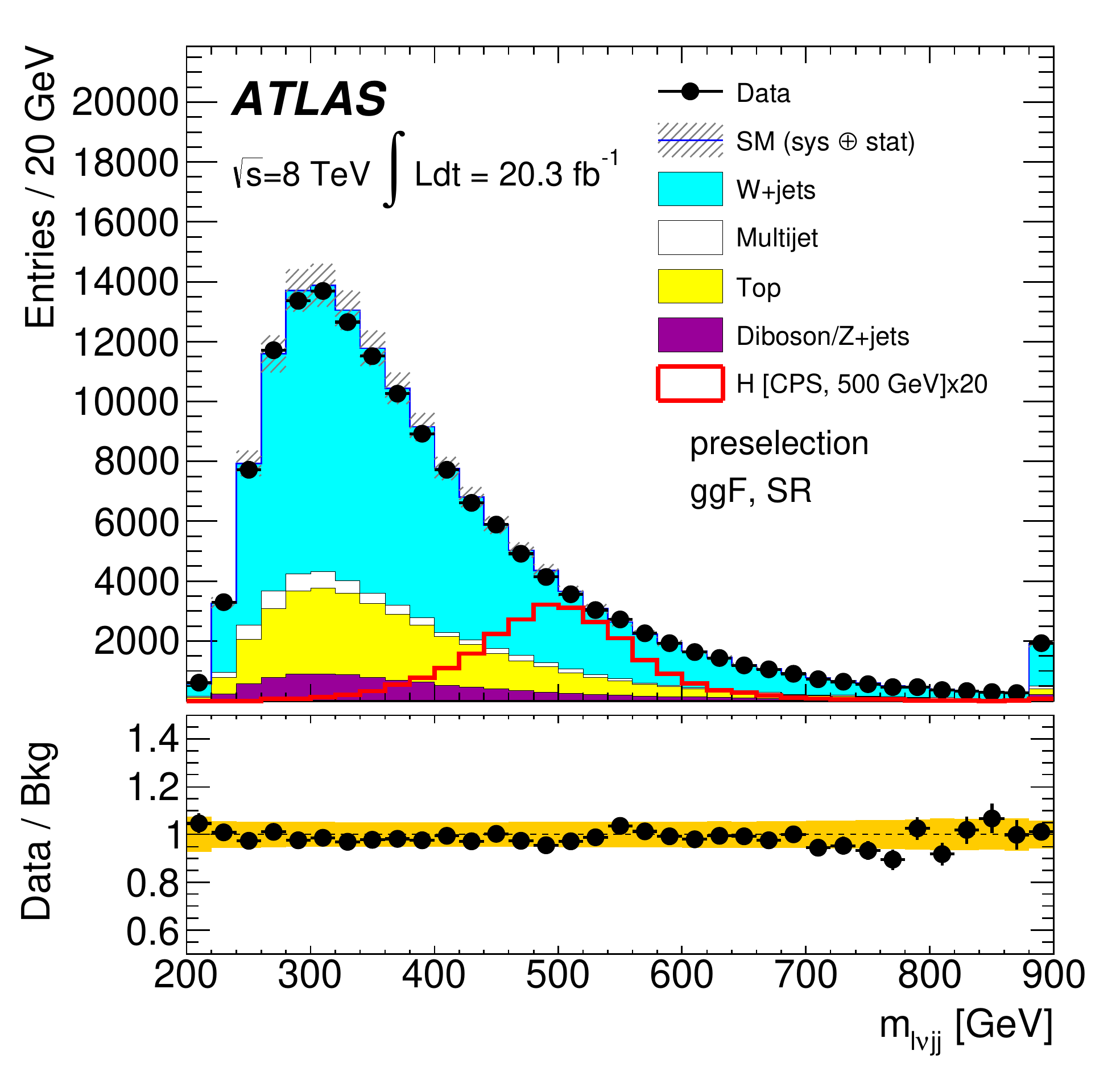}
\includegraphics[width=0.43\textwidth]{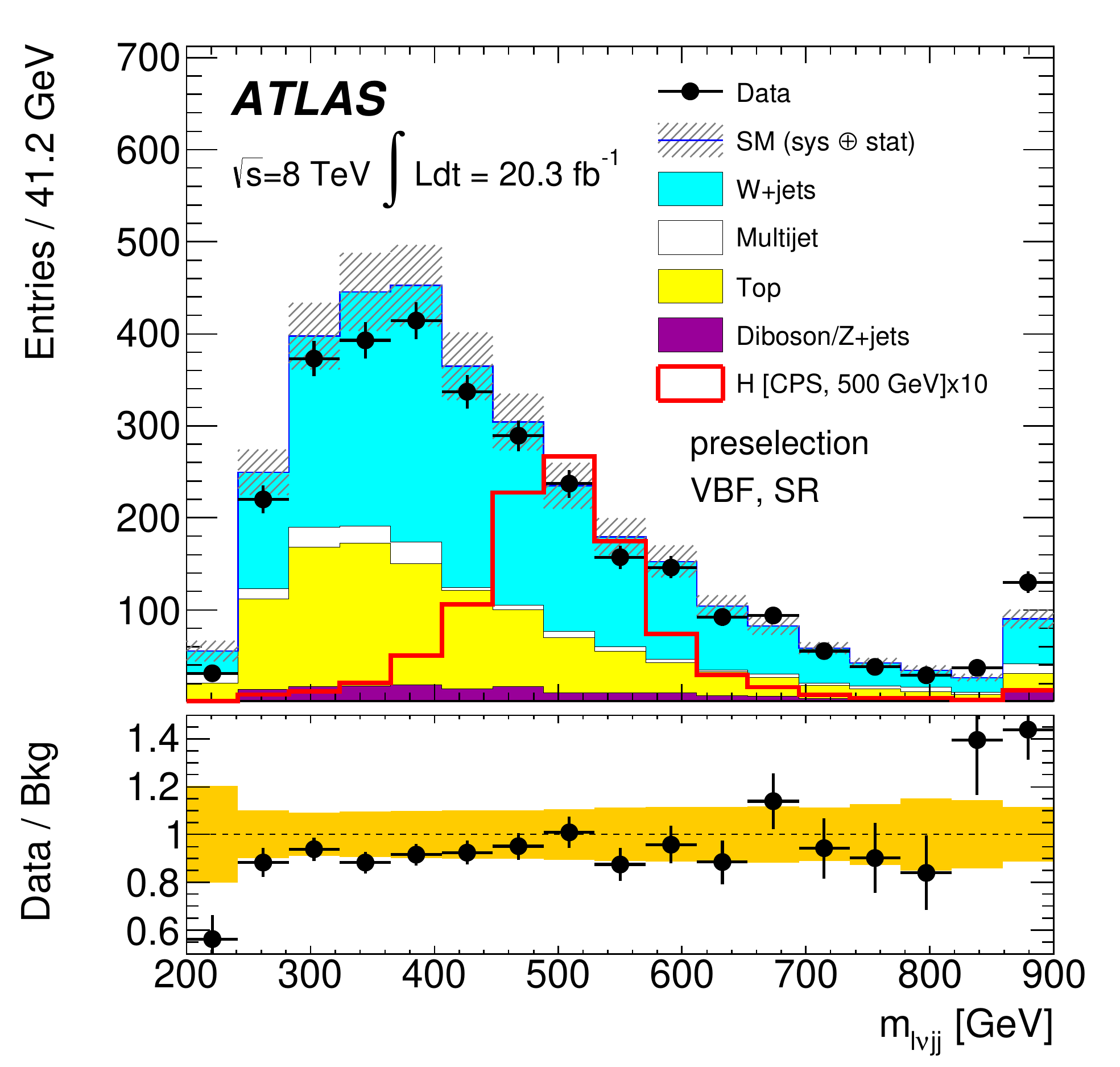}
\caption{Distributions of the invariant mass $m_{\ell\nu jj}$ of the $WW$ system 
and the ratio of data to background expectation for the WCR (top), 
TopCR (middle), and SR (bottom) after the ggF
preselection (left) and the VBF preselection (right) for the $\HWWlvqq$ analysis.
The histograms for the background processes are shown stacked, and the distribution for a hypothetical 
  CPS signal process with $\mH=500\GeV$ and the cross section predicted by the SM for that mass is overlaid. 
All the flavour and charge categories are summed in each plot.  
No normalisation scale factors are applied to the top-quark or $W$ background samples. 
  The combined statistical and systematic uncertainties on the 
  prediction are shown by the hatched band in the upper pane and the shaded band in the lower pane.
  In each figure, the last bin contains the overflow.
}
\label{fig:PRESEL_LNUJJ_h_m}
\end{figure}

The Higgs boson signal yield in each final state is determined using a
binned maximum likelihood fit to the observed $\mlvjj$
distribution in the range $200\GeV < \mlvjj < 2000\GeV$. For
the $\mH=500\GeV$ selection, the control and signal regions distributions are shown in
Figure~\ref{fig:POSTFIT_LNUJJ_h_m}. In these distributions, the $\mlvjj$ reweighting is applied and
the background normalisations
are corrected using the results of the fit to the signal and control regions. There is no indication of a
significant excess of data above the background expectation. A slight deficit can be seen in the VBF
channel in the centre and lower panels of Figure~\ref{fig:PRESEL_LNUJJ_h_m} but its effect is mitigated
by the mass-dependent selection.

\begin{figure}[hbpt!]
\centering                                                                   
\includegraphics[width=0.43\textwidth]{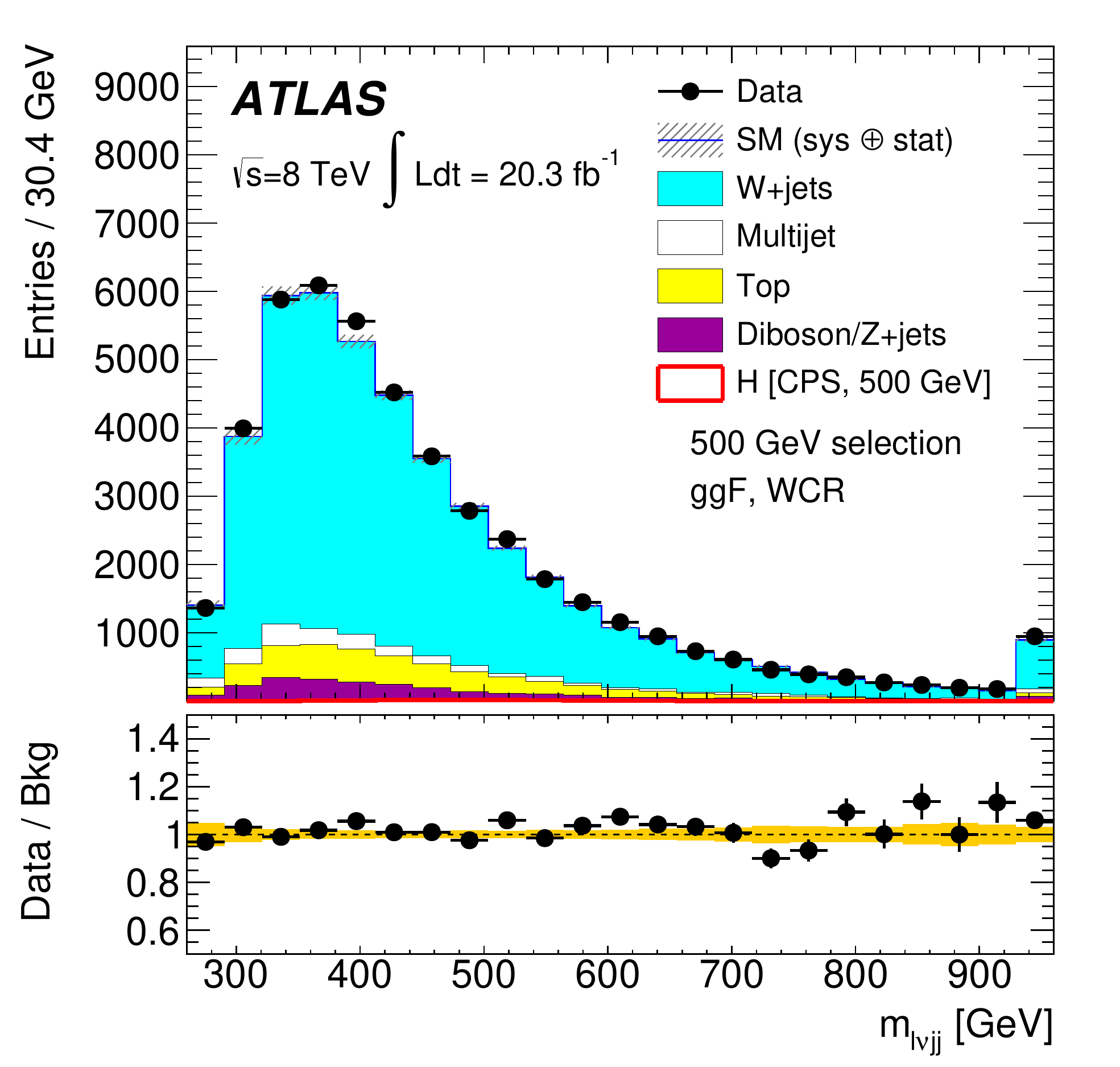}
\includegraphics[width=0.43\textwidth]{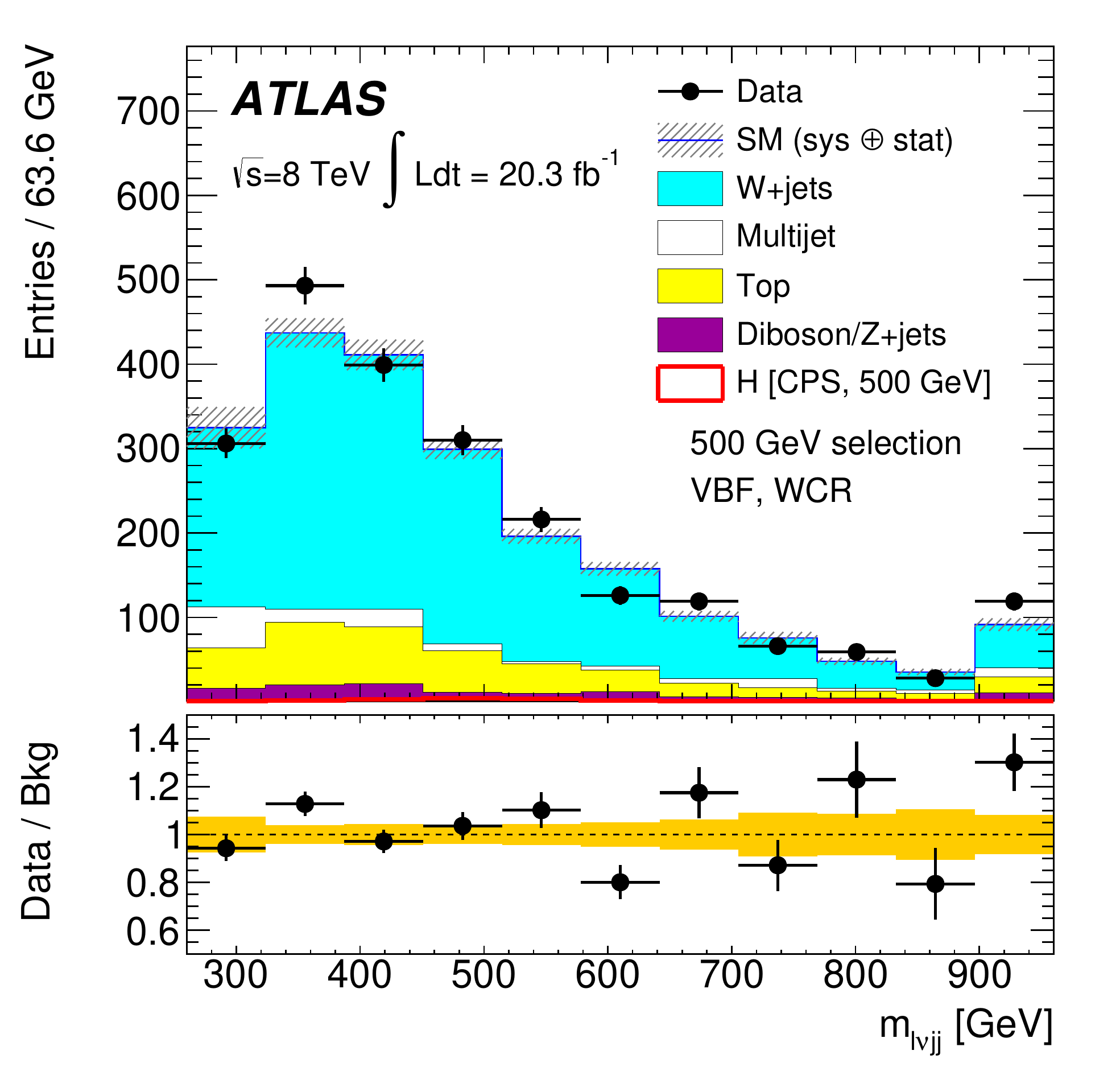}
\includegraphics[width=0.43\textwidth]{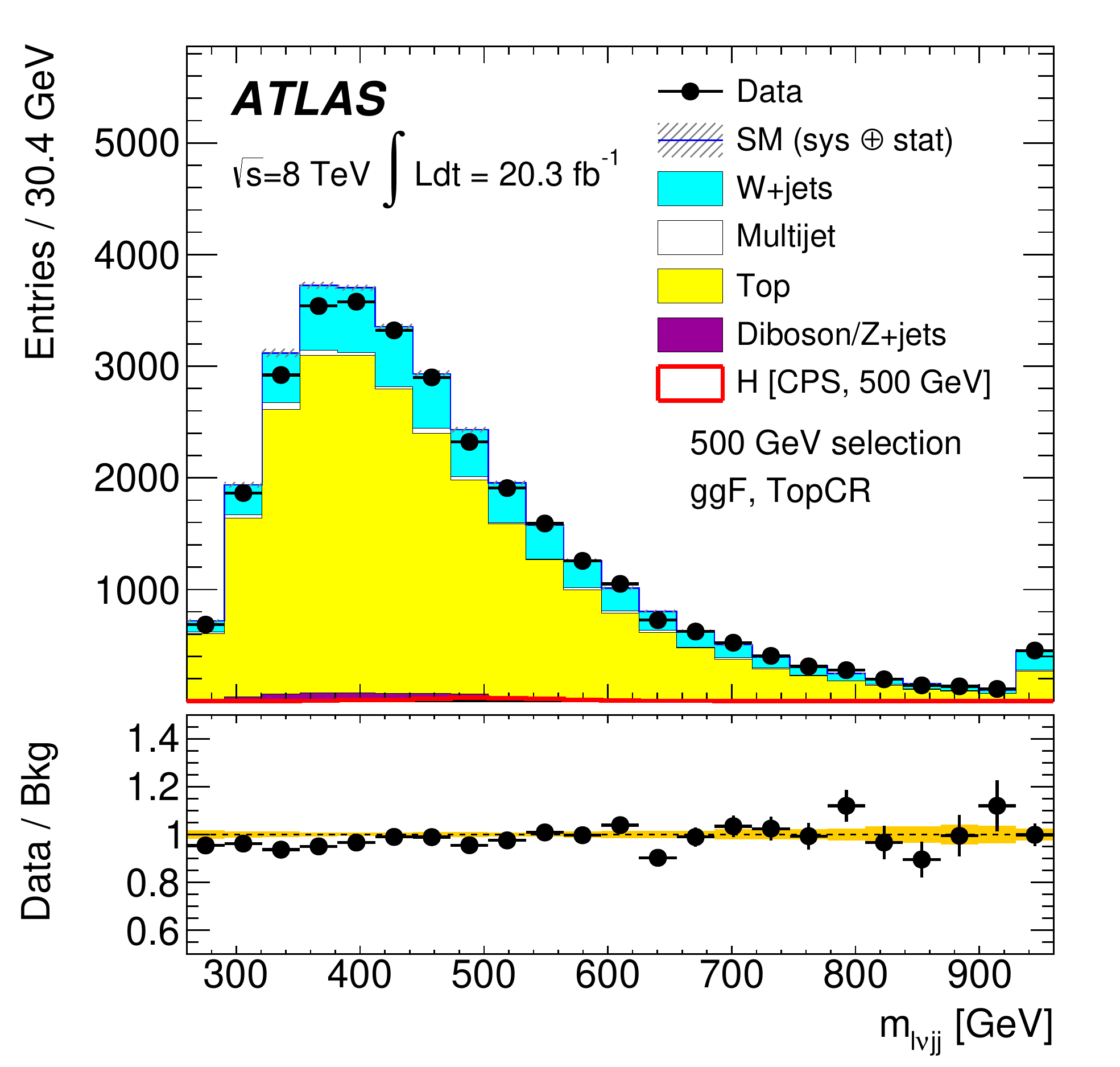}
\includegraphics[width=0.43\textwidth]{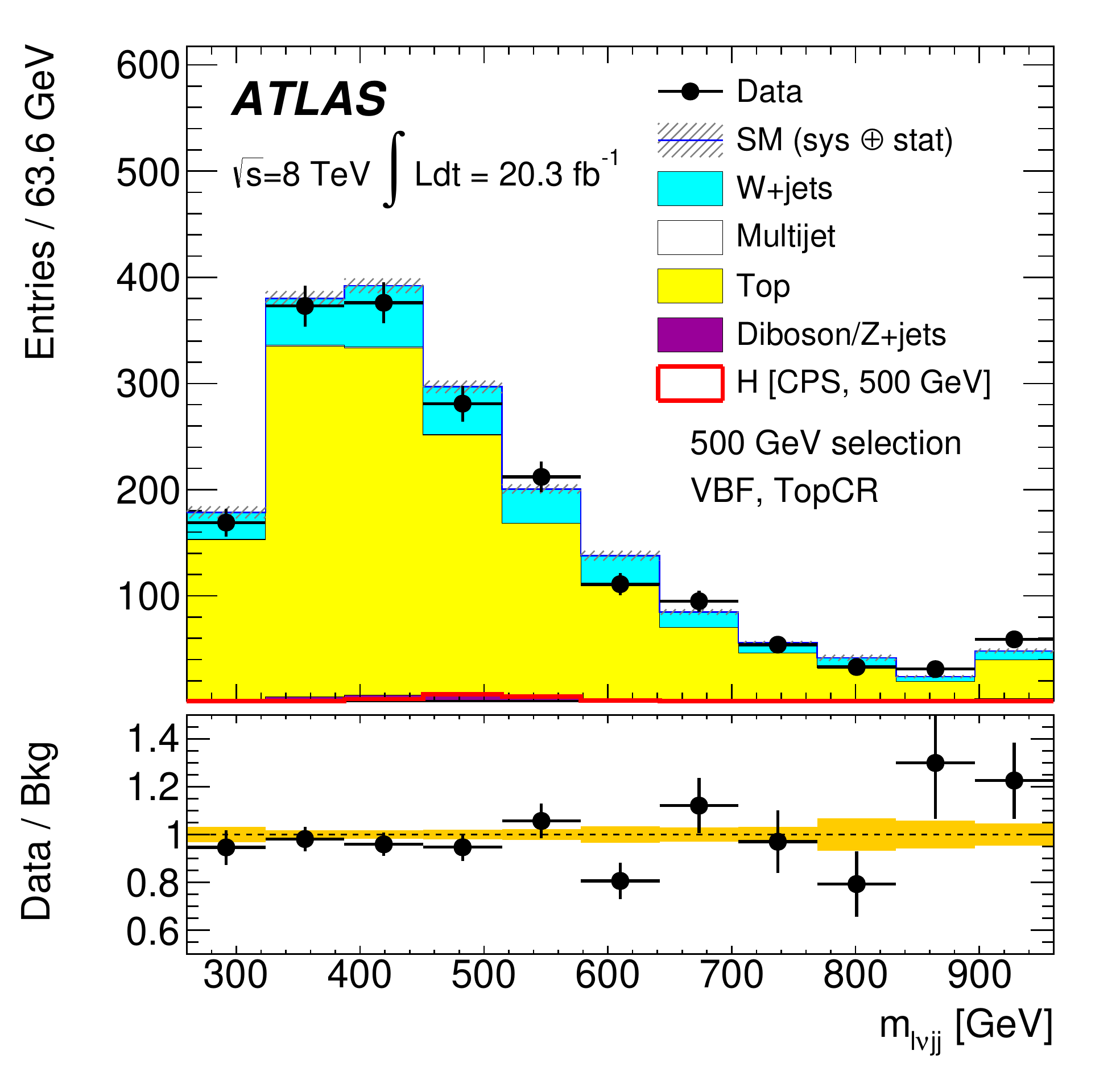}
\includegraphics[width=0.43\textwidth]{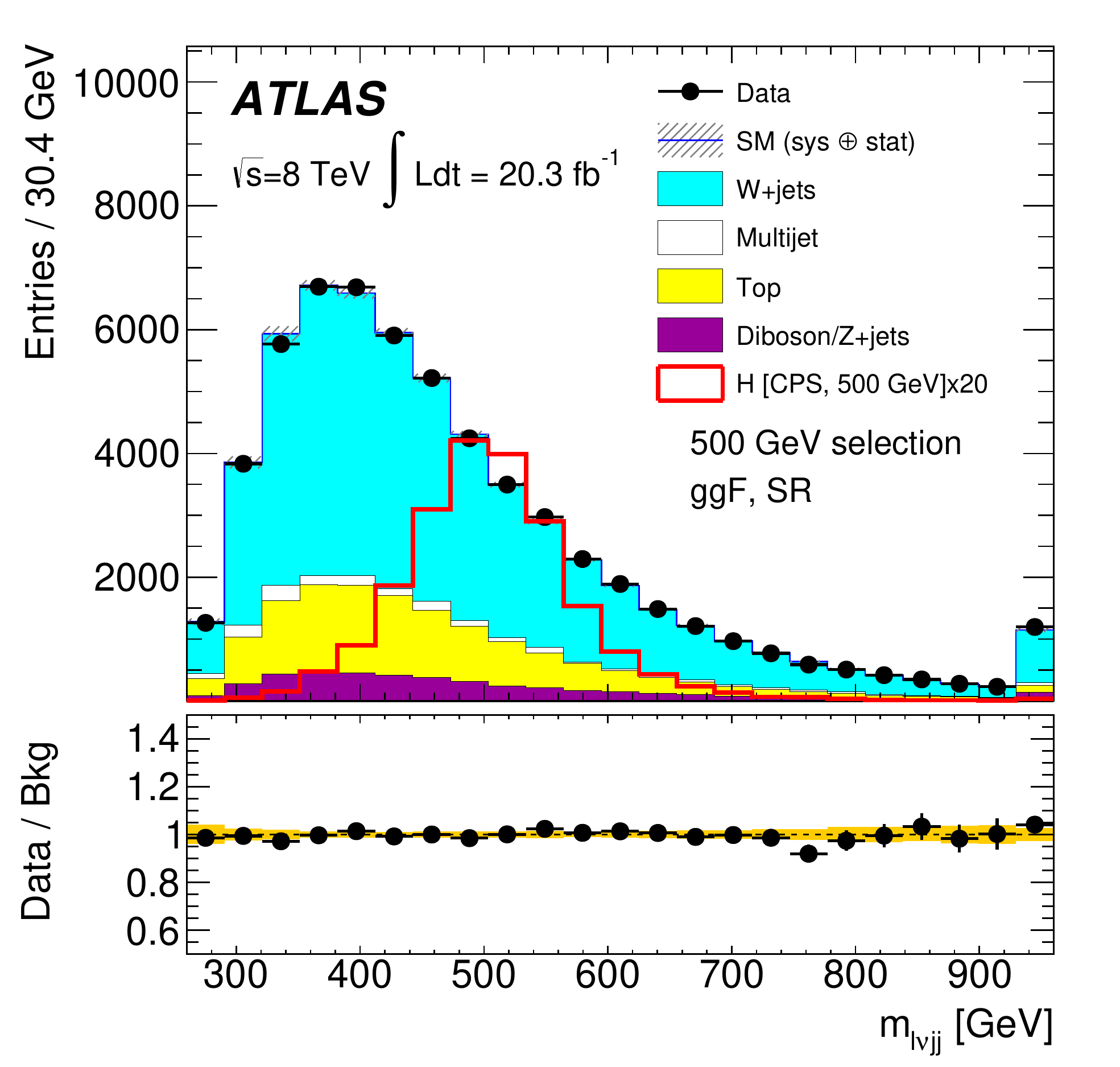}
\includegraphics[width=0.43\textwidth]{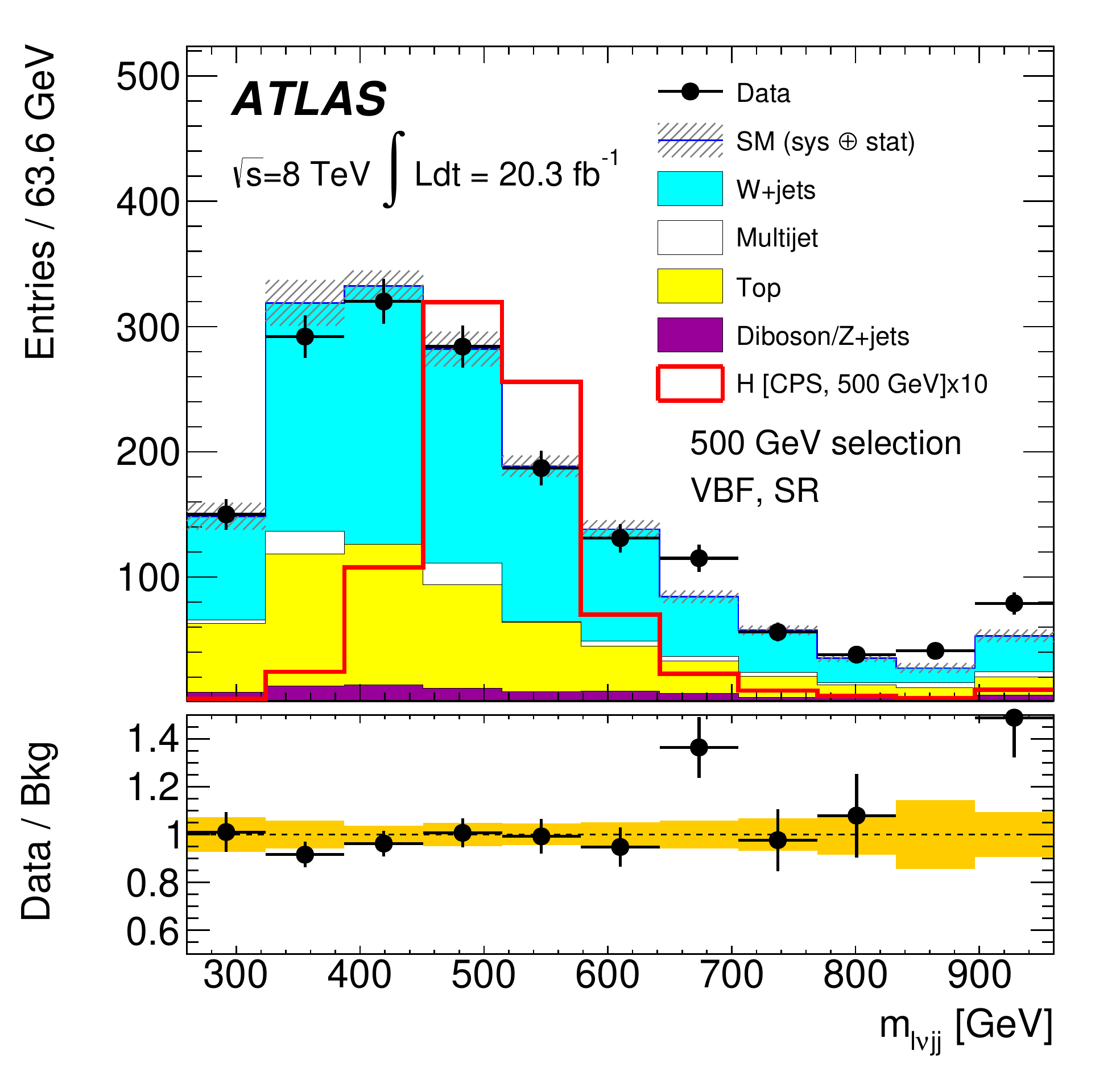}
\caption{Distributions of the invariant mass $m_{\ell\nu jj}$ of the $WW$ system and the ratio of data to
  background expectation for the WCR (top), 
  TopCR (middle), and SR (bottom) after the ggF
  500 GeV selection (left) and the VBF 500 GeV selection (right) in the $\HWWlvqq$ analysis. 
  These plots are after fitting for the $\mH=500\GeV$ hypothesis, and 
  the binning used is identical to the binning used in the fit. 
  The histograms for the background processes are shown stacked, and the distribution for a hypothetical 
  CPS signal process with $\mH=500\GeV$ and the cross section predicted by the SM for that mass is overlaid. 
  All the flavour and charge categories are summed in each plot.
  The combined statistical and systematic uncertainties on the 
  prediction are shown by the hatched band in the upper pane and the shaded band in the lower pane.
  In each figure, the last bin contains the overflow.
}
\label{fig:POSTFIT_LNUJJ_h_m}
\end{figure}

\section{Results and Interpretations}
\label{sec:interp}

\subsection{Statistical methodology}
\label{subsec:stats}

The methodology used to derive statistical results is described in detail
in Ref.~\cite{Aad:2012an}. A likelihood function $\mathcal{L}$ is
defined using the distributions of the discriminant for events
in the signal region of each analysis category, namely, the 0-, 1- and $\ge{2}$-jet 
categories in the \hwwlnln\ analysis and
the ggF and VBF categories in the \hwwlnqq\ analysis. The likelihood is a
product of Poisson functions over the bins of the discriminant in the
signal regions and ones describing the total yield in each control region. 
Each systematic uncertainty is
parameterised by a corresponding nuisance parameter $\theta$ modelled by a Gaussian
function (the set of all such nuisance parameters is $\boldsymbol\theta$). 
The modification of affected event yields is implemented as a log-normal
distribution parameterised by $\theta$ to prevent predicted event yields from taking
unphysical values.

In the \hwwlnln\ analysis, the $\mT$ distributions in the signal
regions are divided into ten, six and four bins, respectively, for
\AllJet. The bins are of variable widths such as to have the same
number of expected signal events in each bin. 
Because the signal is peaked in $\mT$ and the background is not, this 
binning strategy improves the sensitivity by producing bins with different signal-to-background
ratios and is robust against statistical fluctuations in the background model. 

In the $\HWWlvqq$ final state, the statistical analysis is performed using a variable
fit range and number of bins, adapted to each $m_H$ hypothesis and production
mode.  The bins are always of equal width, because the $\mlvjj$ distribution for signal
peaks more strongly than the signal $\mT$ distribution in the \hwwlnln\ channel, 
so the discriminant produces bins with sufficiently different signal-to-background ratios 
without further optimisation.
The bin width is chosen to ensure adequate statistical precision for the background predictions.
The search is always preformed in a $700\GeV$-wide window in $\mlvjj$, enclosing 
the resonance peak and as much of the tails as feasible for non-NWA signal models.  
The range \mbox{$200$--$900\GeV$} is used for the $300\GeV$ hypothesis, and 
\mbox{$500$--$1200\GeV$} for the $1000\GeV$ hypothesis.  The ggF (VBF) search uses 35 (17) bins
for the $300\GeV$ hypothesis.  The number of bins decreases with increasing $m_H$, and
12 (6) bins are used for the ggF (VBF) search for $m_H = 1000\GeV$.  Mass hypotheses above
$1\TeV$ are also tested, but the binning and fit range for the $1\TeV$ mass hypothesis
are maintained, because there is insufficient data and simulated events to populate the background
model at higher values of $\mlvjj$.

Both the $\hwwlnln$ and $\hwwlnqq$ analyses have signal regions optimised
for the VBF and ggF signal production modes, but the presence of both signal processes
is accounted for in all signal regions.  Limits are obtained separately for ggF and 
VBF production in all interpretations. To derive the expected limits on the ggF production mode, the VBF
production cross section is set to zero, so that the expected limits
correspond to the background-only hypothesis. To derive the
observed limits on the ggF (VBF) production mode, the VBF (ggF) production cross
section is treated as a nuisance parameter in the fit and
profiled using a flat prior, as is used for the normalisation of backgrounds using CRs.
This approach avoids making any assumption on the presence or absence of the 
signal in other production modes, by using the signal regions themselves to set 
the normalisation of the production mode not being tested.

The modified frequentist method known as CL$_\textrm{s}$, combined with the asymptotic
approximation, is used to
compute $95\%$ CL upper limits~\cite{CLs_2002, asymptotics}.  The method uses a test statistic
$q_\mu$, a function of the signal strength $\mu$ which is defined as
the ratio of the measured $\sigma_H\times\mathrm{BR}(H\rightarrow WW)$ to that
predicted.\footnote{The SM cross-section prediction is used to define $\mu$ for the NWA and 
intermediate-width scenarios.} The test statistic is defined as: 
\begin{equation}
q_\mu\,{=}-2\ln\bigl(\mathcal{L}(\mu;
\hat{\boldsymbol{\theta}}_\mu)/\mathcal{L}(\hat{\mu};
\hat{\boldsymbol{\theta}})\bigr)
\label{eq:qmu}
\end{equation}
The denominator does not depend on $\mu$. The quantities $\hat{\mu}$
and $\hat{\boldsymbol{\theta}}$ are the values of $\mu$ and $\boldsymbol{\theta}$, respectively,
that unconditionally maximise $\mathcal{L}$. The numerator depends on
the values $\hat{\boldsymbol{\theta}}_\mu$ that maximise $\mathcal{L}$
for a given value of $\mu$.

\subsection{Upper limits from the \hwwlnln\ analysis}

Figure~\ref{fig:lvlv_abs_limits_cps} shows the 95\% CL upper
limits on $\sigma_H\times\mathrm{BR}(H\rightarrow WW)$ as a function of $\mH$ 
for a Higgs boson in the CPS scenario, separately for ggF and VBF production, in the mass
range $220\GeV \le \mH \le 1000\GeV$. 
Figure~\ref{fig:lvlv_abs_limits_nwa} shows the upper limits on a
Higgs boson with a narrow width in the range 
$300\GeV \le \mH \le 1500\GeV$, separately for ggF and VBF production. 
Below $300\GeV$, the limits in the NWA scenario are expected to be
similar to the CPS scenario, as the width is small enough in the latter 
case to have a negligible effect.

The systematic uncertainties with the largest effect on the observed limits at $\mH=300\GeV$,
in approximate order of importance, are those related to the modelling of the $WW$ background, 
the $b$-jet tagging efficiency, the jet energy scale and resolution, the top-quark background modelling 
in the $\ZeroJet$ category, the QCD scale uncertainties on the signal from the exclusive jet multiplicity
categories, and the jet energy scale and resolution.  As the mass hypothesis increases,
the experimental systematic uncertainties diminish in relative importance and the 
leading sources of uncertainty are on the signal cross section in exclusive $\Njet$ categories 
and the $WW$ background model.  Also, for the CPS scenario, the uncertainty on the interference 
weighting becomes important at high $\mH$.  

The relative importance of systematic and statistical
uncertainties for the analysis can be illustrated by recalculating the limits with the systematic
uncertainties omitted.  For the CPS scenario, the ggF limits decrease by about 20\% for $\mH=300\GeV$ and
by about 10\% for $\mH=1000\GeV$. For high mass hypotheses, higher values of $\mT$, where there 
are fewer events, are implicitly tested. Similarly, there are 
fewer candidates in the $\TwoJet$ category, so that the limits are less sensitive to systematic uncertainties than
the corresponding ggF limits. For the CPS VBF limit, removing the systematic uncertainties has
about an 8\% effect on the observed limit at $\mH=300\GeV$ and a negligible effect at $\mH=1000\GeV$.

The deficit at $\mT \gtrsim 450\GeV$ observed in Figure~\ref{fig:SR_low} results 
in a stronger limit than that predicted for background-only in the 
ggF production mode in both signal models, although they are consistent
within the given uncertainties.  For signal, the relation
$\mT\lesssim\mH$ holds, so the observed limits are stronger than the 
expected ones above this threshold of about $450\GeV$.

\begin{figure}[t!]
\begin{center}
\includegraphics[width=0.49\textwidth]{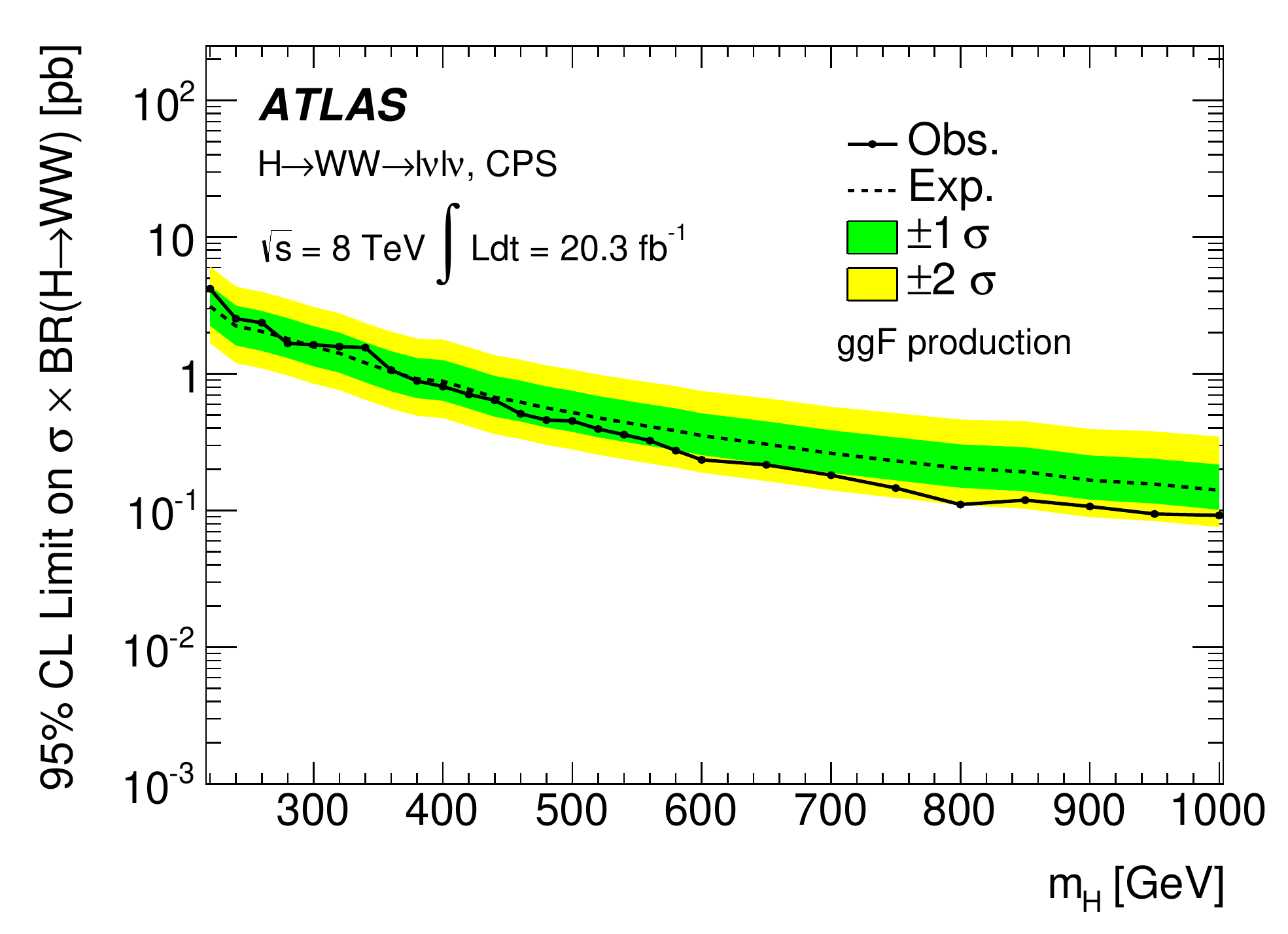}
\includegraphics[width=0.49\textwidth]{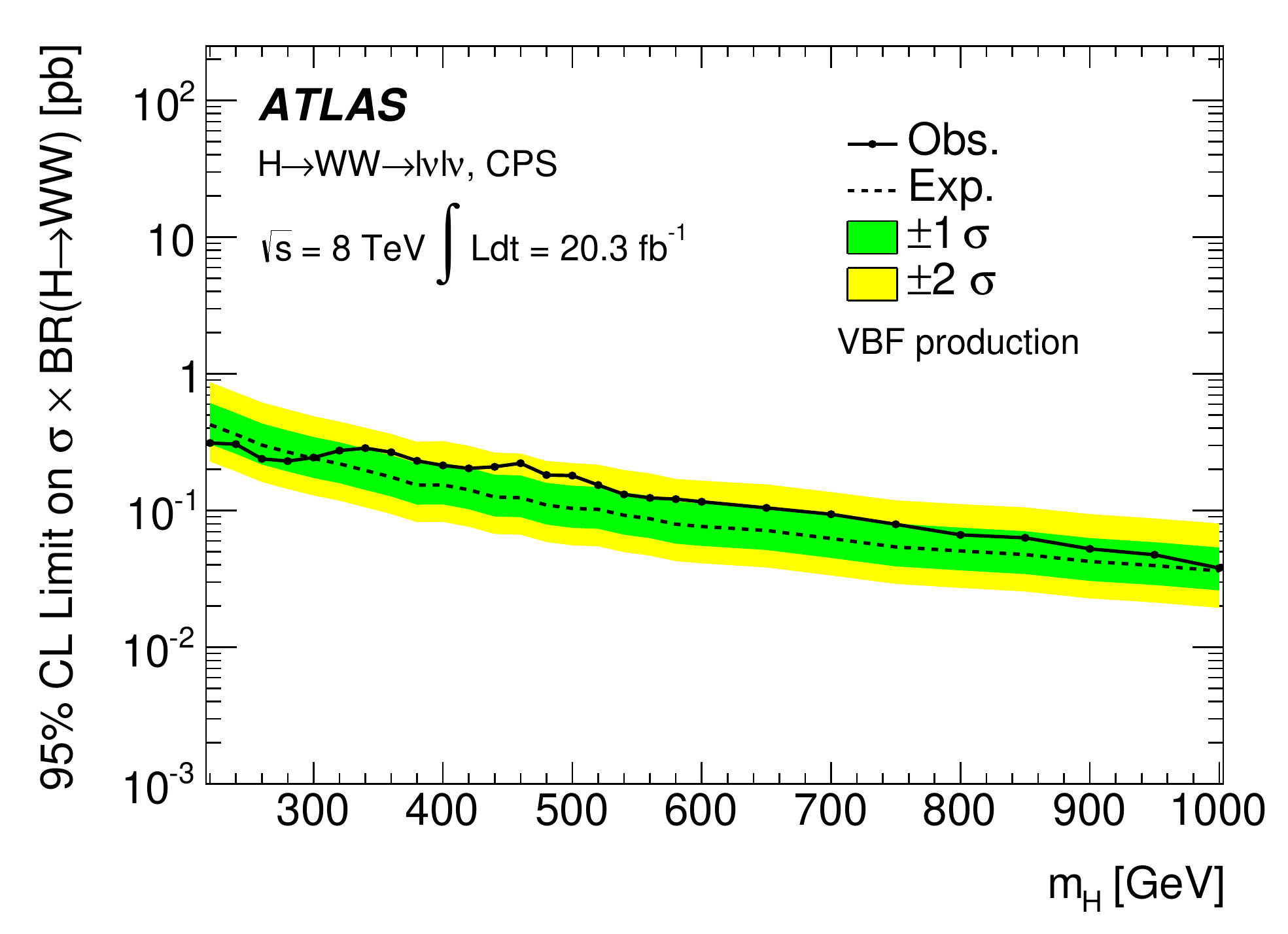}
\caption{95\% CL upper limits on $\sigma_H\times\mathrm{BR}(H\rightarrow
  WW)$ from the \hwwlnln\ analysis for the CPS scenario. Limits for
  ggF production (left) and VBF production (right) are shown. The
  green and yellow bands show the $\pm 1\sigma$ and $\pm 2\sigma$
  uncertainties on the expected limit.}
\label{fig:lvlv_abs_limits_cps}
\end{center}
\end{figure}

\begin{figure}[t!]
\begin{center}
\includegraphics[width=0.49\textwidth]{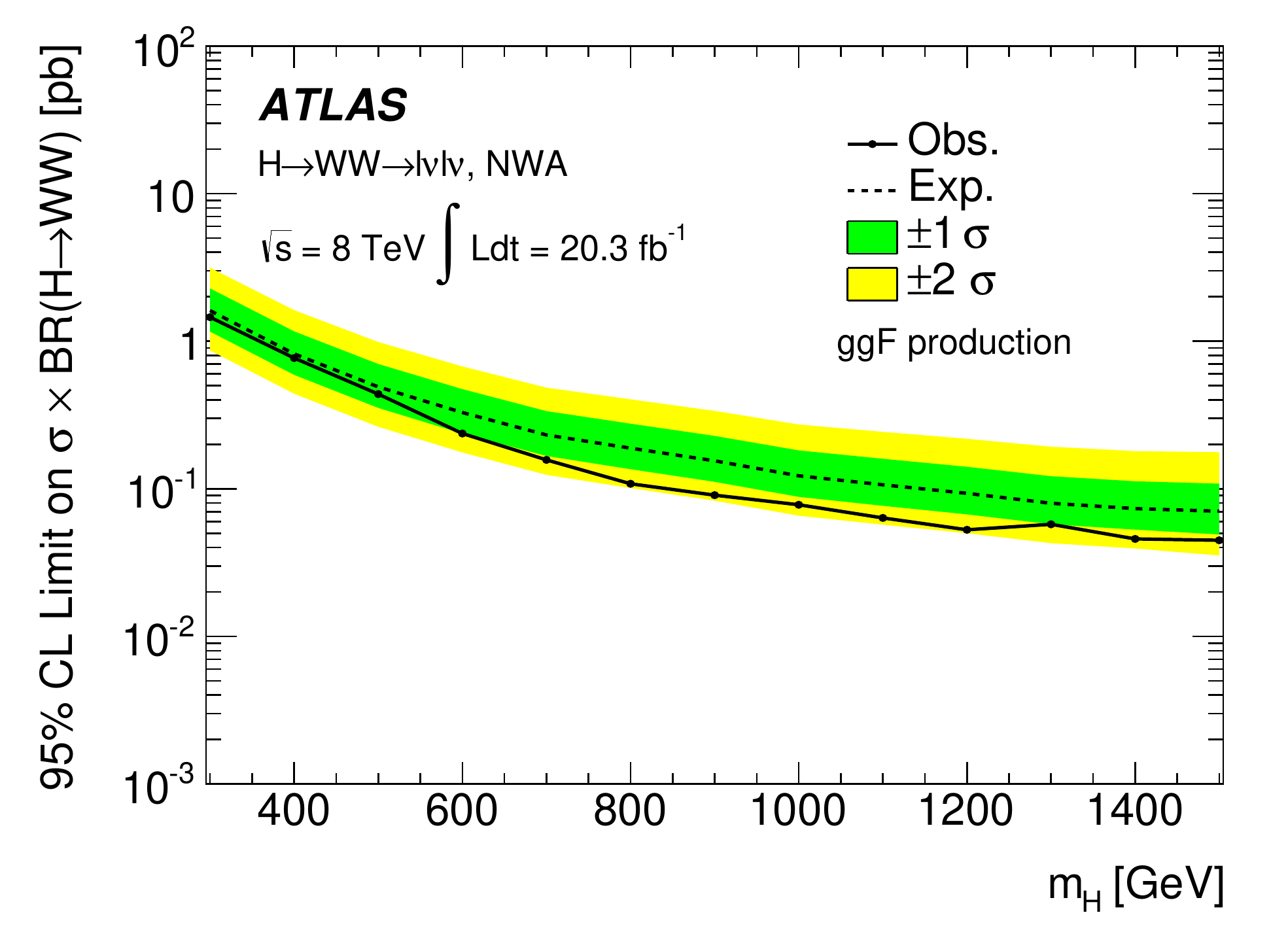}
\includegraphics[width=0.49\textwidth]{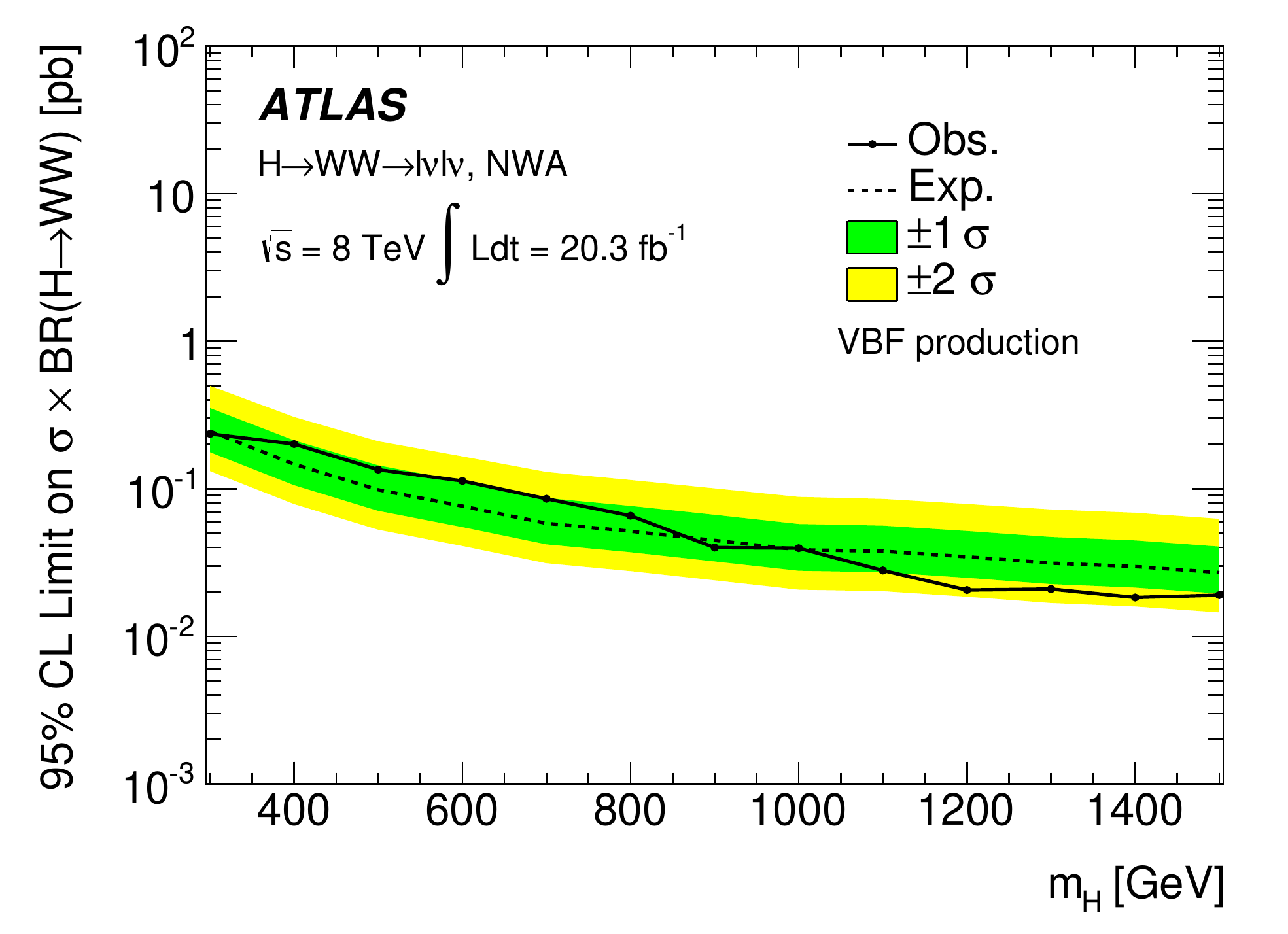}
\caption{95\% CL upper limits on $\sigma_H\times\mathrm{BR}(H\rightarrow
  WW)$ from the \hwwlnln\ analysis for a signal with a narrow width.
  Limits for ggF production (left) and VBF production
  (right) are shown. The green and yellow bands show the $\pm 1\sigma$
  and $\pm 2\sigma$ uncertainties on the expected limit.}
\label{fig:lvlv_abs_limits_nwa}
\end{center}
\end{figure}

\subsection{Upper limits from the $\HWWlvqq$ analysis}

Limits are derived following the same procedure as for the \hwwlnln\ channel.
Figure~\ref{fig:lvqq_abs_limits_cps} shows the 95\% CL upper
limits on $\sigma_H\times\mathrm{BR}(H\rightarrow WW)$ as a function of $\mH$ 
for the CPS scenario, separately for ggF and VBF
production, in the mass range $300\GeV \le \mH \le 1000\GeV$. 
The limits derived from the $\HWWlvqq$ analysis are comparable to 
those derived from the \hwwlnln\ analysis.
Figure~\ref{fig:lvqq_abs_limits_nwa} shows the upper
limits  on a Higgs boson with a narrow width in the range 
$300\GeV \le \mH \le 1500\GeV$, separately for ggF and VBF production. 

The systematic uncertainties with the largest effect on the observed limits at $\mH=300\GeV$
are those related to the jet energy scale and resolution, the multijet background estimation,
and the $b$-jet tagging, particularly the uncertainty on the rate for mistakenly
tagging a light-quark jet as a $b$-jet.  As the tested mass hypothesis increases, the
$\mlvjj$ shape uncertainties on the $\Wjets$ and top backgrounds become the leading sources
of uncertainty and the multijet background systematic uncertainties become negligible.  For the
CPS scenario, the uncertainty on the interference weighting also becomes a
dominant systematic uncertainty at high $\mH$.  

To show the overall effect of systematic uncertainties, the exercise done 
for the $\hwwlnln$ analysis is repeated.  If systematic uncertainties are omitted,
the observed ggF limits in the CPS scenario decrease by 66\% at $\mH=300\GeV$ and 40\% at $\mH=1000\GeV$.
The corresponding VBF limits decrease by about 40\% and 20\%, respectively.  The trends relative
to $\mH$ and ggF vs.~VBF are similar to what is seen in the dilepton final state, but
systematic uncertainties have a larger effect on the $\HWWlvqq$ limits than on the $\hwwlnln$ 
limits because the larger candidate event samples in the former analysis result in smaller statistical uncertainties.

The downward excursions of the observed limits compared to the expected ones seen 
for $\mH\gtrsim 600\GeV$ in Figure~\ref{fig:lvqq_abs_limits_cps}
in the ggF category and for $\mH\approx 750\GeV$ in both categories in Figure~\ref{fig:lvqq_abs_limits_nwa}
have been investigated and no underlying systematic effect identified.  
In particular, the simultaneous dip in the ggF 
and VBF NWA limits at $\mH\approx 750\GeV$ is attributable to a coincidence of 
deficits in the data in the statistically independent ggF and VBF SRs at that
value of $\mlvjj$.

\begin{figure}[t!]
\begin{center}
\includegraphics[width=0.49\textwidth]{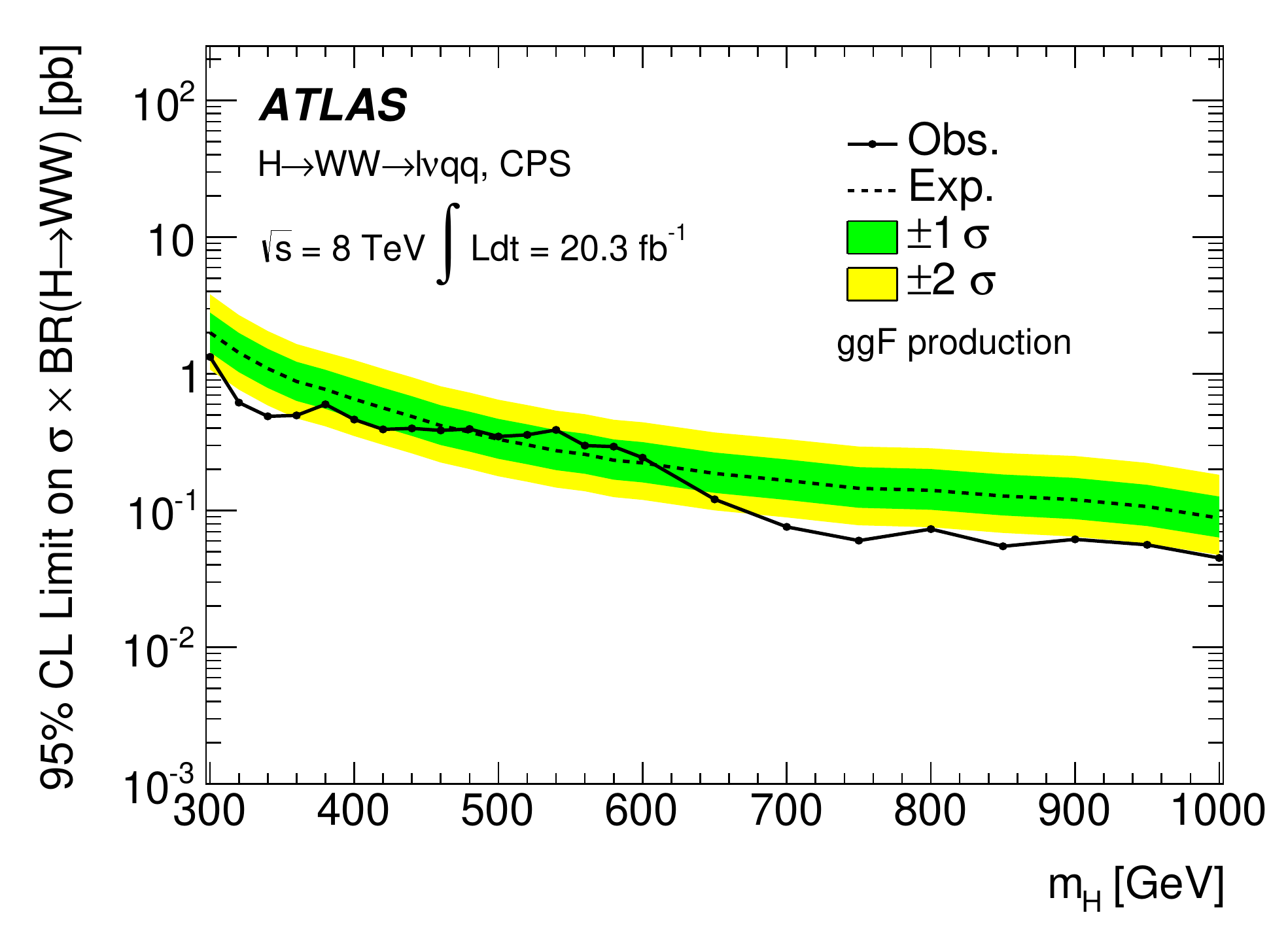}
\includegraphics[width=0.49\textwidth]{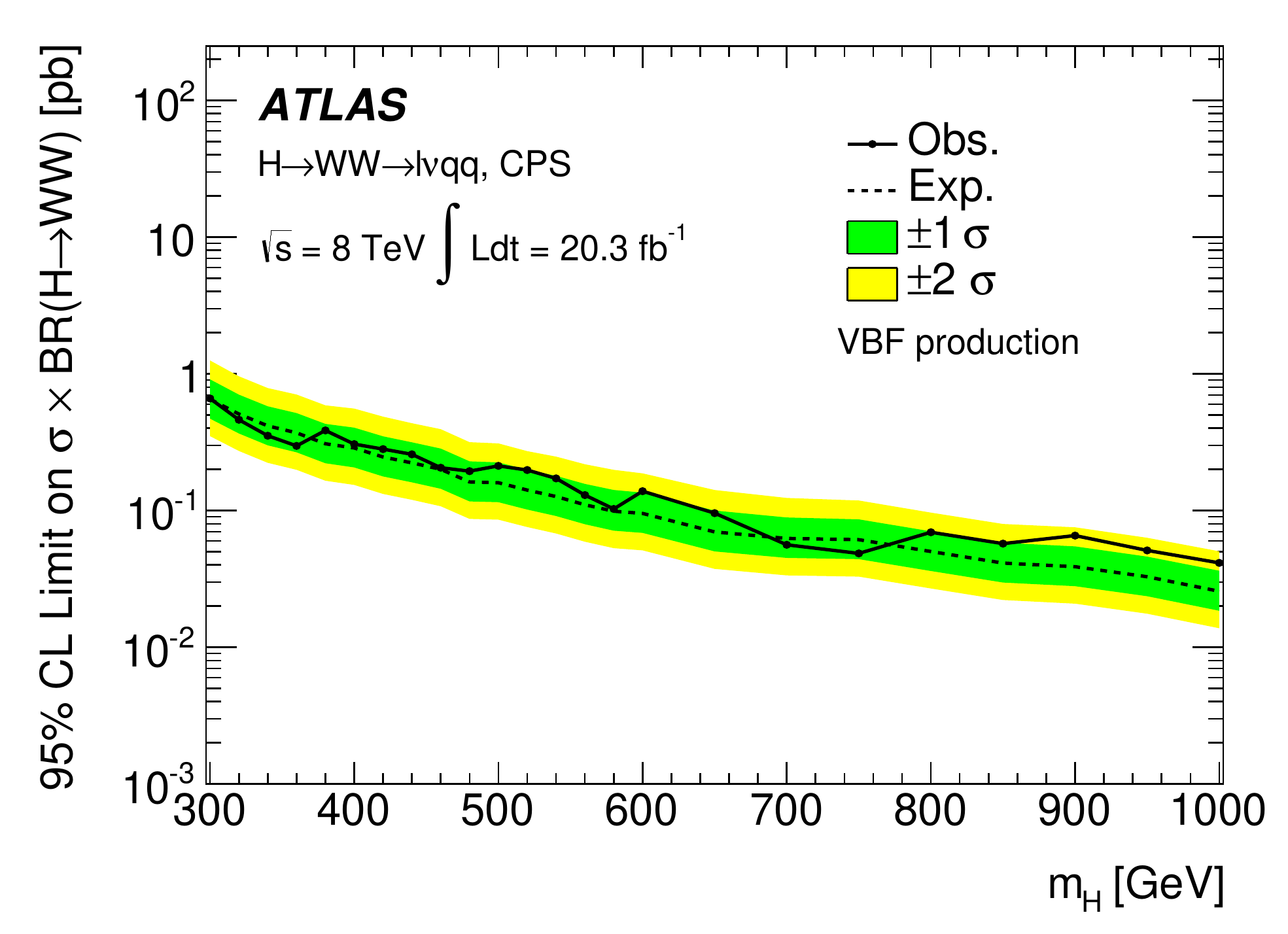}
\caption{95\% CL upper limits on $\sigma_H\times\mathrm{BR}(H\rightarrow
  WW)$ from the $\HWWlvqq$ analysis for the CPS scenario. Limits for
  ggF production (left) and VBF production (right) are shown. The
  green and yellow bands show the $\pm 1\sigma$ and $\pm 2\sigma$
  uncertainties on the expected limit.}
\label{fig:lvqq_abs_limits_cps}
\end{center}
\end{figure}

\begin{figure}[t!]
\begin{center}
\includegraphics[width=0.49\textwidth]{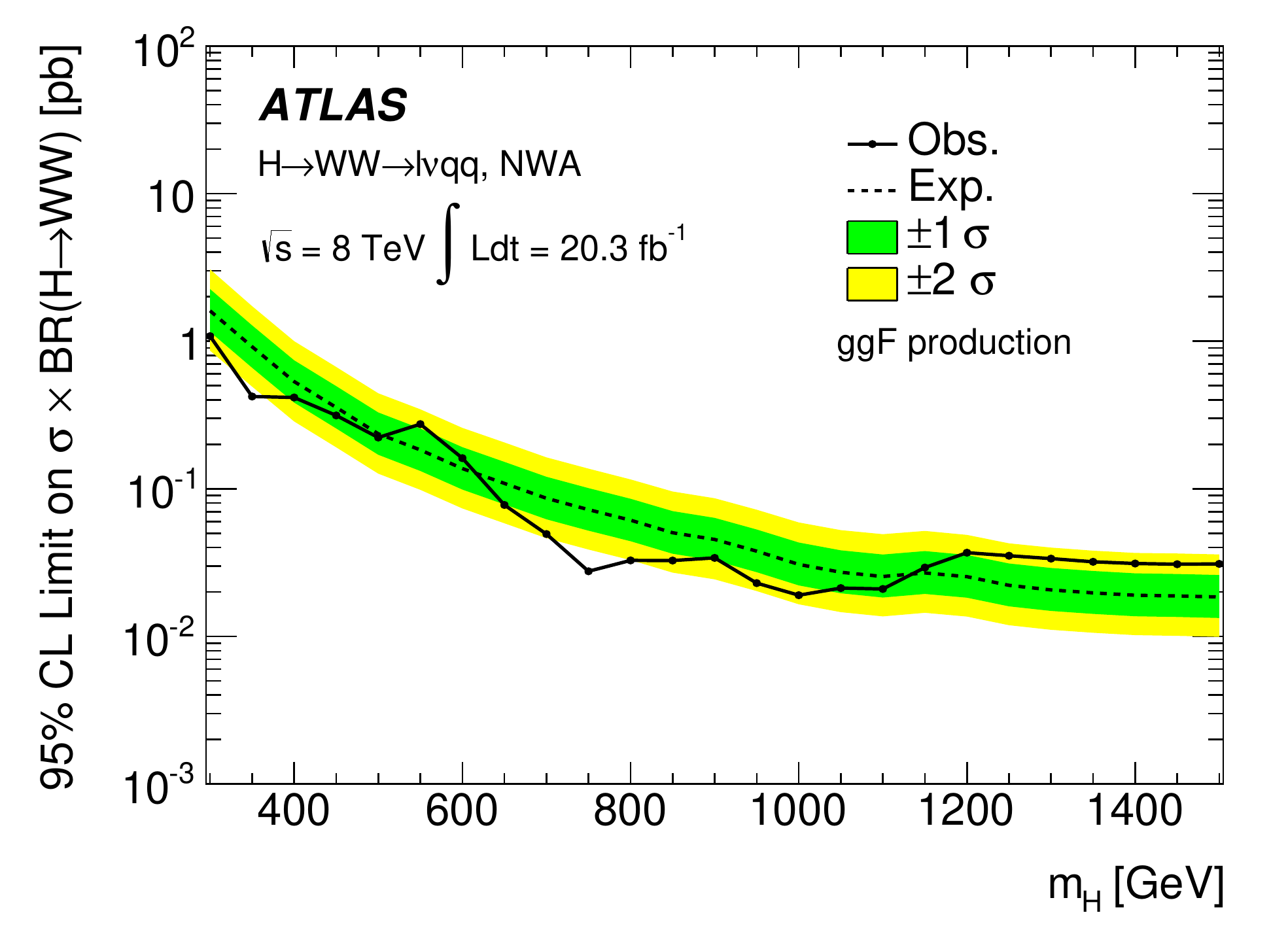}
\includegraphics[width=0.49\textwidth]{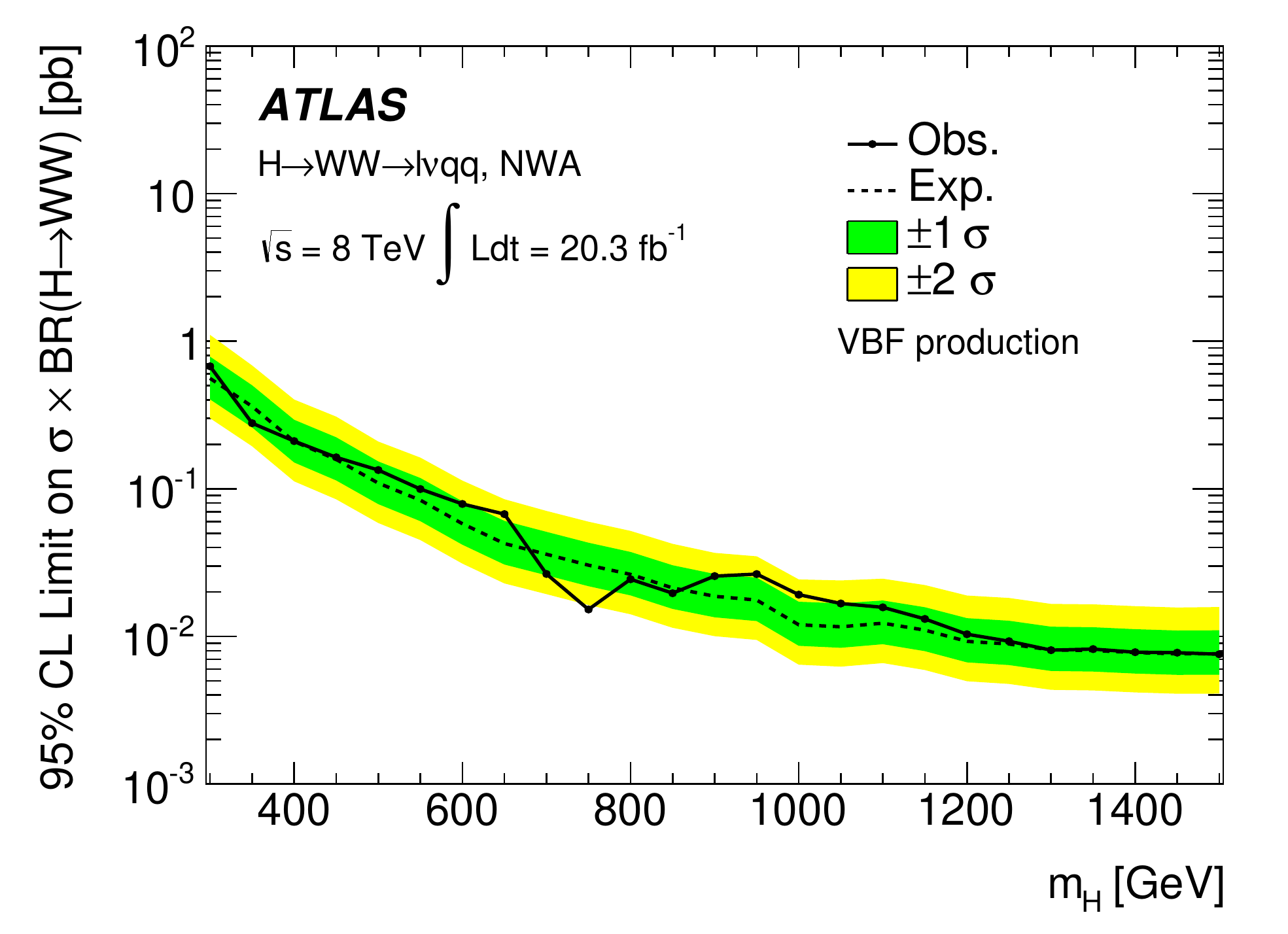}
\caption{95\% CL upper limits on $\sigma_H\times\mathrm{BR}(H\rightarrow
  WW)$ from the $\HWWlvqq$ analysis for a signal with a narrow width.
  Limits for ggF production (left) and VBF production
  (right) are shown. The green and yellow bands show the $\pm 1\sigma$
  and $\pm 2\sigma$ uncertainties on the expected limit.}
\label{fig:lvqq_abs_limits_nwa}
\end{center}
\end{figure}

\subsection{Combined upper limits}
\label{sec:comb}

This section presents 95\% CL upper limits on the production of high-mass Higgs bosons 
in the CPS and NWA scenarios from a combination of the
\hwwlnln\ and \hwwlnqq\ final states. In the statistical combination, the
likelihood function is constructed from
the signal and background probability density functions
from the two analyses. The combination
takes into account all statistical and systematic uncertainties in
both analyses. In particular, correlated effects of given sources of
systematic uncertainties in the two final states are taken into
account correctly. These correlated effects arise from sources of
uncertainty common to the final states, for example, those related to
detector response affecting the reconstruction, identification and
calibration of electrons, muons, jets, \MET\ and $b$-tagging, as well as
the integrated luminosity. Systematic uncertainties that affect both
final states are correlated in the combination unless there is a
specific reason not to correlate them.

Since the \hwwlnqq\ analysis sets upper limits starting at a Higgs boson
mass hypothesis of 300 GeV, the combination is performed starting at
$m_H = 300\GeV$. In the
CPS scenario, the upper range of the combination is
$m_H = 1000\GeV$ since neither analysis performs the search above this
mass because of the large width. In the NWA case, the upper range of the combination extends to
$m_H = 1500\GeV$.

Figure~\ref{fig:comb_abs_limits_cps} shows combined upper limits
separately on the ggF and VBF production modes for a Higgs boson in the CPS scenario. As 
in the case of the \hwwlnln\ and \hwwlnqq\ final states, when expected limits 
on a given production mode are extracted, the cross section of the other
production mode is set to zero, while for deriving observed limits on
the production mode, the cross section of the other mode is profiled
using data. 

\begin{figure}[t!]
\begin{center}
\includegraphics[width=0.49\textwidth]{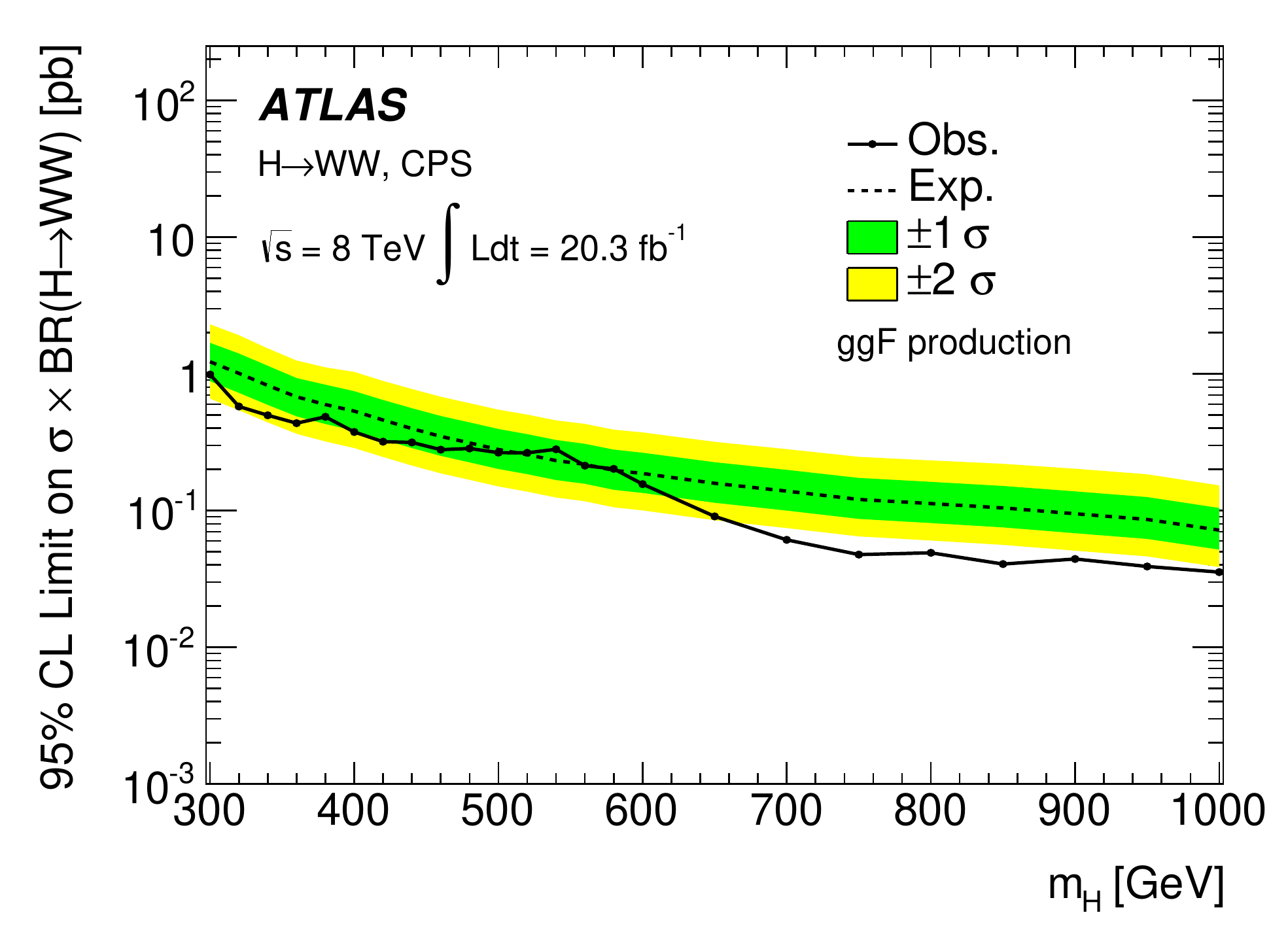}
\includegraphics[width=0.49\textwidth]{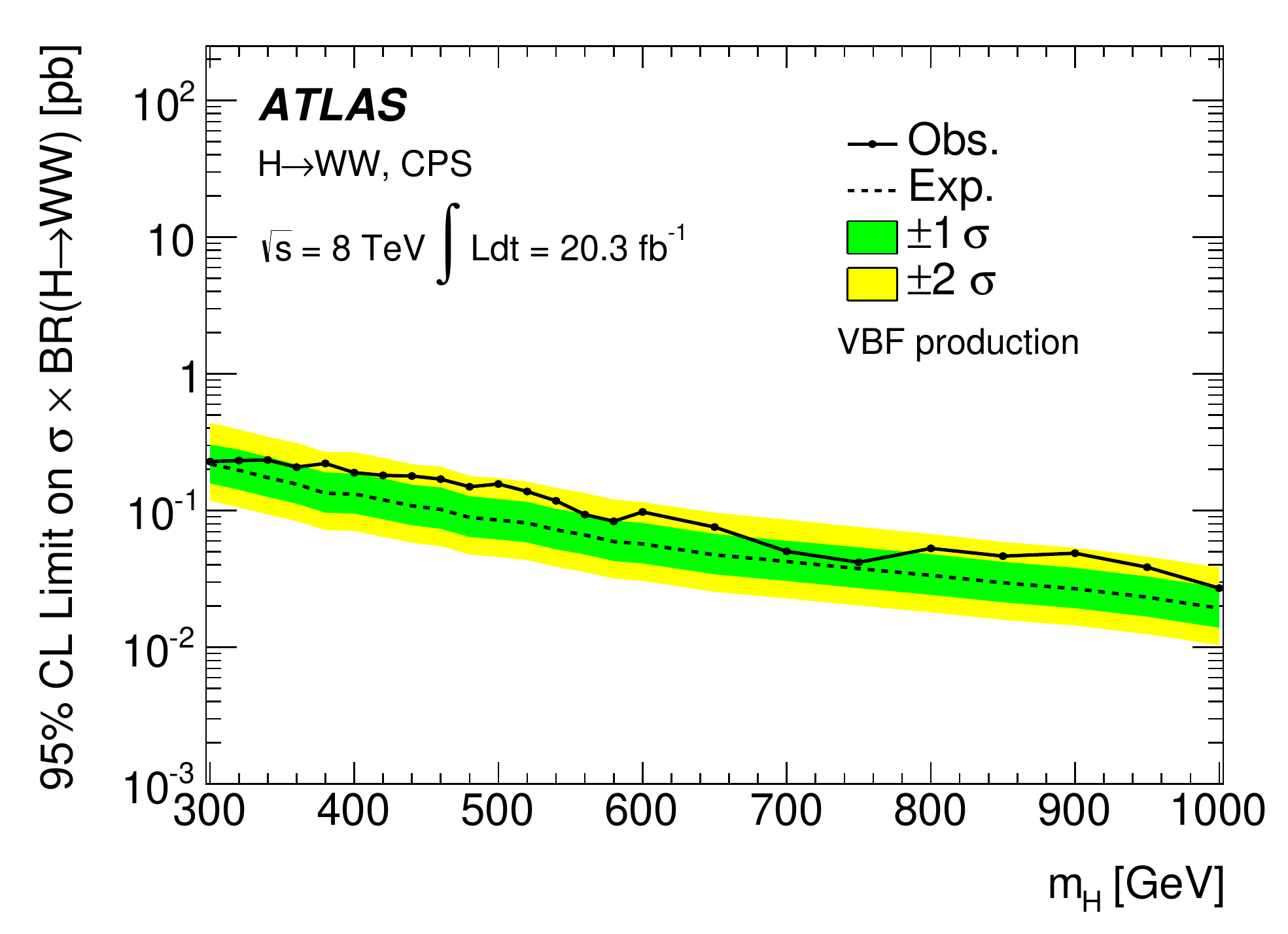}
\caption{95\% CL upper limits on $\sigma_H\times\mathrm{BR}(H\rightarrow
  WW)$ for the CPS scenario from the combination of the \hwwlnln\ and
  \hwwlnqq\ final states. Limits for ggF production (left) and VBF
  production (right) are shown. The green and yellow bands show the
  $\pm 1\sigma$ and $\pm 2\sigma$ uncertainties on the expected limit.}
\label{fig:comb_abs_limits_cps}
\end{center}
\end{figure}

Figure~\ref{fig:comb_abs_limits_nwa} shows the limits on
$\sigma_H\times\mathrm{BR}(H\rightarrow WW)$ as a function of $\mH$ for a
narrow-width Higgs boson, separately for the ggF and VBF production
modes. As in the CPS scenario, when observed limits on a given production mode
are extracted, the strength parameter of the other production mode is
profiled as a nuisance parameter in the fit.

\begin{figure}[t!]
\begin{center}
\includegraphics[width=0.49\textwidth]{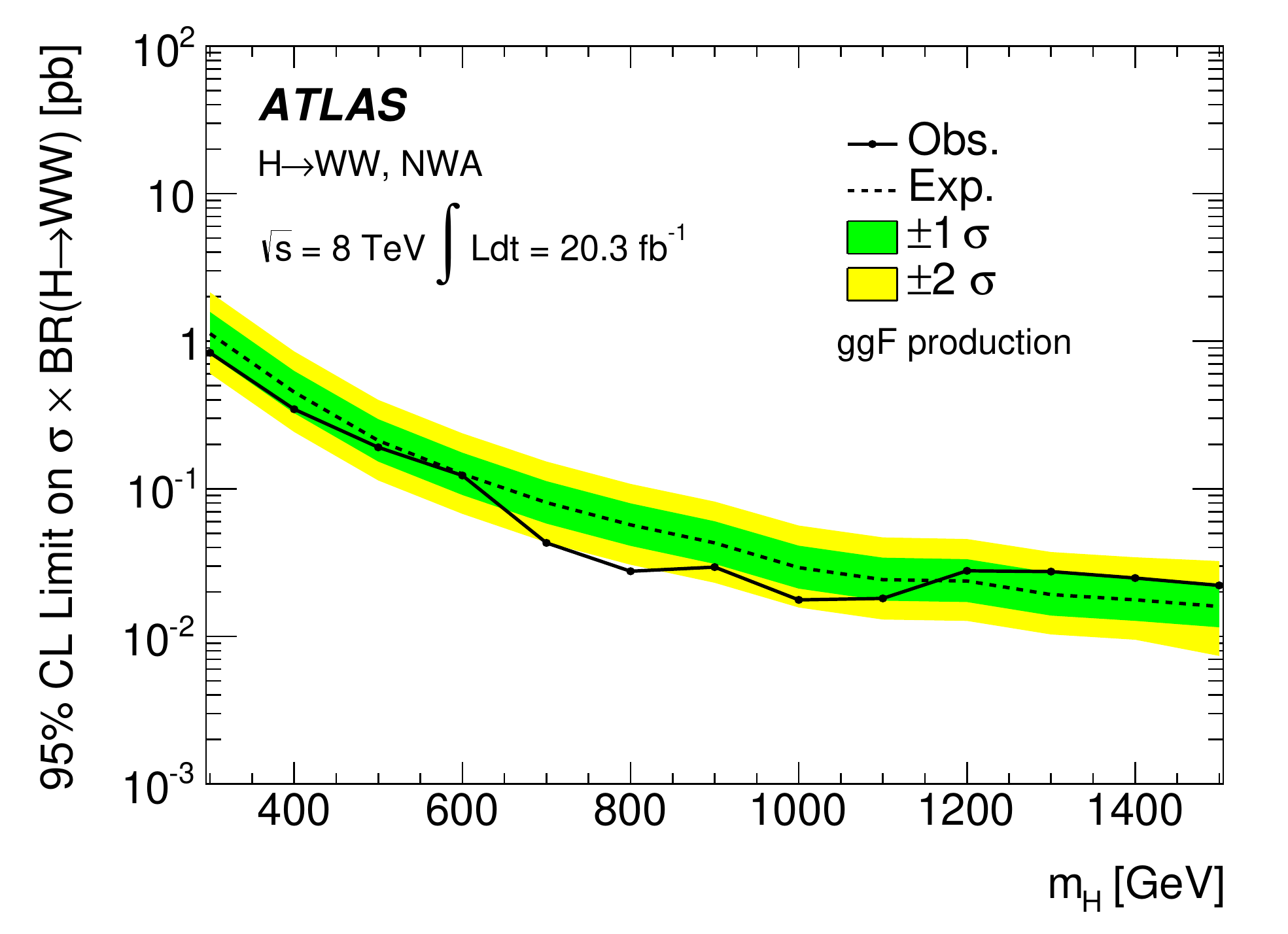}
\includegraphics[width=0.49\textwidth]{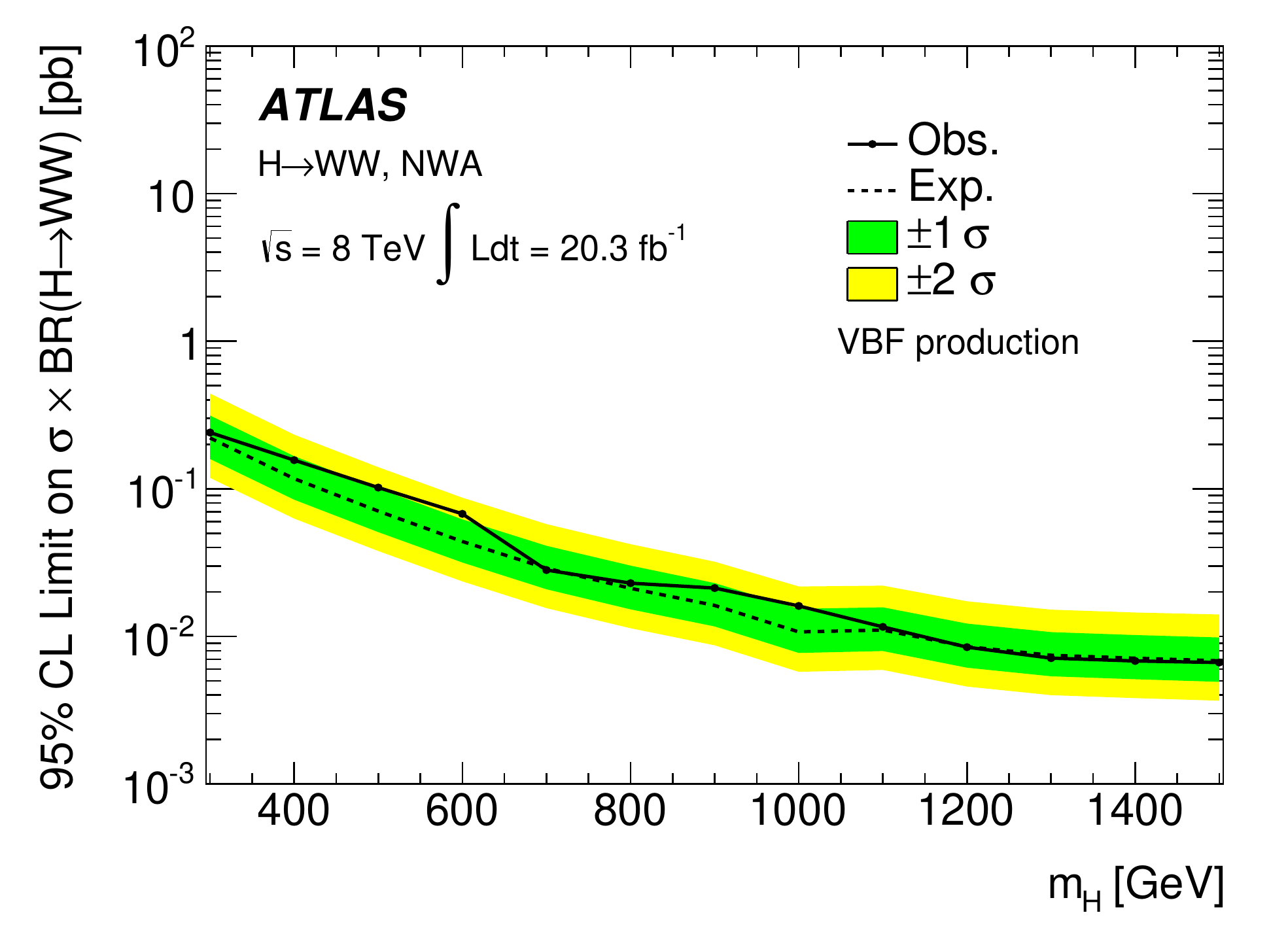}
\caption{95\% CL upper limits on $\sigma_H\times\mathrm{BR}(H\rightarrow
  WW)$ for a signal with a narrow width from the combination
  of the \hwwlnln\ and \hwwlnqq\ final states. Limits for ggF
  production (left) and VBF production (right) are shown. The green
  and yellow bands show the $\pm 1\sigma$ and $\pm 2\sigma$
  uncertainties on the expected limit. }
\label{fig:comb_abs_limits_nwa}
\end{center}
\end{figure}

\subsection{Results in the intermediate-width scenario}
\label{sec:IW}

The data can also be interpreted in terms of an additional Higgs boson 
with a width intermediate between the narrow-width approximation and the 
CPS scenario.  This interpretation is motivated by
the electroweak singlet (EWS) model, and assumes that the production cross sections and 
partial widths of the heavy Higgs boson are related to those of the SM Higgs boson 
by a single, constant scale factor $(\kappa')^2$.  This allows combination
of the ggF and VBF production modes as well as the two $WW$ decay channels considered here.
Non-SM decay modes, possible in the EWS model, are not considered in this analysis 
and the branching ratios of the heavy Higgs boson are the same as for a hypothetical
SM Higgs boson of the same mass.
The cross section, width and branching ratio of the heavy Higgs boson can be expressed as follows:
 \begin{equation}
 \begin{array}{lcl}
 \vspace{0.2cm}
    \sigma_{H}  & = & \kappa'^2\times \sigma_{H, \mathrm{SM}} \\
 \vspace{0.2cm}
   \Gamma_{H}  & = & \kappa'^2\times \Gamma_{H, \mathrm{SM}} \\
   \mathrm{BR}_{i} & = & \mathrm{BR}_{\mathrm{SM}, i} .
 \end{array}
 \label{eqn:EWSinglet_H}
 \end{equation}
 \noindent where $\sigma_{H, \mathrm{SM}}$, $\Gamma_{H, \mathrm{SM}}$, and $\mathrm{BR}_{\mathrm{SM}, i}$ are the cross section, total
 width and branching ratio to decay mode $i$ of a SM Higgs boson with mass $m_{H}$, respectively.
The parameters of a true electroweak singlet model are substantially constrained 
by measurements of the Higgs boson at $\mH\approx 125\GeV$~\cite{atlas:combined-paper-run1}.
The treatment described here allows a greater spectrum of possible widths to be explored.

Figure~\ref{fig:ews_lim_vs_mh} shows upper limits in the $\hwwlnln$ channel as a function of
$m_H$ in the intermediate-width scenario for widths in the range 
$0.2\Gamma_{H, \mathrm{SM}} \le \Gamma_{H} \le 0.8\Gamma_{H, \mathrm{SM}}$.  
Limits are shown on $\sigma_H\times\mathrm{BR}(H\rightarrow WW)$ divided by 
$\kappa'^2$ to facilitate readability, since otherwise the limit curves 
corresponding to the various $\kappa'^2$ values approximately coincide. This feature
indicates that the \hwwlnln\ channel has little sensitivity to the width of the
resonance.  Similarly, Figure~\ref{fig:lvqq_ews_lim_vs_mh} shows corresponding limits for the
$\HWWlvqq$ analysis, and Figure~\ref{fig:comb_limits_ews} shows the combination of the
$\hwwlnln$ and $\HWWlvqq$ analyses.

\begin{figure}[t!]
\begin{center}
\includegraphics[width=0.65\textwidth]{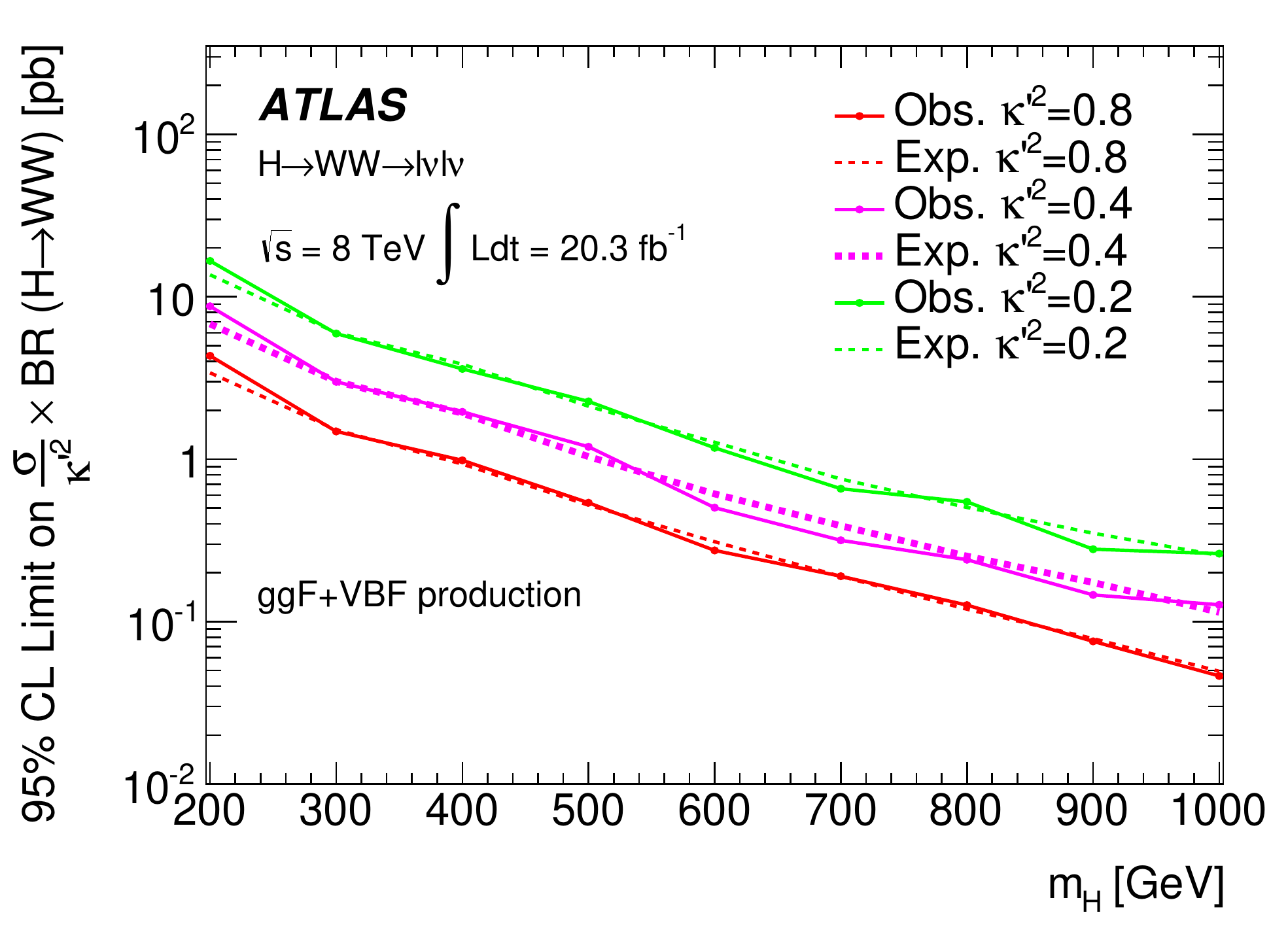}
\caption{95\% CL upper limits in the intermediate-width scenario on $\frac{\sigma_H}{\kappa'^{2}}\times\mathrm{BR}
  (H\rightarrow WW)$ from the \hwwlnln\ analysis for a heavy scalar resonance 
  with width in the range $0.2\Gamma_{H, \mathrm{SM}} \le \Gamma_{H} \le
  0.8\Gamma_{H, \mathrm{SM}}$. 
  The ggF and VBF production modes have been combined.
}
\label{fig:ews_lim_vs_mh}
\end{center}
\end{figure}

\begin{figure}[t!]
\begin{center}
\includegraphics[width=0.65\textwidth]{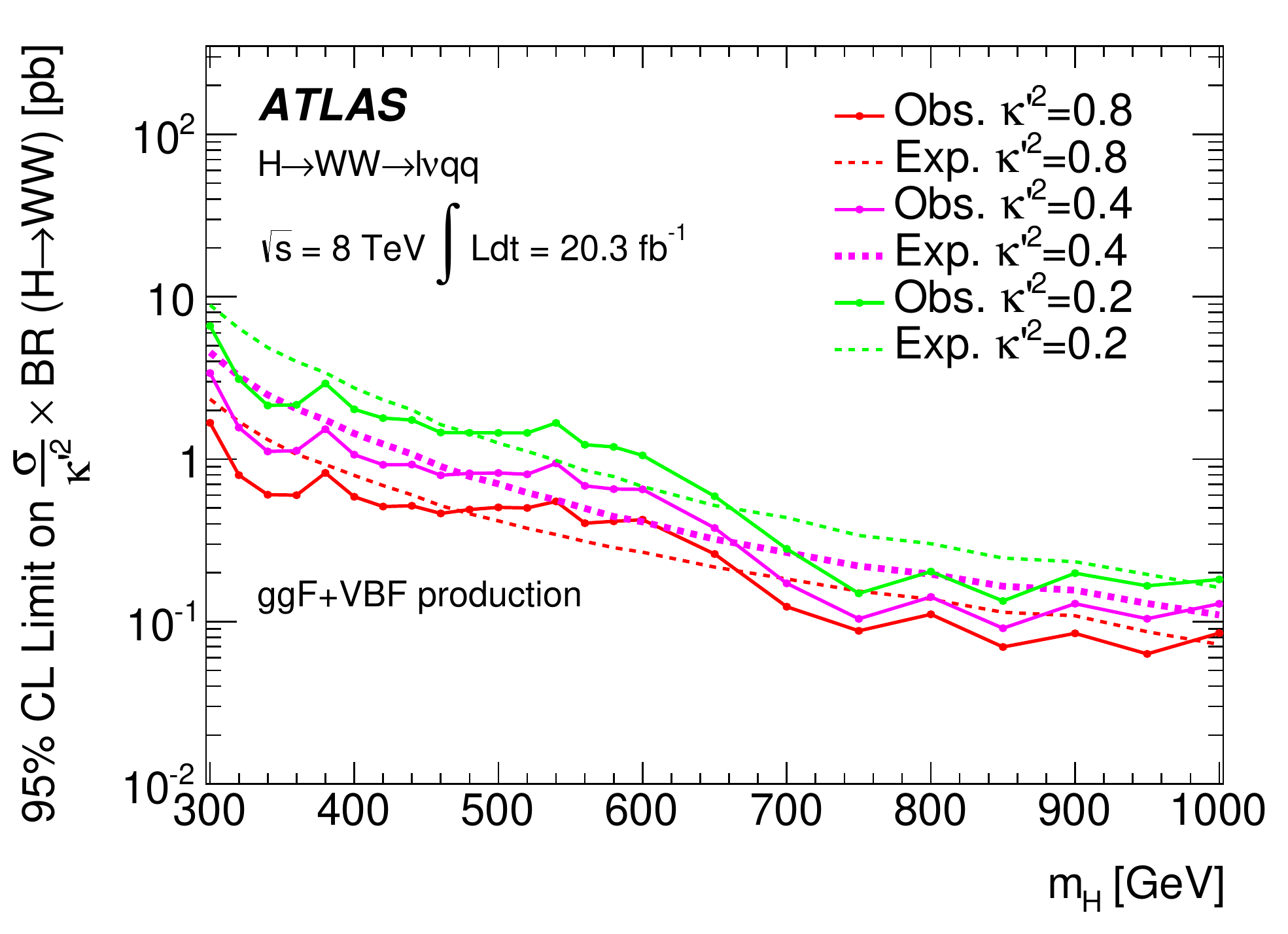}
\caption{95\% CL upper limits in the intermediate-width scenario on $\frac{\sigma_H}{\kappa'^{2}}\times\mathrm{BR}
  (H\rightarrow WW)$ from the \hwwlnqq\ analysis for a heavy scalar resonance 
  with width in the range $0.2\Gamma_{H, \mathrm{SM}} \le \Gamma_{H} \le
  0.8 \Gamma_{H, \mathrm{SM}}$. 
  The ggF and VBF production modes have been combined.}
\label{fig:lvqq_ews_lim_vs_mh}
\end{center}
\end{figure}

\begin{figure}[t!]
\begin{center}
\includegraphics[width=0.65\textwidth]{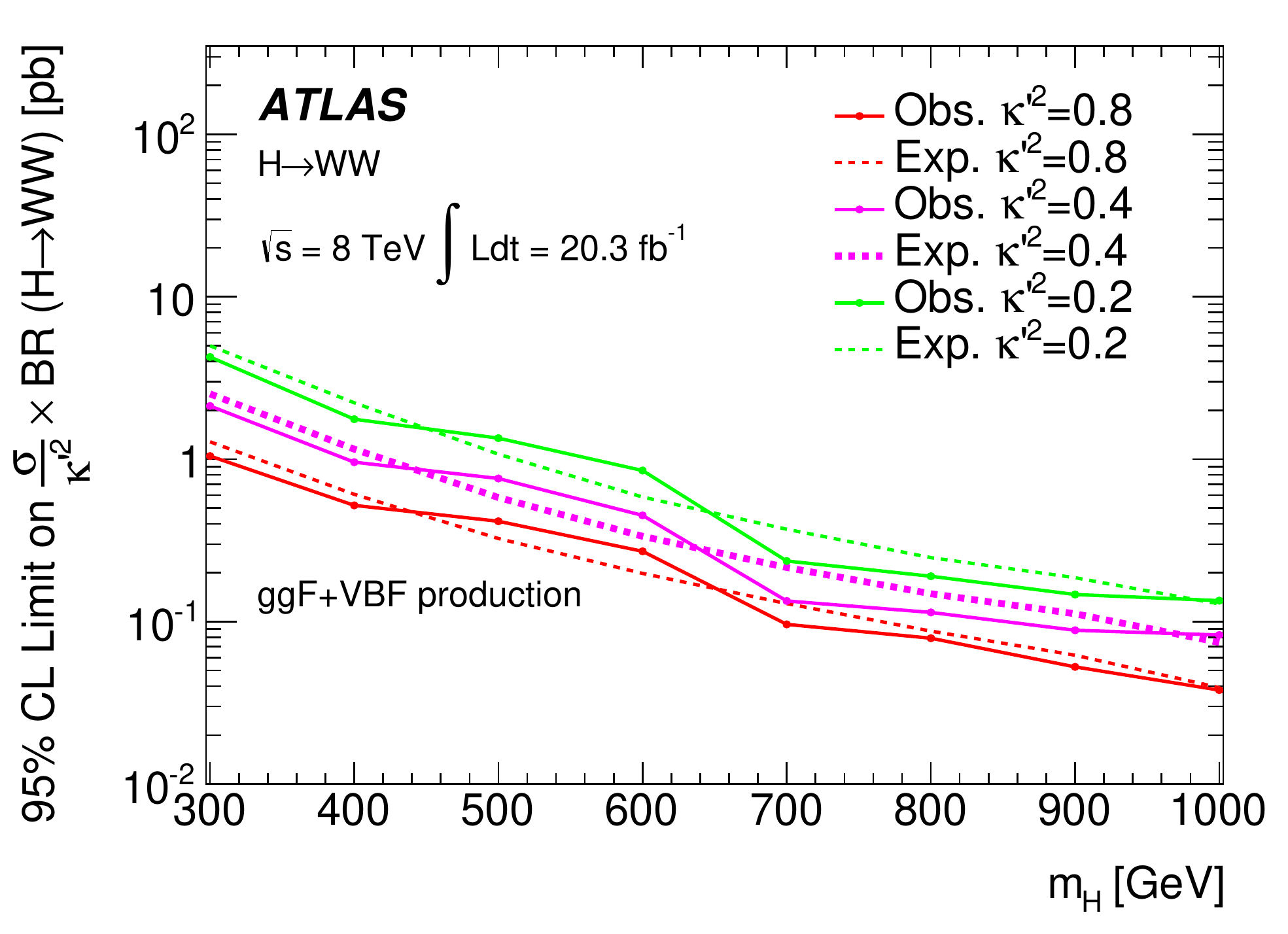}
\caption{95\% CL upper limits in the intermediate-width scenario from the combination
  of the \hwwlnln\ and \hwwlnqq\ final states. Limits are set on
  $\frac{\sigma_H}{\kappa'^{2}}\times\mathrm{BR}(H\rightarrow WW)$ 
  for a heavy scalar resonance with a width in the range $0.2\Gamma_{H, \mathrm{SM}}
  \le \Gamma_{H} \le 0.8\Gamma_{H, \mathrm{SM}}$. 
    The ggF and VBF production modes have been combined.}
\label{fig:comb_limits_ews}
\end{center}
\end{figure}

\FloatBarrier

\section{Conclusion}
\label{sec:conc}

  A search for a high-mass Higgs boson in the \hwwlnln{} and \hwwlnqq{} channels by the ATLAS experiment at the LHC is
  reported. The search uses proton--proton collision data at a centre-of-mass energy of 8 TeV corresponding to an integrated luminosity of
  \lumi\ fb$^{-1}$. No excess of events beyond the Standard Model background prediction is found. Upper limits are set on the product of
  the production cross section and the $H \rightarrow WW$ branching ratio in three different scenarios: a high-mass Higgs boson with
  a CPS lineshape and the width predicted for a SM Higgs boson, one with a narrow width, and one with an intermediate
  width. 

  For all mass hypotheses tested, the strongest upper limits are observed for the narrow-width scenario. 
  At $\mH=300\GeV$, these are $\sigma_H\times\mathrm{BR}(H\rightarrow WW) < 830$ fb at $95\%$ CL for the gluon-fusion production mode
  and $\sigma_H\times\mathrm{BR}(H\rightarrow WW) < 240$ fb at $95\%$ CL for the vector-boson fusion production mode.
  For $\mH = 1500\GeV$, the highest mass-point tested, 
  the cross section times branching ratio 
  is constrained to be less than $22$~fb and~$6.6$ fb at the 95\% CL for the ggF and VBF production modes, respectively.

  The limits in the CPS and intermediate-width scenarios are qualitatively similar but somewhat weaker
  due to the increased resonance width, particularly for $\mH$ approaching $1\TeV$.
  For the CPS scenario, the combined 95\% CL upper limits on $\sigma_H\times\mathrm{BR}(H\rightarrow WW)$ for the
  ggF production mode range from
  $990$~fb at $\mH=300\GeV$ to $35$~fb at $\mH=1000\GeV$.  For the VBF production mode, the equivalent 
  values are $230$~fb and $27$~fb, respectively. 
  These results are a substantial improvement over the previous results from the  ATLAS experiment in terms of both
  the range of $\mH$ explored and the cross section times branching ratio values excluded.

% Acknowledgements for papers with collision data
% Version 23-Mar-2015

% Standard acknowledgements start here
%----------------------------------------------
We thank CERN for the very successful operation of the LHC, as well as the
support staff from our institutions without whom ATLAS could not be
operated efficiently.

We acknowledge the support of ANPCyT, Argentina; YerPhI, Armenia; ARC,
Australia; BMWFW and FWF, Austria; ANAS, Azerbaijan; SSTC, Belarus; CNPq and FAPESP,
Brazil; NSERC, NRC and CFI, Canada; CERN; CONICYT, Chile; CAS, MOST and NSFC,
China; COLCIENCIAS, Colombia; MSMT CR, MPO CR and VSC CR, Czech Republic;
DNRF, DNSRC and Lundbeck Foundation, Denmark; EPLANET, ERC and NSRF, European Union;
IN2P3-CNRS, CEA-DSM/IRFU, France; GNSF, Georgia; BMBF, DFG, HGF, MPG and AvH
Foundation, Germany; GSRT and NSRF, Greece; RGC, Hong Kong SAR, China; ISF, MINERVA, GIF, I-CORE and Benoziyo Center, Israel; INFN, Italy; MEXT and JSPS, Japan; CNRST, Morocco; FOM and NWO, Netherlands; BRF and RCN, Norway; MNiSW and NCN, Poland; GRICES and FCT, Portugal; MNE/IFA, Romania; MES of Russia and NRC KI, Russian Federation; JINR; MSTD,
Serbia; MSSR, Slovakia; ARRS and MIZ\v{S}, Slovenia; DST/NRF, South Africa;
MINECO, Spain; SRC and Wallenberg Foundation, Sweden; SER, SNSF and Cantons of
Bern and Geneva, Switzerland; NSC, Taiwan; TAEK, Turkey; STFC, the Royal
Society and Leverhulme Trust, United Kingdom; DOE and NSF, United States of
America.

The crucial computing support from all WLCG partners is acknowledged
gratefully, in particular from CERN and the ATLAS Tier-1 facilities at
TRIUMF (Canada), NDGF (Denmark, Norway, Sweden), CC-IN2P3 (France),
KIT/GridKA (Germany), INFN-CNAF (Italy), NL-T1 (Netherlands), PIC (Spain),
ASGC (Taiwan), RAL (UK) and BNL (USA) and in the Tier-2 facilities
worldwide.
%----------------------------------------------

\clearpage
%zz\bibliographystyle{atlasBibStyleWithTitle}
% standard bibtex bibliography
%zz\bibliography{HWWHighMassPaper_2014}
\printbibliography

\clearpage
\appendix

\newpage 

% ATLAS Collaboration author list
% Data extracted on 11-Aug-2015 for paper reference HIGG-2013-19
%\documentclass[11pt]{article}
%\usepackage{a4wide}\begin{document}
\begin{flushleft}
{\Large The ATLAS Collaboration}

\bigskip

G.~Aad$^{\rm 85}$,
B.~Abbott$^{\rm 113}$,
J.~Abdallah$^{\rm 151}$,
O.~Abdinov$^{\rm 11}$,
R.~Aben$^{\rm 107}$,
M.~Abolins$^{\rm 90}$,
O.S.~AbouZeid$^{\rm 158}$,
H.~Abramowicz$^{\rm 153}$,
H.~Abreu$^{\rm 152}$,
R.~Abreu$^{\rm 116}$
% ...
\bigskip
\\
$^{1}$ Department of Physics, University of Adelaide, Adelaide, Australia\\
% ...
$^{11}$ Institute of Physics, Azerbaijan Academy of Sciences, Baku, Azerbaijan\\
% ...
$^{85}$ CPPM, Aix-Marseille Universit{\'e} and CNRS/IN2P3, Marseille, France\\
% ...
$^{90}$ Department of Physics and Astronomy, Michigan State University, East Lansing MI, United States of America\\
% ...
$^{107}$ Nikhef National Institute for Subatomic Physics and University of Amsterdam, Amsterdam, Netherlands\\
% ...
$^{113}$ Homer L. Dodge Department of Physics and Astronomy, University of Oklahoma, Norman OK, United States of America\\
% ...
$^{116}$ Center for High Energy Physics, University of Oregon, Eugene OR, United States of America\\
% ...
$^{151}$ Institute of Physics, Academia Sinica, Taipei, Taiwan\\
$^{152}$ Department of Physics, Technion: Israel Institute of Technology, Haifa, Israel\\
$^{153}$ Raymond and Beverly Sackler School of Physics and Astronomy, Tel Aviv University, Tel Aviv, Israel\\
% ...
$^{158}$ Department of Physics, University of Toronto, Toronto ON, Canada\\
% ...
\end{flushleft}

%\end{document}
% Created with ./xml2latex.py

\end{document}
\bye